\documentclass[fleqn,usenatbib]{mnras}
\usepackage{natbib}
\usepackage{amsmath,amssymb}
\usepackage{ulem}
\usepackage{adjustbox}
\usepackage[version=3]{mhchem}
\usepackage{pdflscape}
\usepackage{subcaption}
\usepackage{hyperref}
\usepackage{deluxetable}
\usepackage{caption}
\usepackage{graphicx}
\usepackage[displaymath, mathlines]{lineno}
\usepackage[title]{appendix}
\usepackage{xcolor} 

\title[VLBI Images \& $\gamma$-Rays of V3890~Sgr]{Shocks in the Symbiotic Recurrent Nova V3890~Sgr:\\
VLBI Radio Imaging and {\it Fermi} GeV Gamma-Rays}
\author[I. Molina et al.]{Isabella Molina$^{1}$\thanks{E-mail: molinai1@msu.edu},
Peter Craig$^{1}$,
Rebecca Diesing$^{2,3}$,
Laura Chomiuk$^{1}$,
Justin D.\ Linford$^{4}$,
\newauthor
Brian D.\ Metzger$^{3,5}$,
Jun Yang$^6$,
Brandon Benavente$^{1}$,
Kim L. Page$^7$,
Kirill V.\ Sokolovsky$^{9}$,
\newauthor
Elias Aydi$^{8}$,
Amy J.\ Mioduszewski$^{4}$,
Koji Mukai$^{10,11}$,
Miriam M.\ Nyamai$^{12}$,
Michael P.\ Rupen$^{13}$,
\newauthor
J. L.\ Sokoloski$^{3}$, 
and
Montana N.\ Williams$^{14}$
\\
$^{1}$Center for Data Intensive and Time Domain Astronomy, Department of Physics and Astronomy, Michigan State University,\\
East Lansing, MI 48824, USA\\
$^{2}$School of Natural Sciences, Institute for Advanced Study, Princeton, NJ 08540, USA\\
$^{3}$Department of Physics and Columbia Astrophysics Laboratory, Columbia University, New York, NY 10027, USA\\
$^{4}$National Radio Astronomy Observatory, P.O. Box O, Socorro, NM 87801, USA\\
$^{5}$ Center for Computational Astrophysics, Flatiron Institute, 162 5th Ave, New York, NY 10010, USA\\
$^6$Department of Space, Earth and Environment, Chalmers University of Technology, Onsala Space Observatory, SE-43992 Onsala, Sweden\\
$^7$School of Physics and Astronomy, University of Leicester, Leicester LE1 7RH, UK\\
$^8$Department of Physics \& Astronomy, Texas Tech University, Box 41051, Lubbock, TX, 79409-1051, USA\\
$^{9}$Department of Astronomy, University of Illinois at Urbana-Champaign, 1002 W. Green Street, Urbana, IL 61801, US\\
$^{10}$Center for Space Science and Technology, University of Maryland Baltimore County, Baltimore, MD 21250, USA\\
$^{11}$CRESST and X-ray Astrophysics Laboratory, NASA/GSFC, Greenbelt MD 20771 USA\\
$^{12}$Joint Institute for VLBI ERIC (JIVE), Oude Hoogeveensedijk 4, 7991 PD,
Dwingeloo, The Netherlands\\
$^{13}$Herzberg Institute of Astrophysics, National Research Council of Canada, Penticton, BC V2A 6J9, Canada\\
$^{14}$Department of Physics, New Mexico Tech, 801 Leroy Pl., Socorro, NM 87801, USA}

\begin{document}
\label{firstpage}
\pagerange{\pageref{firstpage}--\pageref{lastpage}}
\maketitle
\begin{abstract}
We present very long baseline interferometric (VLBI) radio imaging and {\it Fermi}/LAT GeV $\gamma$-ray observations of the 2019 eruption of the symbiotic recurrent nova V3890~Sgr.
 The VLBI imaging spans 8 -- 51 days after eruption, synchronous with the detected $\gamma$-rays. 
VLBI imaging shows the eruption starts out asymmetric on day 8 with an eastern component brighter than a western component. 
By day 32 the blast is rather circularly symmetric, and on day 49, the nova shell is brighter along the north--south axis. 
This morphological evolution is explained by interaction with circumstellar material (CSM) comprised of a spherical wind plus an over-density in the orbital plane.
Comparing radio images to optical line widths gives an expansion parallax distance of 6.8~kpc. 
In the first 32 days or eruption, VLBI images capture $>$80 per cent of the integrated flux (as measured by the VLA), implying that synchrotron emission dominates.
A second peak in the VLA light curve is explained by an image on day 48 that reveals the nova shell surrounded by a diffuse halo, powered by synchrotron emission from particles that have diffused upstream of the shock. 
The $\gamma$-rays appear around optical maximum and remain detectable for 23 days; marginally significant $\gamma$-rays reappear around day 60, concurrent with the second radio peak. 
Modelling indicates radio and $\gamma$-ray emission arise in distinct shock regions: $\gamma$-rays from dense CSM in the orbital plane, radio from the more spherical CSM component.
X-ray observations constrain the spherical CSM density, which is higher than in other symbiotic recurrent novae.
Assuming equipartition, we estimate the fraction of the post-shock pressure in magnetic fields, $\epsilon_B  = 3 \times 10^{-4} - 2 \times 10^{-3}$.
\end{abstract}


\section{Introduction} \label{intro}

Novae are thermonuclear eruptions that occur on the surfaces of accreting white dwarfs in binary systems 
\citep{Gallagher&Starrfield78, Bode&Evans08, Chomiuk+21}. 
The white dwarf accumulates hydrogen rich material on its surface from the companion star, 
which may range from a main sequence star to a giant. As this material builds up, the bottom layer compresses, 
temperature increases and the rate of nuclear reactions increase, leading to a thermonuclear runaway, 
seen as an increase in luminosity \citep{Starrfield+16}. The eruption blows off the accreted envelope
mixed with some white dwarf material at velocities ranging from 
$500$-$10,000$\,km\,s$^{-1}$ \citep[e.g.,][]{Yaron+05,Anupama+13,Ozdonmez+18}. 
As the ejected mass is relatively small \citep[$10^{-3}$ to $10^{-7} M_\odot$;][]{1998PASP..110....3G,Yaron+05}, 
the binary system survives the eruption and accretion can continue, 
accumulating hydrogen-rich matter for the next nova eruption 
\citep{2014A&A...563L...9D,2025ApJ...994..229B}. 

In some cases, subsequent eruptions have been observed in the same system; these novae are called recurrent novae. 
The majority of novae have recurrence times longer than a human lifespan and have only had one observed eruption. 
However, 10 recurrent novae have been observed in the Milky Way \citep{Schaefer10,2021gacv.workE..44D}, with four of them being 
symbiotic binaries \citep{Kenyon86}, including V3890~Sgr. The prevalence of giant companions among recurrent novae 
is often explained by accretion fed by red giant winds reaching higher rates than accretion fed by a main sequence companion 
filling its Roche Lobe \citep{Chen+16, Kemp+21}.

\subsection{V3890~Sgr}
V3890~Sgr is a symbiotic recurrent nova \citep{Schaefer10} with an M5 III giant companion \citep{Harrison+93}. It has had three observed eruptions:  2019 August 27.87 \citep{Strader+19}, 1990 April 27.7 \citep{Buckley+90, Anupama_Sethi94} and 1962 June 2 \citep{Wenzel90, Miller91}. The recurrence time is short, only about 28 years between eruptions. It is possible that we have missed an eruption as the nova is behind the sun every December, and the optical light curve shows a rapid decline,  $t_3 = 14$ days \citep{Mikolajewska+21}. The distance is constrained to 6--9 kpc \citep{Mikolajewska+21}, and is discussed further in section \ref{sec:dist}. 
The mass of the white dwarf is estimated at $1.35\, M_{\odot}$, 
and the orbital period is $747.6$ d \citep{Mikolajewska+21}.

The most recent 2019 eruption of V3890~Sgr was observed in $\gamma$-ray, X-ray, UV, optical, IR, and radio wavelengths \citep{Buson+19,Nyamai+19, Polisensky+19,Orio+20, Evans+22,Kaminsky+22, Ness+22}. Here we are focused on high-resolution very long baseline interferometry (VLBI) radio observations and GeV $\gamma$-ray observations of the eruption. 

Radio observations of the 2019 eruption were reported by \cite{Nyamai+23}. Observations began about 7 days after discovery of the eruption and were done using MeerKAT and the Karl G.\ Jansky Very Large Array (VLA), with frequencies ranging from  $1 - 37$ GHz. The $5$ and $7$ GHz light curves peak on day $11.2$ at $\sim$50 mJy, while the $1.26-1.78$ GHz light curves peak around day $15$ in the range of $44.2$ -- $49.0$ mJy.   
\cite{Nyamai+23}  concluded that the radio light curve is dominated by synchrotron emission based on brightness temperature arguments. There is another increase in radio flux density on day $45$ seen in frequency bands $1.26$ to $7.0$ GHz. This second peak in the radio light curve is difficult to explain as thermal  emission (as in other novae; \citealt{Weston+16a, Finzell+18}). This second bump has been interpreted instead as synchrotron emission resulting from interaction with circumstellar material (CSM) with a complex structure. \cite{Nyamai+23} suggest that the CSM around the nova is complex in shape with perhaps a denser region at a distance from the white dwarf, $\sim 10^{15}$ cm. 

V3890~Sgr was also detected in GeV $\gamma$-rays by the Large Area Telescope (LAT) on the {\it Fermi Gamma Ray Space Telescope} shortly after the start of eruption \citep{Buson+19}. This makes it the third nova with a giant companion to be significantly detected by {\it Fermi}, joining the ranks of V407 Cyg \citep{Abdo+10} and RS Oph \citep{Cheung+22} (V745 Sco was also a marginal detection; \citealt{Franckowiak+18}).
According to this Astronomer's Telegram on V3890 Sgr, the {\it Fermi} onset began at the time of optical peak, and lasted 12 days.


\subsection{Shocks in Nova Eruptions}

This paper focuses on the shock signatures from the 2019 eruption of V3890~Sgr, in both radio and $\gamma$-ray bands. In the last decade, shocks in novae have been recognized as energetically important and key for shaping their electromagnetic emission \citep{Chomiuk+21}. To date, 26 Galactic novae have been detected in GeV $\gamma$-rays by {\it Fermi}/LAT, which signals that novae efficiently accelerate particles to relativistic speeds through diffusive shock acceleration \citep{Craig+26}. Most of these novae have main-sequence companions and are surrounded by low-density environments born of conservative mass transfer \citep{Ackermann+14}, so the shocks are internal to the nova ejecta \citep{Chomiuk+14}. However, a handful of {\it Fermi}-detected novae have giant companions, where mass transfer is expected to be messier and the environment is polluted by the red giant wind \citep{Abdo+10, Diesing+23}. In these systems, the shock is expected to be external with pre-existing CSM. This dichotomy is further supported by X-ray observations of novae; in {\it Fermi}-detected novae with main-sequence companions, hot shocked gas is not detected contemporaneous with $\gamma$-rays despite numerous observations with {\it Swift}/XRT, implying high absorption columns \citep{Nelson+19, Gordon+21} or suppression of hot gas behind a radiative shock due to mixing with cooler gas \citep{Steinberg&Metzger18,Metzger+25,Mitrani+25}. On the other hand, novae with giant companions often show hot X-ray emission from shocked gas from the earliest phases of their eruption \citep{Sokoloski+06, Nelson+12, Page+15, Page+20}, implying significantly lower absorption columns.

The CSM is often modelled as a spherical wind with a $r^{-2}$ density profile, and an equatorial density enhancement (EDE) in the orbital plane, driven by gravitational focusing of the mass loss by the white dwarf \citep{Mohamed&Podsiadlowski12, Orlando+09, Orlando+17, Orlando+25}. This is the scenario we have in mind in studying V3890~Sgr's 2019 eruption: external shocks between the nova ejecta and CSM, and denser interaction in the orbital plane.

\subsection{This Paper}

Here we present high-resolution radio images and analysis of the 2019 eruption of V3890~Sgr using data obtained with the Very Long Baseline Array (VLBA) and the European VLBI Network (EVN) plus the 
Enhanced Multi-Element Remotely Linked Interferometer Network (e-MERLIN). We also analyse data from {\it Fermi}/LAT to characterize the GeV $\gamma$-ray light curve. With observations in both wavelengths we are able to compare results, 
investigate the efficiency of particle acceleration and discuss the evolution of the blast and the structure of the surrounding CSM. 

In \S \ref{sec:data}, we describe the data sets under consideration. Section \ref{sec:vlbi} contains analyses of the VLBA and EVN+e-MERLIN images, using them to test the origin of the radio emission, to measure the magnetic field strength, and to estimate an expansion parallax distance of 6.8 kpc. In \S \ref{sec:fermi}, we analyse the {\it Fermi} data, presenting the light curve over the first 75 days of eruption and demonstrating a curious-but-marginal reappearance of $\gamma$-rays around the time of the second radio bump. We discuss and model our findings in \S \ref{sec:model}, estimating the efficiency of magnetic field amplification and particle acceleration, describing efforts to simultaneously model the radio and $\gamma$-ray emission, and exploring the origin of the second bump in the radio and $\gamma$-ray light curves. We conclude in \S \ref{sec:concl}.

\begin{table*}
    \centering
    \renewcommand{\arraystretch}{1.1}
    \caption{VLBI and Corresponding VLA Observations of V3890~Sgr \label{tab:radio}}
    \begin{tabular}{|l|l|l|l|l|l|l|l|l|}
    \hline
    \hline
        UT Date & $t-t_0$\tablenotemark{a} &Freq &Flux $\pm$  Error (VLA) & Flux $\pm$ Error (VLBI)\tablenotemark{b} &Major Axis\tablenotemark{c} & Minor Axis\tablenotemark{d} & RMS noise & Panel\tablenotemark{e} \\
       & (days) & (GHz) & (mJy) & (mJy) & (mas) & (mas) & ($\mu$Jy bm$^{-1}$) & \\
        \hline
        2019 Sep 5.0& 8.1 & 4.87&$43.5\pm 2.2$ & $37.75 \pm 4.17$  &  &  & 115 & (a)\\ 
        2019 Sep 5.0 & 8.1 & 8.37& $44.6\pm 2.2$ &$36.91\pm2.37$  & $8.4$& $8.0$ & 82 & (b)\\
        2019 Sep 13.0 & 16.1 & 4.87& $44.0\pm 2.2$ &$36.42\pm2.02$ &  &   & 116 & (c)\\
        2019 Sep 12.9 & 16.0 & 8.37& $39.9\pm 2.0 $ &$32.68\pm1.75$& $13.7$& $13.3$ & 58 & (d) \\
        2019 Sep 28.9 & 32.0 & 4.87& $24.8\pm 1.2$ &$24. 61\pm 1.37$& $22.5$& $21.5$ & 60 & (e)\\
        2019 Oct 15.6 & 48.7 & 4.93 & $17.3\pm 0.9$ & $12.44 \pm 0.64$\tablenotemark{f} & 46.0 & 43.2 & 31 & (g)\\
        2019 Oct 15.6 & 48.7 & 4.93 & $17.3\pm 0.9$ & $17.15 \pm 0.94$\tablenotemark{g} & 157 & 96 & 31 & (h)\\
        2019 Oct 17.9 & 51.0 & 4.87& $17.3\pm 0.9$ &$14.00\pm0.80$ & $34.0$& $31.9$ & 22 & (f) \\
        \hline
    \end{tabular}
\tablenotetext{a}{We take the start of eruption $t_0$ to be 2019 August 27.9 UT.}
\tablenotetext{b}{All VLBI images were obtained with the VLBA, except the 2019 Oct 15 epoch, which was obtained with the EVN+e-MERLIN.}
\tablenotetext{c}{Oriented east--west in all images, except the diffuse flux on 2019 Oct 15, when the major axis has a PA = 50$^{\circ}$.}
\tablenotetext{d}{Oriented north--south in images,  except the diffuse flux on 2019 Oct 15.}
\tablenotetext{e}{Images are shown in these corresponding panels of Figure \ref{fig:bigcomp}.}
\tablenotetext{f}{Measurements for the bright nova shell, equivalent to what is captured in VLBA images.}
\tablenotetext{g}{Measurements including the diffuse emission surrounding the nova shell.}
\end{table*}

\section{Observations and Data Reduction} \label{sec:data}

\subsection{VLBA Radio Observations of the 2019 Eruption}

High-resolution imaging observations of V3890~Sgr with the VLBA started on 2019 September 5.0  (8.1 days after start of eruption) and ended 2020 February 8 (164 days after eruption). Nine observations were obtained under VLBA project code BL267 (PI Linford).  For observations taken on 2019 September 28 and October 17  in X band (center frequency $8.37$ GHz) and February 8 in C band (center frequency $4.87$ GHz), we were unable to properly calibrate the data. The phases were scattered and the check source when imaged was not a point source indicating that there was an issue with the calibration, perhaps due to the southerly declination of the source and resulting low elevation of observations. We made several attempts to calibrate the data but unfortunately were unable to get usable results. Because of this we decided to cut out these data and do not include them in any of the following plots, images, or analysis. The six observations that yielded good images are listed in Table \ref{tab:radio}. VLA observations, which are useful for determining V3890~Sgr's integrated flux density, began on August 30th \citep{Nyamai+23}. For each VLBA observation there were near-contemporaneous VLA observations at 5 GHz and 7 GHz. 

The September 5 and September 13 observations included both $4.87$ GHz and $8.37$ GHz, with observations switching between bands on average every 42 minutes. The September 28 and October 17 observations only had usable data at $4.87$ GHz.  The phase reference calibrator was J1820$-$2528, and the check source was J1819$-$2036. For the final two observations a fringe finder (J1924$-$2914) was included. 
Each observation has 4 spectral windows, each 128 MHz wide, resulting in a total bandwidth of 512 MHz. 
There are 128 spectral channels per spectral window.  
All data were obtained in full Stokes mode. 
The exposure time of each epoch was about 1.8 hours on V3890~Sgr, 
except for the October 17th $4.87$ GHz observation which was 3.6 hours. 
On the September 5 observation, Mauna Kea did not join until about 2 hours after the start of the observation.  On the September 28 observation, Brewster was not recording data.


After flagging poor data and radio frequency interference on the calibrator scans, we noticed the phase calibrator J1820$-$2528 showed complex structure, so we split it into 
its own FITS file and loaded it into \textsc{difmap} \citep{Shepherd_Difmap_1997} in order to improve our model. There we iteratively self calibrated and imaged the calibrator data, recovering some diffuse flux surrounding a point-like source. This improved model was loaded back into \textsc{aips} and then used in the calibration process. 

The data were calibrated using standard routines in \textsc{AIPS} 31DEC22 \citep{Greisen03}. Instrumental delay calibration and bandpass calibration were performed using J1820$-$2528 for the observations taken on September 5 and 13. For observations taken on September 28 and October 17 both J1820$-$2528 and J1924$-$2914 were used. Global fringe fitting  was then carried out with solution intervals of 1 minute over the course of the observation. Again, for the last two observations J1924$-$2914 was included in the global fringe fitting. Flux calibration was carried out using VLBAAMP in \textsc{aips}. In \textsc{aips} phase and amplitude self calibration were performed on the calibrator with a solution interval of 0.5 and 1 minute respectively. 


The science target was then flagged, calibrated, and split. It was exported into \textsc{difmap} and imaged in Stokes I. We used \textsc{radplot} to further flag the data before cleaning the image. Natural weighting was used for all epochs. Images of all VLBA observations  were made with untapered data to obtain maximum angular resolution. We boxed regions with significantly higher flux density than the background and used the \textsc{CLEAN} algorithm to clean the image. This was done repeatedly, applying phase \textsc{selfcal} of decreasing integration intervals until we reached a solution interval of 2 minutes. The final image was viewed in \textsc{mapl cln}.

\subsection{EVN plus e-MERLIN Observations}

To reveal more details about  V3890~Sgr as it expands and fades, we also performed real-time $e$-VLBI observations (project code: RY008, PI: Jun Yang) with the EVN plus e-MERLIN at 4.9~GHz on 2019 October 15 (48.7 days after the start of eruption). The target-of-opportunity observations lasted about four hours (UT: 14:30--18:30). The participating stations were Jodrell Bank, Westerbork, Effelsberg, Medicina, Onsala, Yebes, Hartesbeesthoek, Zelenchukskaya, Irebene, Cambridge, Darnhall, Pickmere, Defford, and Knockin. The observations used the default frequency setup at C band for the EVN (16 subbands in dual polarization, 32 MHz per subband, 2-bit quantization) and e-MERLIN (2 subbands in dual polarization, 64 MHz per subband, 2-bit quantization) stations.  

We used a different source, J1828$-$2417, as the phase-referencing calibrator during the EVN and e-MERLIN observations of V3890~Sgr. The calibrator is only half degree apart from V3890~Sgr and has a compact core-jet VLBI structure with a correlation amplitude of 21 mJy on the long baselines at 8.4 GHz. The nodding observations used a cycle time of three minutes: 20~s for the calibrator, 100~s for the target, and 60~s for the antenna slewing time. To derive the relative position of J1828$-$2417 with respect to J1820$-$2528, i.e. the bright calibrator used in the above VLBA observations, we also observed the pair of calibrators for 11 phase-referencing cycles. 

The correlation data were calibrated in \textsc{AIPS} 31DEC22 \citep{Greisen03}. When the visibility data were loaded into \textsc{aips} discs, we ran \textsc{aips} task \texttt{ACCOR} to slightly correct cross-correlation amplitudes. Before general a-priori amplitude calibration, the antenna system temperature data were significantly smoothed to minimize their random errors. When antenna system temperatures and gain curves were not available, we used nominal system equivalent flux densities and a flat gain curve to do a-priori amplitude calibration. The task \texttt{TECOR} was used to remove ionospheric dispersive delays calculated according to maps of total electron content provided by Global Positioning System (GPS) satellite observations. The phase errors resulting from the antenna parallactic angle variations were removed. We corrected the instrumental phases and delays across the subbands via running fringe-fitting with the data from the bright fringe finder J1800$+$3848. After removing the instrumental phase errors, all subband data of were combined to run more accurate fringe fitting with a solution interval of 20~s. The most sensitive telescope Effelsberg was used as the reference station. The solutions were also applied by two-point linear interpolation from J1828$-$2417 to V3890~Sgr and J1820$-$2528. The phase connections were also quite smooth at all stations. Bandpass calibration was also performed using the fringe finder J1800$+$3848. 

The deconvolution was performed in \textsc{difmap} \citep{Shepherd_Difmap_1997}. The calibrator J0834$-$0417 shows a one-sided core-jet structure with a total flux density of $\sim$38~mJy at 4.93~GHz. With the input image, we re-ran the fringe-fitting and the amplitude and phase self-calibration in \textsc{aips}. All of these solutions were also transferred to the target data by linear interpolation. The target V3890~Sgr was imaged without self-calibration. The low-elevation data in the first and last hour hours were excluded in the imaging process. The small positional offset of the weak calibrator J1828$-$2417 was calculated with respect to the bright calibrator J1820$+$2528, and was corrected in the VLBI map of V3890~Sgr.

\subsection{{\it Fermi}/LAT $\gamma$-ray Observations of the 2019 Eruption}


V3890~Sgr was observed at GeV $\gamma$-ray energies using {\it Fermi}/LAT. 
For our $\gamma$-ray analysis of V3890~Sgr, we used \texttt{Fermitools} (version 2.4.0). 
Our analysis is based on the binned likelihood tutorial provided by 
the {\it Fermi} team\footnote{\url{https://fermi.gsfc.nasa.gov/ssc/data/analysis/scitools/binned_likelihood_tutorial.html}}. 
We begin by selecting events based on their class and type, selecting only events with \texttt{evclass}=128 and \texttt{evtype}=3. 
We also select for good time intervals using the logical expression \texttt{(DATA\_QUAL>0)\&\&(LAT\_CONFIG==1)}. 
The selected observations contain all the Pass~8 (P8R3\_V3) $\gamma$-ray events from 100 MeV to 300 GeV within 10 degrees of the nova. 
We use a model including all the known sources in the 4FGL-DR3 catalogue \citep{Abdollahi_etal_2022} within 20 degrees of V3890~Sgr. 
Also included are the Galactic diffuse emission and the isotropic emission models 
(i.e. \texttt{gll\_iem\_v07} and \texttt{iso\_P8R3\_SOURCE\_V3\_v1}). 
A point source model for V3890~Sgr is added as well, assuming a power law spectrum with an exponential cutoff, 
as is frequently used to model the $\gamma$-ray spectra of novae \citep{Franckowiak+18}. 
This spectral model is shown in equation \ref{eqn:powerlawexp}, where $N_0$ is a normalization parameter, 
$\gamma_1$ is the power law index, $E_c$ is the energy cutoff, and $E_0$ is an energy scale fixed to 200 MeV.
\begin{equation}\label{eqn:powerlawexp}
\frac{dN}{dE} = N_{0}\left( \frac{E}{E_{0}} \right)^{-\gamma_1}e^{-\frac{E}{E_{\rm c}}}
\end{equation}

We analysed the {\it Fermi}/LAT data to obtain V3890~Sgr's GeV $\gamma$-ray flux, fitting this model to data ranging from 2 days before the optical peak to 20 days after the peak. This 22-day window was used to obtain the average flux for this nova, and to obtain spectral parameters. As is typical for a {\it Fermi} analysis, we estimate our detection significance using the test statistic, which is $-2$ times the likelihood ratio between the best fitting model including the nova and a background only model. This can be converted to a typical significance level by taking the square root of the test statistic. In our case, this time window yields a $> 5 \sigma$ detection, making the detection of $\gamma$-ray emission from V3890~Sgr unambiguous. 
Due to the relatively large distance, and correspondingly small observed $\gamma$-ray flux of V3890~Sgr, 
the spectral parameters remain highly uncertain. We recover a spectral index of $\gamma_1 = 1.4 \pm 1.8$ with a cutoff energy of $2$ GeV, though the spectral model is not well constrained.

After obtaining the average flux and associated test statistic for V3890~Sgr we built a light curve starting 60 days before the optical peak, with each point using 15 days of data (as in \citealt{Franckowiak+18}). The resulting light curve is discussed in \S \ref{sec:fermi}.
Throughout this light curve, we fix the spectral parameters to a set that are typical for the $\gamma$-ray emission from novae, assuming a power law index of 1.9 and a cutoff energy of 4.3 GeV \citep{Franckowiak+18}. In many of the light curve bins the data are insufficient to meaningfully constrain the spectral parameters, so the only model parameter for V3890 Sgr allowed to vary is the normalization.

\subsection{\textit{Swift} XRT X-ray Observations}
X-ray Telescope (XRT) observations with the \textit{Neil Gehrels Swift Observatory} began about 0.7 days after eruption, revealing a bright hard X-ray source \citep{Sokolovsky+19}. Data were taken from August 28 to September 5, took a brief hiatus for proximity to the Moon, and proceeded at decreasing cadence from September 10 until V3890~Sgr entered solar conjunction on November 11. Once the nova was no longer blocked by the Sun observations resumed on 2020 February 17 until May 4. 

\textit{Swift} data were processed using \textsc{HEASoft} and the most recent calibration files available for late 2019. Here we use the same spectral fits as \citet{Page+19a} for constraining the X-ray emission at the epochs of our VLBI imaging. On days 7.7 and 51.4, the spectrum is fit with a single optically thin thermal component (i.e. APEC, which represents hot shocked gas). On day 15.5, the fit is improved by the addition of a second APEC component and also a blackbody component (which represents the supersoft source). By day 31.9, the supersoft source had faded, but the optically thin thermal emission is still better fit by two APEC components than one. Absorption is held fixed at $N_H = 5.1 \times 10^{21}$ cm$^{-2}$. The APEC parameters (kT and normalization) are listed in Table \ref{tab:Xray}.

\subsection{Optical Spectroscopy of the 2019 Eruption}\label{sec:smarts}
As an additional constraint on the ejecta velocities in V3890~Sgr, we considered optical spectroscopy of the 2019 eruption obtained using the Small and Moderate Aperture Research Telescope System (SMARTS) 1.5 m telescope and its CHIRON optical spectrograph \citep{Walter+12}. The observations covered the wavelength range  $4082 - 8770$ \textup{\AA} with a resolution of $R=27,800$. Observations post eruption began on August 28, 2019 and ended November 4, 2019, with some spectra obtained pre-eruption. 

\newpage

  \begin{figure*}
  \begin{subfigure}{0.95\columnwidth}
  \includegraphics[trim={0 3.8cm 0 2cm},clip, width=\textwidth]{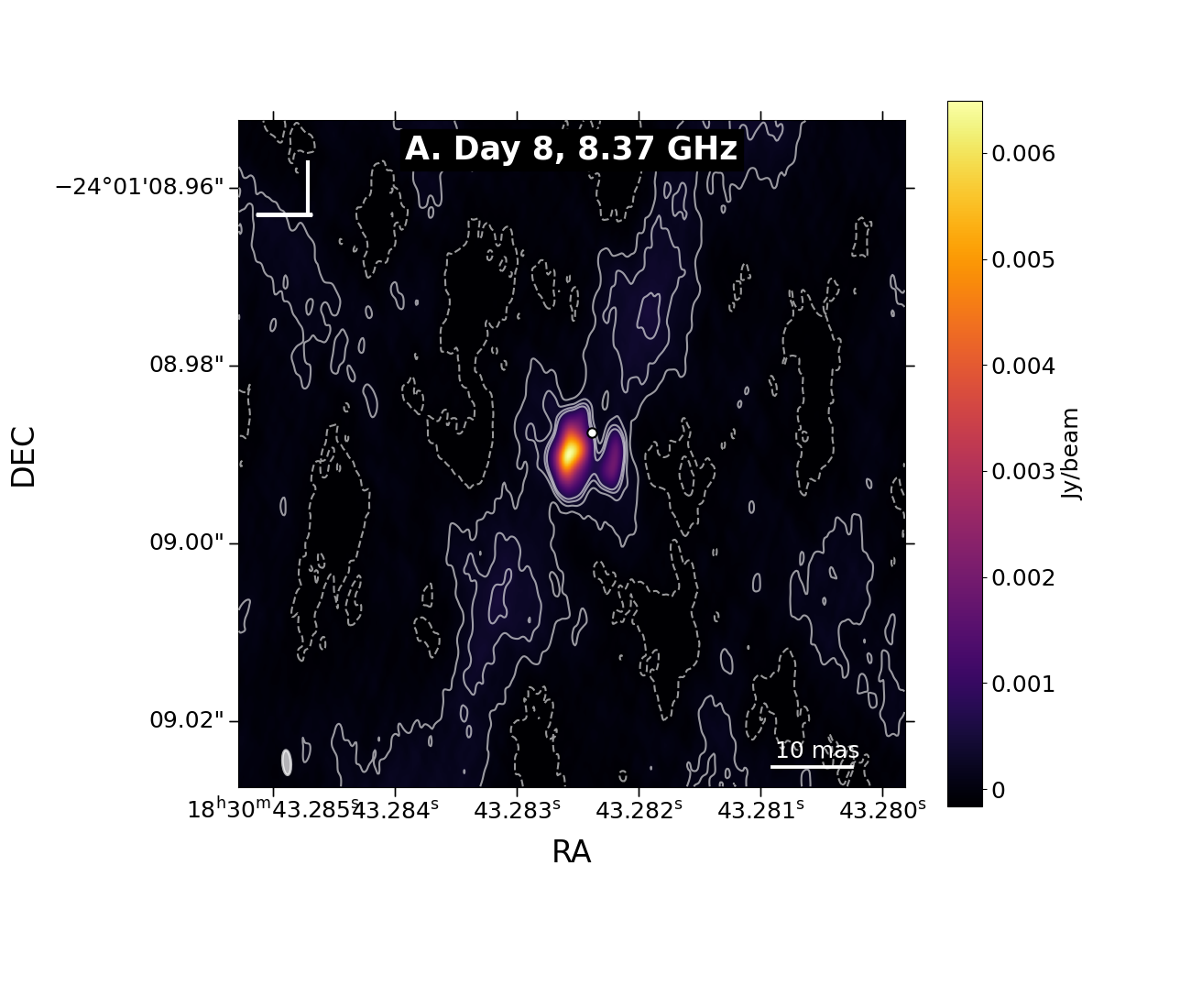}
  \end{subfigure}
  \hfill
  \begin{subfigure}{0.95\columnwidth}
  \includegraphics[trim={0 3.8cm 0 2cm},clip, width=\textwidth]{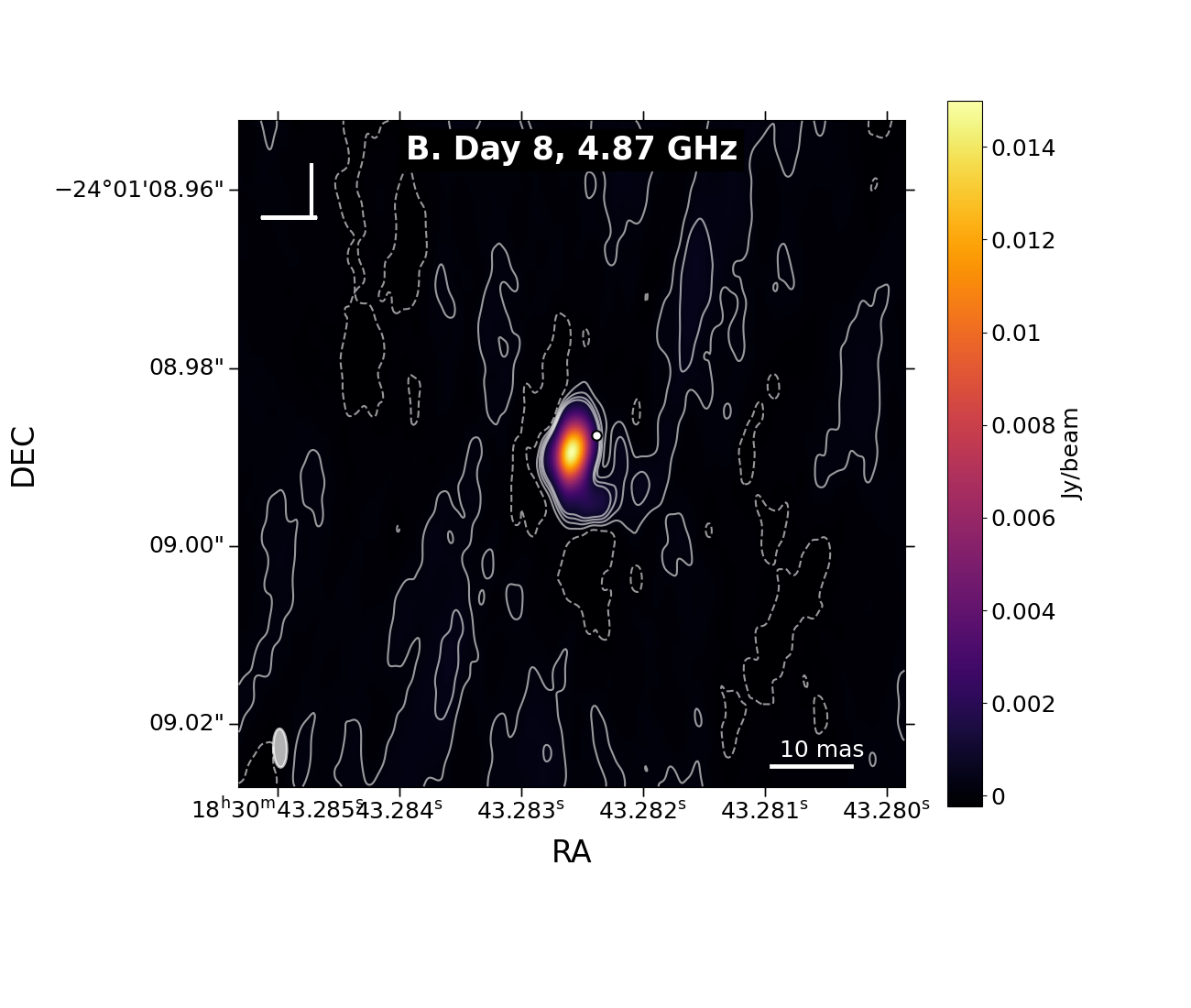}
  \end{subfigure} 
  \begin{subfigure}{0.95\columnwidth} 
  \includegraphics[trim={0 3.8cm 0 2cm},clip, width=\textwidth]{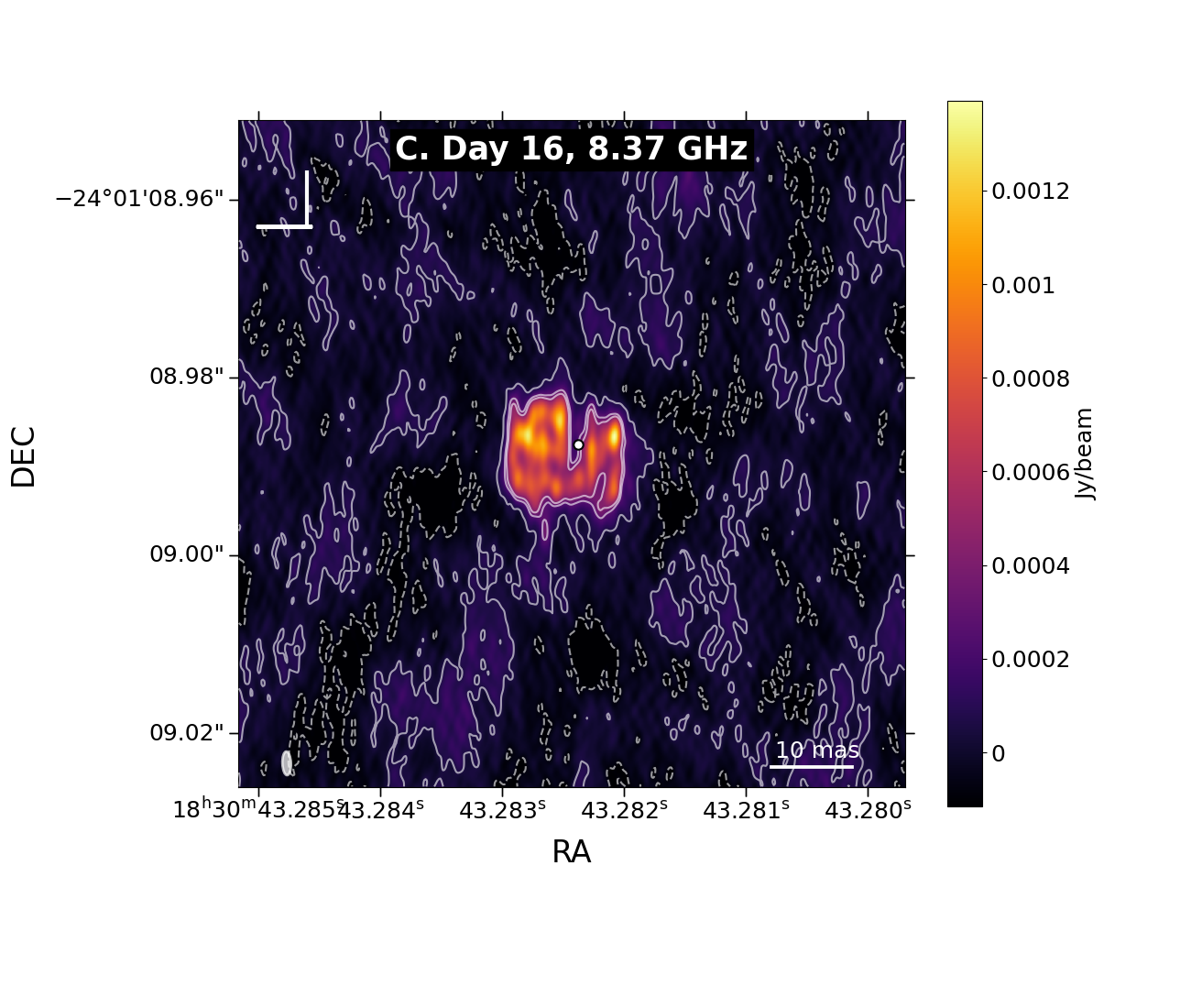} 
  \end{subfigure}  
  \hfill 
  \begin{subfigure}{0.95\columnwidth} 
  \includegraphics[trim={0 3.8cm 0 2cm},clip, width=\textwidth]{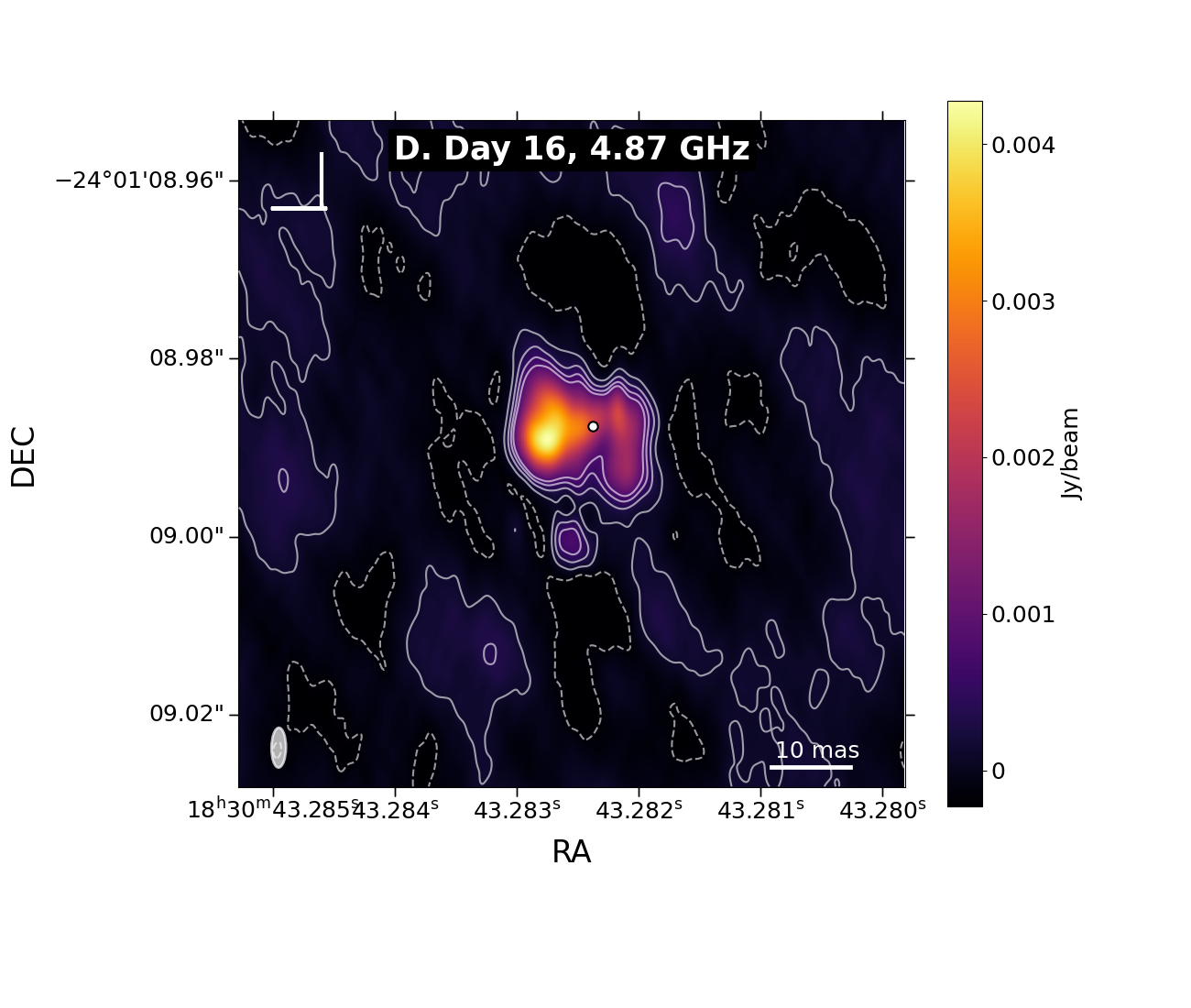} 
  \end{subfigure}
  \begin{subfigure}{0.95\columnwidth} 
  \includegraphics[trim={0 3.8cm 0 2cm},clip, width=\textwidth]{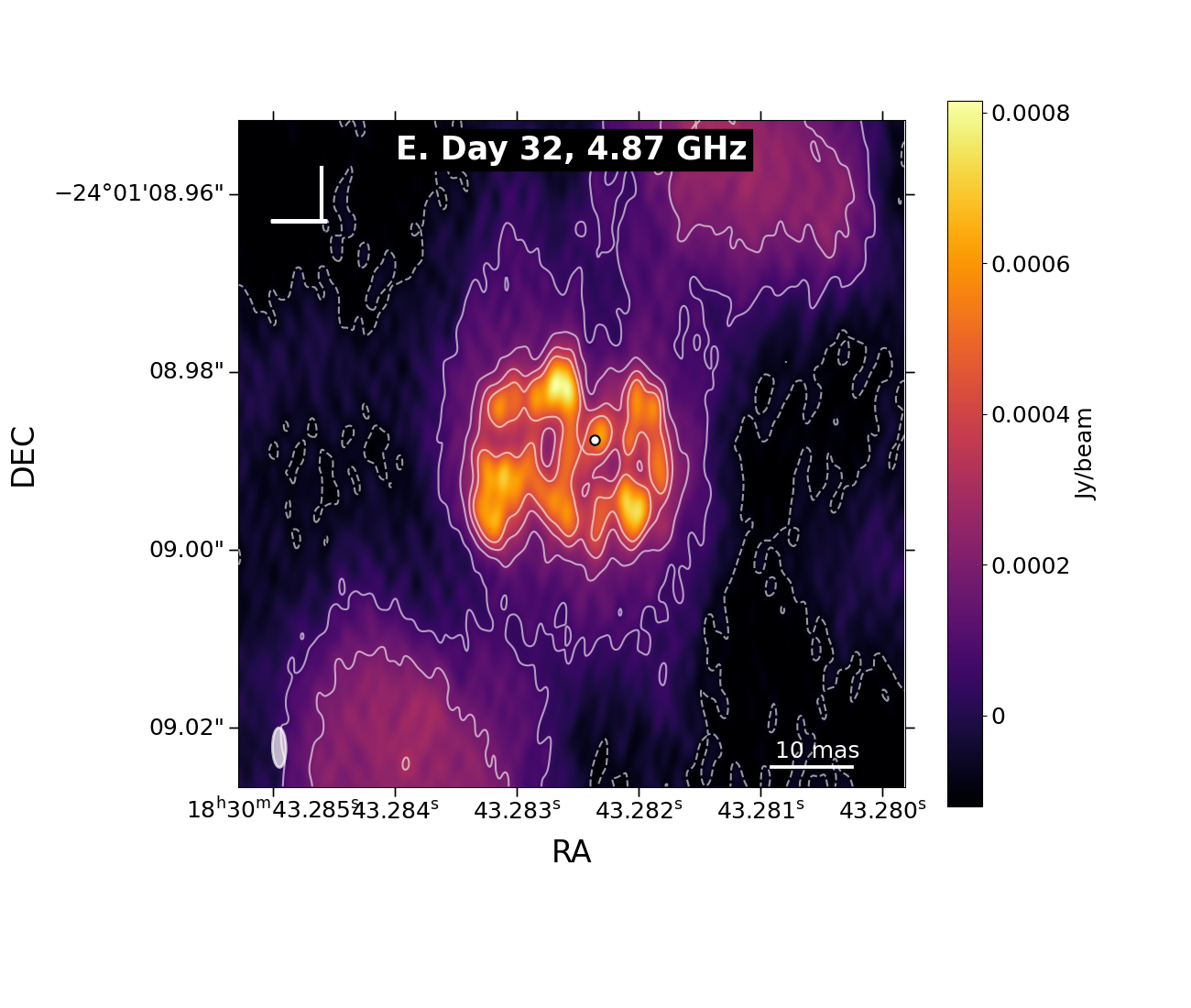} 
  \end{subfigure}  
  \hfill 
  \begin{subfigure}{0.95\columnwidth} 
  \includegraphics[trim={0 3.8cm 0 2cm},clip, width=\textwidth]{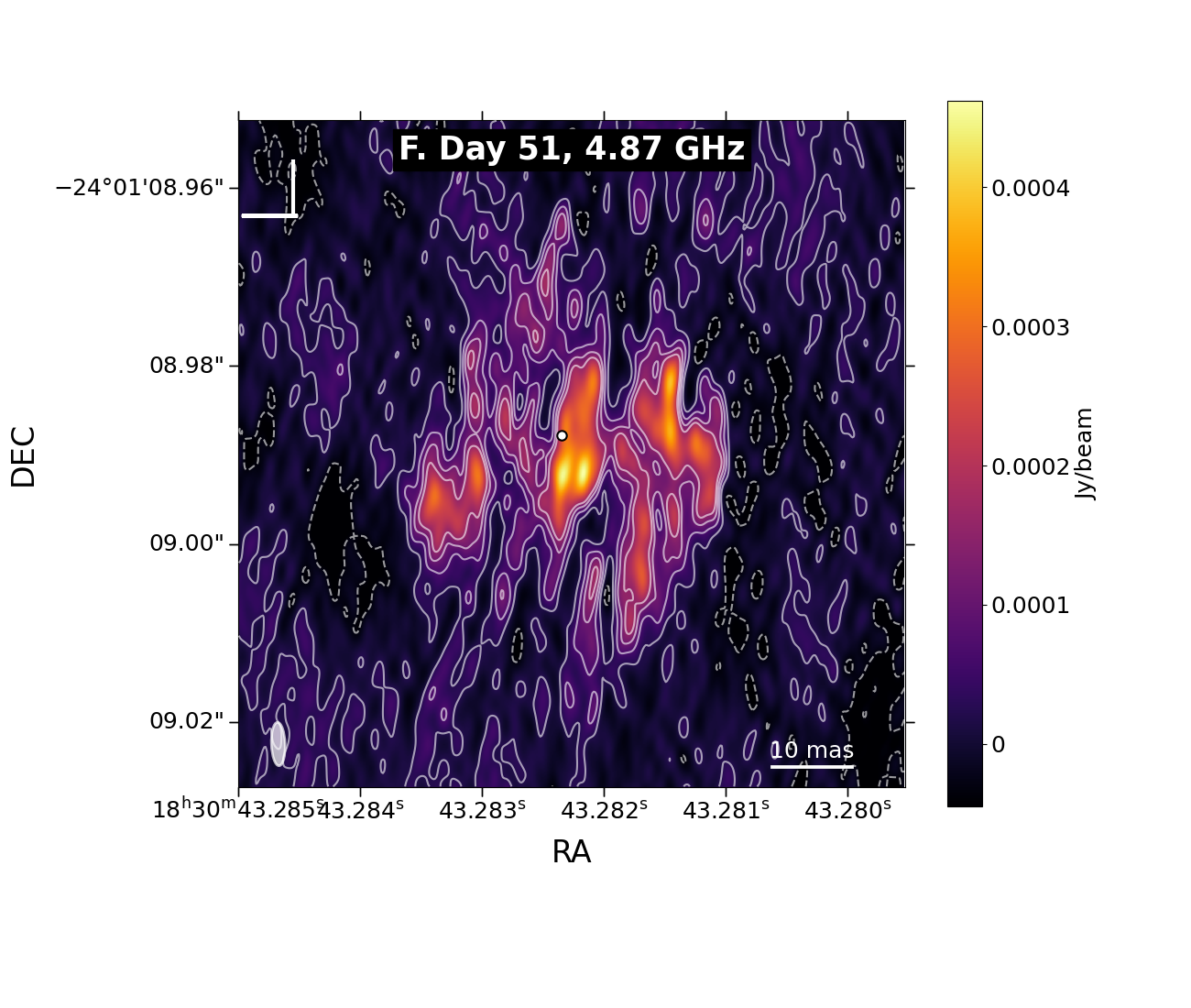} 
  \end{subfigure}
  \begin{subfigure}{0.95\columnwidth} 
  \includegraphics[trim={0 3.8cm 0 2cm},clip, width=\textwidth]{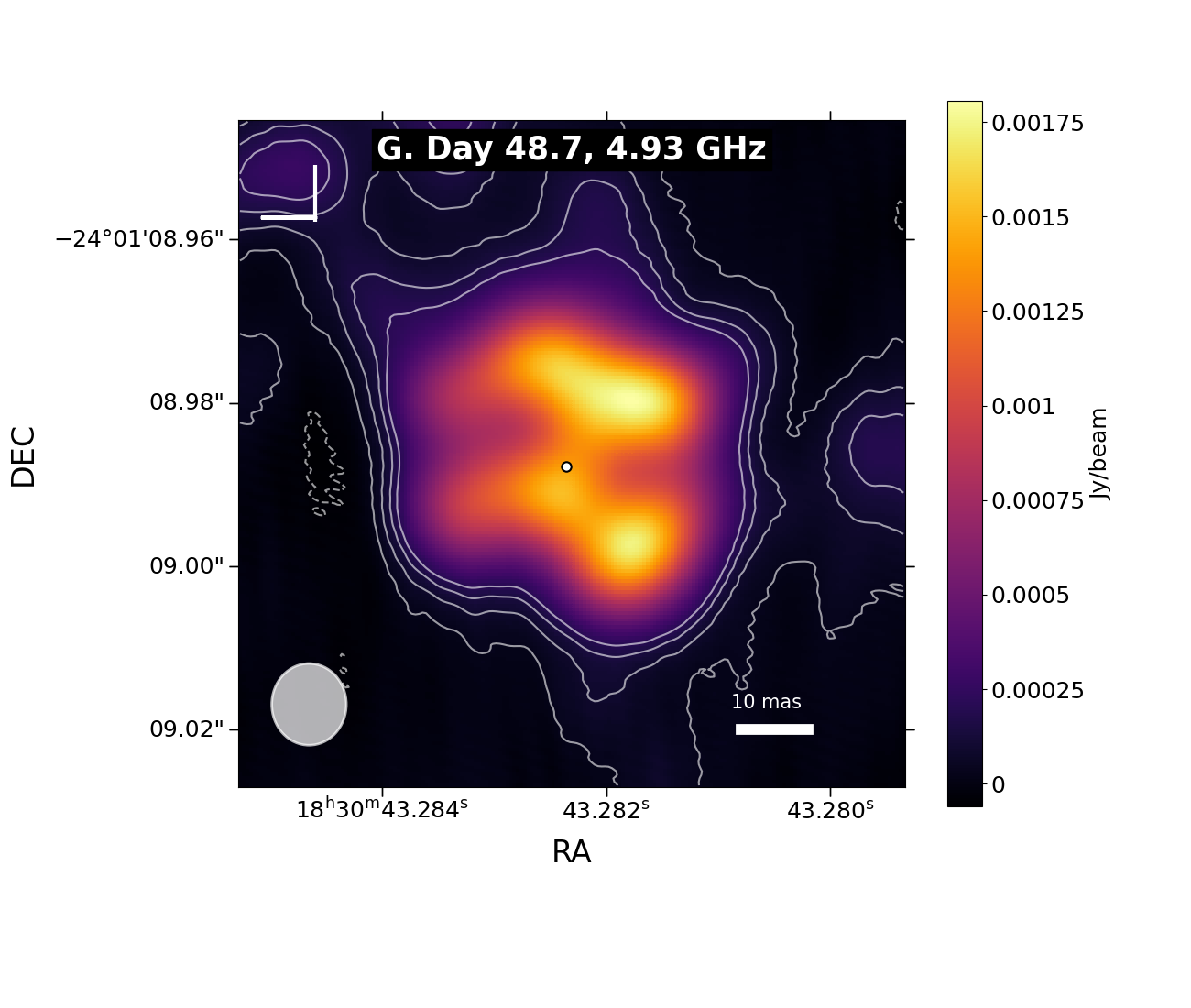} 
  \end{subfigure}  
  \hfill 
  \begin{subfigure}{0.95\columnwidth} 
  \includegraphics[trim={0 3.8cm 0 2cm},clip, width=\textwidth]{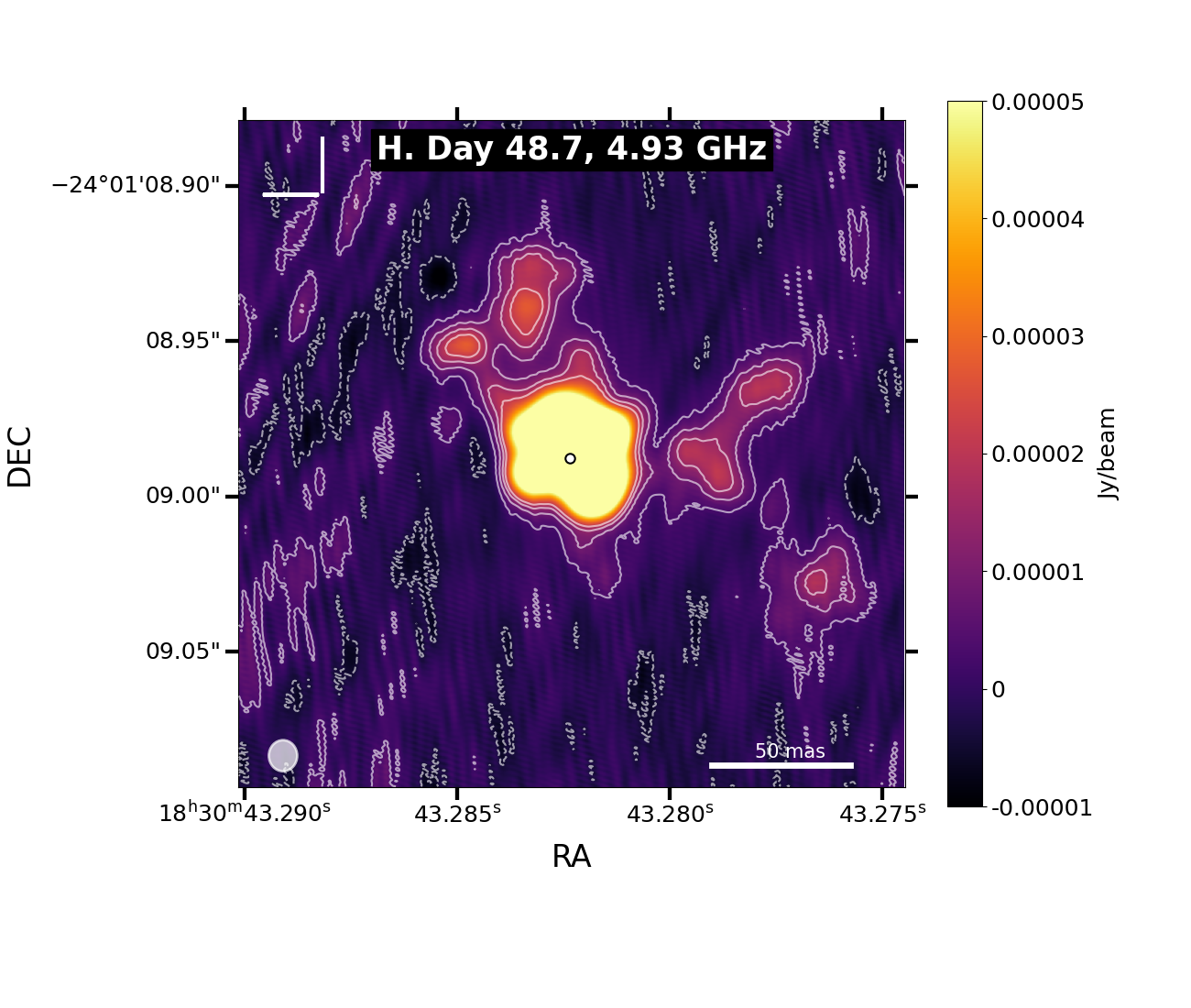} 
  \end{subfigure}
  \caption{VLBI radio images of the 2019 eruption of V3890 Sgr. The top two panels, (a) and (b), show 4.87 and 8.37 GHz VLBA images on day 8.1. The second row---panels (c) and (d)---show 4.87 and 8.37 GHz VLBA images on day 16. The third row are 4.87 GHz VLBA images from day 32.0 (panel (e)) and day 51.0 (panel (f)). The bottom row show the EVN+e-MERLIN image from day 48.7; panel (g) is on the same scale as panels (a--f), while panel (h) zooms out to show a larger field of view. In each panel, the contour levels are set to $-1.5, 1, 3, 5$ and $7 \sigma$ (see Table \ref{tab:radio} for $\sigma$ values). The white dot is located at the {\it Gaia} position of V3890~Sgr. In the upper left corner is a compass showing the North (up) and East (left) directions. The synthesized beam is plotted in the bottom left corner. The field of view is $80$ mas across for panels (a--g), and $225$ mas across for panel (h). Note that each panel has its own color scale, as denoted in its color bar.}
   \label{fig:bigcomp}
  \end{figure*}

\section{VLBI Radio Image Analysis}\label{sec:vlbi}

Our VLBA radio images of V3890~Sgr's 2019 eruption are plotted in panels (a--f) of Figure~\ref{fig:bigcomp} in J2000 equatorial coordinates. The EVN+e-MERLIN image is shown in panels (g) and (h) of Figure~\ref{fig:bigcomp}. 
In each image, the position 
derived from 
{\it Gaia} Data Release 3 \citep{Gaia16, Gaia23} is marked using a white dot and error bars; 
the {\it Gaia} position was corrected for proper motion to the observation time of each image. 
The {\it Gaia} position at the average observing epoch is RA = 18h30m43.2824s and Dec = $-24^{\circ}01^{\prime}08.988^{\prime\prime}$, and the error bars are smaller than the marker and are not visible in the images. The mean {\it Gaia} epoch is 2019.81.


The background RMS noise was found in \textsc{difmap} by calculating statistics on  a blank image region
away from the nova. The rms noise levels in the 8.37 GHz images were 58--82 $\mu$Jy beam$^{-1}$ and the noise values in the VLBA 4.87 GHz images were 22--116 $\mu$Jy beam$^{-1}$. The specific image noise values are listed in Table \ref{tab:radio}. The rms noise in the 4.93 GHz EVN+e-MERLIN image is 31 $\mu$Jy beam$^{-1}$.
 The image RMS values were used to set the contour levels in the images to $-1.5, 1, 3, 5$ and $7 \sigma$ (Figure~\ref{fig:bigcomp}).
These noise levels can be compared with the theoretical thermal noise of the VLBA, as calculated using the EVN calculator. 
This led to theoretical noise estimates of $18.1\ \mu$Jy beam$^{-1}$ at $8.37$ GHz with the VLBA, $15.4\ \mu$Jy beam$^{-1}$ at $4.87$ GHz with the VLBA, and $13.0\ \mu$Jy beam$^{-1}$ for the EVN+e-MERLIN observation. It is not unusual for VLBI images to have observed noise values several times that theoretically expected, and reaching theoretical noise was made unusually difficult by the low elevation of V3890~Sgr as observed by these radio facilities.

All VLBA images are centered at the same location (RA= 18h30m43.2826s, Dec = $-24^{\circ}01^{\prime}08.990^{\prime\prime}$) with the same map size (2048 pixels on a side) and pixel size (0.15 mas). 
For the $4.87$ GHz VLBA images, the average FWHM resolution is $B_{\rm maj} = 4.5$ mas along the major axis and $B_{min} = 1.7$ mas along the minor axis with an average position angle of $PA = -1.3^{\circ}$. The $8.37$ GHz images have an average resolution $B_{\rm maj} = 2.7$ mas and $B_{\rm min} = 1.0$ mas. The $4.93$ GHz EVN+e-MERLIN image is restored with a symmetric beam of FWHM $= 10$ mas.

\subsection{Morphology and Evolution}

The images in Figure~\ref{fig:bigcomp} show how the shape of the nova changes over time.  
Eight days after eruption we see an elliptical compact source offset to the east---likely a lobe of a bipolar outflow---and another small region to the west of source center (panels (a) and (b) of Figure~\ref{fig:bigcomp}). This fainter western component is more apparent in the $8.37$ GHz image than the $4.87$ GHz image (panel (b) compared with panel (a)). This likely implies that the western structure is absorbed by intervening CSM in this first epoch.


16 days after eruption, radio emission from V3890~Sgr has become more spread out (panels (c) and (d) of Figure~\ref{fig:bigcomp}), and both lobes are now more apparent.  In the $4.87$ GHz image the brightest region remains the eastern lobe. The small structure beneath the nova, located at Dec of about $-$24:01:09.00 in panel (c), is likely an 
artefact 
as we do not see it in later epochs. At 8.37 GHz, the radio emission is noisier, as expected from the higher angular resolution sampling the nova with more synthesized beams. The eastern and western lobes are of more comparable brightness at $8.37$ GHz, implying that optical depth continues to affect the western lobe at 4.87 GHz. 


The shape of the early radio emission could be explained if the CSM is composed of a spherical wind and a region of increased density in the orbital plane, (i.e., the EDE; \citealt{Orlando+17, Walder+08, Mohamed+2013, Booth+14}). The position angle of the EDE could range from $0$ to $-15^{\circ}$ (oriented roughly up--down in Figure \ref{fig:bigcomp}). 
The EDE impedes expansion in the north--south direction and confines the ejecta into two lobes oriented mostly east--west. 
The lobes being imaged in VLBA suggest that the synchrotron emission is coming from the lobes and not the central disc.  The synchrotron emission would then be coming from interactions with the spherical component of the CSM.  
The inclination angle of the nova is well constrained to be $67- 69^{\circ}$ by \cite{Mikolajewska+21}, and our VLBA imaging shows that the eastern side is tilted towards us, while the western side is tilted away from us and undergoes more absorption from the EDE in early imaging epochs. 

During the third epoch, 32 days after eruption (panel (e) of Figure~\ref{fig:bigcomp}), the radio emitting region continues to expand and now looks quite symmetric east--west. The nova no longer looks so bipolar, and has similar dimensions north--south and east--west (Table \ref{tab:radio}). There are significant deconvolution artifacts visible around the nova in this epoch (visible as blotches at top-right and bottom-left in panel (e)); these should be disregarded.

The time period of day 48--51 is imaged by both the VLBA and EVN; while the VLBA has higher resolution, EVN+e-MERLIN is more sensitive to lower surface brightness emission. 
The VLBA image from this last epoch is quite noisy, implying that the surface brightness of the radio emission has dropped (panel (f) of Figure~\ref{fig:bigcomp}). The EVN+e-MERLIN image, however, captures more of the flux at higher significance, and now shows that the nova is brighter along the north--south axis: a $\sim 90^{\circ}$ flip from what was observed earlier in the eruption (panel (g) of Figure~\ref{fig:bigcomp}). The radio-emitting shocks therefore seem to be primarily interacting with the EDE at this late epoch. 

The EVN+e-MERLIN image from day 48 also shows significant diffuse radio flux external to the primary nova shell (panel (h) of Figure~\ref{fig:bigcomp}). This emission is patchy, elongated, and asymmetrically distributed around the nova shell, with a position angle of $\sim 50^{\circ}$. 

Compared with RS~Oph's 2021 eruption, our VLBA images appear to be missing a central disc component. In addition to this compact central component, RS~Oph shows a bi-lobed structure similar to V3890 Sgr, with one lobe being initially fainter and then increasing in brightness over time, 
which is attributed to absorption by an EDE \citep{Munari+22, Lico+24}. 
However, RS~Oph remains more elongated/bipolar in later epochs (64--65 days after explosion), 
while V3890~Sgr exhibits more circular symmetry in our later imaging epochs (Figure~\ref{fig:bigcomp}). 
V407 Cyg--- a 2010 nova that erupted in a longer period symbiotic binary with an asymptotic giant branch companion  and the first nova detected in GeV $\gamma$-rays \citep{Abdo+10}--- was imaged with VLBI 20 to 203 days 
after eruption. Images revealed an expanding elliptical shell 
with no central brightening \citep{2020A&A...638A.130G}. They also exhibit radio emission at larger radii than can be attributed to the nova eruption \citep{2020A&A...638A.130G}, perhaps similar to the extended diffuse emission we observe in V3890 Sgr with the EVN+e-MERLIN.



\subsection{Angular Size as a Function of Time}
We estimated the size of the radio emission from our VLBI images. In the first two epochs, we use the 8.37 GHz images for size determination, as they are both higher resolution and less affected by absorption than the 4.87 GHz images. As the source morphology is complex and not well described by 
e.g. a Gaussian, we do not fit the source with parametric models in the uv plane; instead, we directly estimate the source size from the image.

For the first epoch the size of the nova was determined using \textsc{JMFIT} in \textsc{AIPS} by fitting two Gaussians to the nova. 
Both lobes of the nova were resolved, so 
we used the deconvolved sizes of the components to estimate the $2 \sigma$ radii of the Gaussians, and then measured the distance between the centres of the Gaussians to get the distance between the two lobes. This was all added together to get the full diameter of the source along the major (roughly east--west) axis. The diameter of the nova along its minor axis was the average of the two deconvolved Gaussian widths in the $\sim$north--south direction.

For the remaining VLBA epochs the size was determined by measuring the diameter of $3\sigma$ contours in \textsc{carta}. 

In the EVN+e-MERLIN image, there is significant diffuse flux external to the nova shell, and we estimated the size in two different ways: the dimensions of the nova shell estimated from $5\sigma$ contours in \textsc{carta}, and the larger dimensions of the diffuse flux estimated from $3\sigma$ contours. We note that the dimensions of the nova shell estimated from the EVN image are significantly larger than in the VLBA image from day 51 (mean diameter of 44.5 mas, compared to 32.9 mas). 
This could be caused by a poorly measured diameter in the noisy VLBA image or the lower resolution of the EVN image.
We note that if we use the EVN dimensions of the nova shell, we find that the expanding shell  would have had to significantly accelerate between day 32 and 48 (\S \ref{sec:dist}), which seems unlikely. We therefore conclude that the EVN dimensions of the nova shell are likely affected by the lower resolution of this image. As the EVN image and day 51 VLBA image yield similar fluxes for the nova shell, but the VLBA image better hones in on the nova shell, we use the VLBA image for estimating the shock velocity and magnetic field (Table \ref{tab:Bfield})

The major and minor axes are listed in Table \ref{tab:radio}. Generally, the major axis of the nova shell is oriented east--west and the minor axis north--south, although we note that the source is not greatly elongated in the east--west direction in any of the epochs.

\subsection{Velocities and Expansion Parallax Distance}\label{sec:dist}

If we assume a distance of 9 kpc to V3890~Sgr, as suggested by \citet{Mikolajewska+21} based on the assumption that the red giant companion is Roche Lobe filling, the diameter of V3890~Sgr's VLBA emission would be $1.1 \times 10^{15}$ cm on day 8, growing to $4.7 \times 10^{15}$ cm on day 51. This implies remarkably high expansion velocities of 8080 km s$^{-1}$ over the first 8 days, and $\sim$5000 km s$^{-1}$ over the following 43 days. 

\begin{figure}
    \centering
    \includegraphics[width=95mm, trim={2.7cm 2.5cm 0 2.5cm},clip]{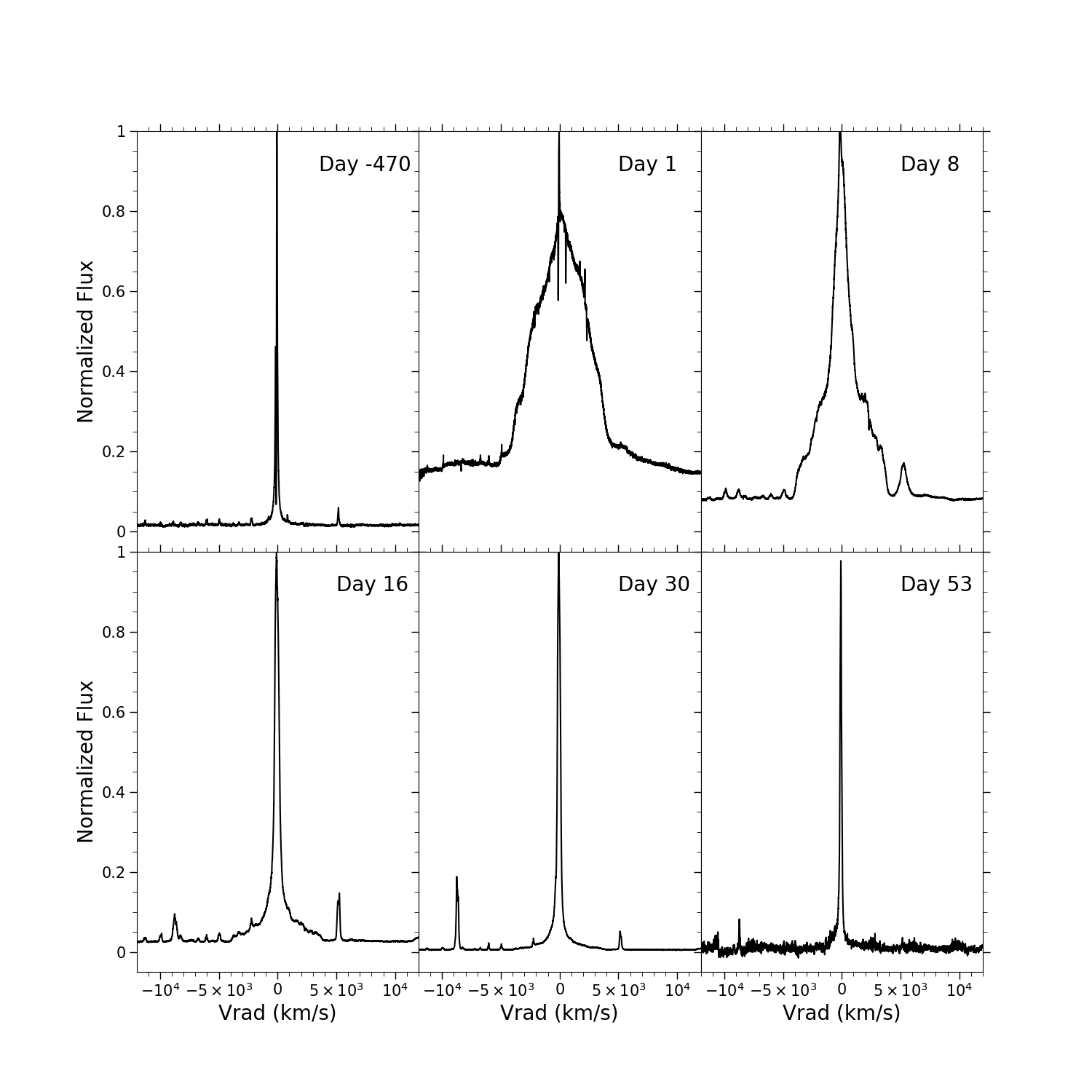}
    \caption{The H$\alpha$ emission line profiles 470 days before V3890~Sgr's 2019 eruption and 1, 8, 16, 30, and 53 days after the eruption. }
    \label{fig:Halpha}
\end{figure}

To test if these high expansion velocities are plausible in V3890~Sgr, we inspected H$\alpha$ line profiles as observed by SMARTS/CHIRON. Six epochs of H$\alpha$ line profiles of the 2019 eruption are plotted in Figure~\ref{fig:Halpha}. The full width at zero intensity (FWZI) was estimated for each of the H$\alpha$ emission lines using \textsc{specplot} in\textsc{ IRAF}. To measure the FWZI we estimated the continuum level of the data and measured where the H$\alpha$ emission line met that continuum level.

A spectrum obtained more than a year before the 2019 eruption shows a very narrow H$\alpha$ emission line with no extended wings. This emission emanates from the circumbinary material, ionized by the accreting white dwarf.
One day following eruption, the H$\alpha$ line shows a much broader emission line with a remnant of the narrow emission line superimposed (implying the presence of unshocked circumstellar material; centered at  $v = -131$ km s$^{-1}$). 
While the bulk of the emission is confined to $\pm$5000 km s$^{-1}$, there may be faint wings on day 1 that extend to about $\pm9600$ km s$^{-1}$. However, these wings are very faint, and sensitive to vagaries of continuum fitting, so we are not certain how reliable the presence of 9600 km s$^{-1}$ gas is in V3890~Sgr. 
Prior to the eruption there is no evidence of these extended wings. By day 8, the narrow component has broadened and the broad wings have disappeared, giving the profile a near-triangular shape.
Although a He\,{\sc i} line at 6678\textup{~\AA} makes it difficult to measure the edge of the H$\alpha$ line profile, we estimate it converges with the continuum around $\pm 4300$ km s$^{-1}$. On day 16, the narrow component is very strong compared to the broad component, and the H$\alpha$ line reaches $-3840$ km s$^{-1}$ and $+3650$ km s$^{-1}$, indicating that the expansion is slowing. The broad component of the H$\alpha$ line has further narrowed by day 30 with  HWZI $v =  2520 $ km s$^{-1}$. By day 53 the line width has decreased to about $v = \pm1440$ km s$^{-1}$.

These spectroscopic velocities are lower than what is implied by our VLBA images, if we take a distance of 9 kpc. \citet{Mikolajewska+21} convincingly argue based on Na~I~D absorption lines that the distance to V3890~Sgr is at least 6 kpc, and a distance of about 9 kpc is likely an upper limit, as it assumes the red giant companion completely fills its Roche Lobe.
We combine the spectroscopic velocities and our VLBA imaging to estimate expansion parallax distances, taking the expansion in the north--south direction, as V3890~Sgr is viewed relatively edge-on ($i \approx 68^{\circ}$; \citealt{Mikolajewska+21}) and our imaging shows the orbital plane is oriented roughly north--south, so the radial velocity will more closely mirror the north--south expansion than east--west. Assuming that the emission began as a point source on day 0 and expanded to a diameter of 7.9 mas by day 8.1, we find a distance of 11.4 kpc if the 9600 km s$^{-1}$ broad wings are reliable, and a distance of 5.9 kpc if the radio emission is characterized by expansion of 5000 km s$^{-1}$, more in line with the bulk of the emission line flux. 
Between days 8 and 16, we estimate an expansion parallax distance of 6.6 kpc (for an average spectroscopic velocity of 4000 km s$^{-1}$ during this time). Between days 16 and 32, we find 6.9 kpc for an average velocity of 3130 km s$^{-1}$ during this time. Choosing to disregard the first epoch due to ambiguity with the FWZI measurement,
 we prefer a distance of 6.8 kpc to V3890~Sgr, and use this throughout the rest of this paper. Velocities estimated from our VLBA images, taking $d = 6.8$ kpc and the geometric mean of the major and minor axis of V3890~Sgr, are listed in Table \ref{tab:Bfield}.

At this distance, the dimensions of the diffuse radio emission detected by EVN+e-MERLIN on day 48 are $8 \times 10^{15}$ cm by $5 \times 10^{15}$ cm. If this material was launched on day 0 of the 2019 eruption, it would have needed to expand at $\sim$15,000 km s$^{-1}$, much faster than is observed in V3890~Sgr (or in any nova that we know). We therefore conclude that this diffuse emission was caused by a relativistic tracer, like photons or cosmic rays, escaping the nova shell and interacting with the environment. We discuss this scenario more in \S \ref{sec:late_halo}.





\subsection{Fraction of Flux `Resolved Out' by VLBA}
\begin{figure}
    \centering
    \includegraphics[width=95mm]{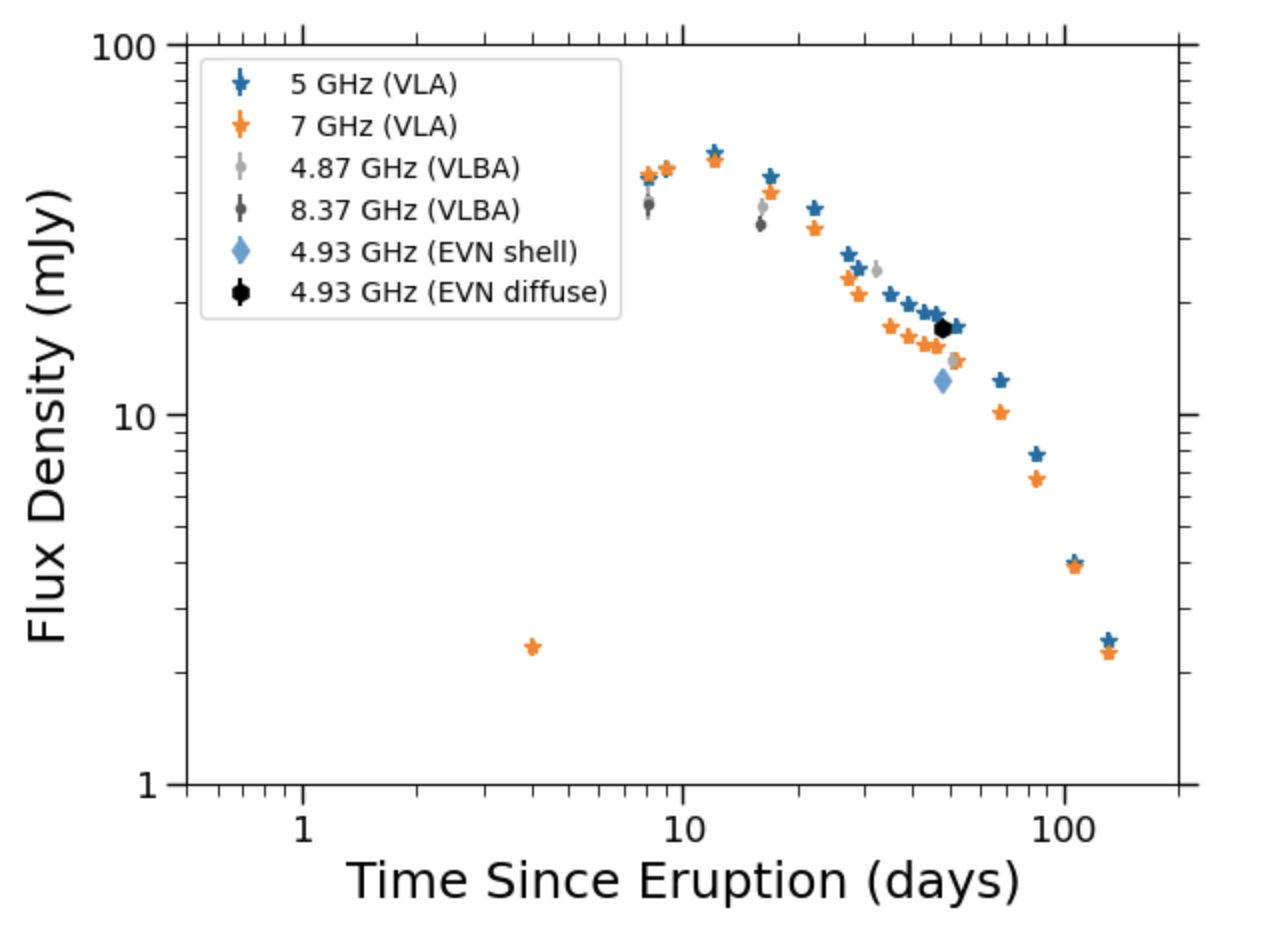}
    \caption{Integrated flux densities from our VLBA observations and the EVN+e-MERLIN observation as a function of time following V3890~Sgr's 2019 eruption, compared to the VLA light curve in similar frequency bands (VLA data published in \citealt{Nyamai+23}). }
    \label{fig:light}
\end{figure}

In Figure~\ref{fig:light} the VLA light curve, which measures V3890~Sgr's integrated radio flux density, is plotted using stars and compared to the VLBI light curve plotted as dots.  The total flux density of V3890~Sgr in our VLBA and EVN+e-MERLIN images is found by boxing a region around the nova in \textsc{carta}. 
We then estimated the associated uncertainty by determining how many synthesized beams fit into this region, and multiplying the image rms noise by the square root of this number of beams.
We added this uncertainty in quadrature with an estimated flux calibration error of 5 per cent, as all frequency bands are below 10 GHz.
The resulting flux densities and errors are listed in Table \ref{tab:radio}. 

We see the VLBA light curve is brightest during our first epoch on day 8, and the 4.87 GHz flux density then declines with a power-law slope of $-0.91\pm 0.45$ (excluding the EVN point that detects the diffuse flux). The VLA light curves peak 11 days after discovery; if we measure the power law slope of the 5.0 GHz band from day 9 to day 51 we find a shallower decline than VLBA with power-law slope of $-0.67\pm 0.05$.

The VLBA observations appear to capture most of the flux density seen in the VLA observations. On day 8 the VLBA captures 87 per cent of the flux density in $4.87$ GHz and on day 16 the VLBA captures 83 per cent. This confirms that the radio emission is synchrotron dominated, as the VLBA is not sensitive to thermal emission (the theoretical 3$\sigma$ brightness temperature sensitivity of the VLBA is $3.1 \times 10^5$ K at 4.87 GHz and $3.5 \times 10^5$ K at 8.37 GHz, for our exposure times; Equation \ref{eq:BT}). 

However, the second bump in the VLA light curve, which happens around day 50, is missed by our day 51 VLBA image, detecting only 13.9 mJy of the 17.3 mJy integrated and explaining the steeper decline of the VLBA light curve, compared to VLA.  We do note the appearance of two significant compact flux components near the center of the nova in our Day 51 image (panel (f) of Figure~\ref{fig:bigcomp}). However,  these two components have a combined flux of $1.8 \pm  0.13$ mJy, which is not enough to explain the increase in flux density in the VLA light curve around this time. 

The flux density measured from the nova shell in the day 48 EVN+e-MERLIN image is similar to the day 51 VLBA image, at 12.4 mJy, but EVN+e-MERLIN recovers significant diffuse flux extended over $\sim$0.1 arcsec which rivals the integrated flux, at 17.2 mJy. We conclude that the origin of the second radio bump is not caused by a brightening at the shock front (as proposed by \citealt{Nyamai+23}), but instead by an enhancement of diffuse radio emission upstream of the shock front. We consider possible physical explanations for this emission in \S \ref{sec:late_halo}.


\subsection{Brightness Temperature}
The brightness temperature was calculated from our VLBA images of the nova shell, assuming a source with a Gaussian shape and the relation from \citep{Synthesis_1999}: 
\begin{equation} \label{eq:BT}
T_{\rm B} = 1224\, {\rm K}\ \frac{S_{\nu}}{\rm mJy}\ \left(\frac{\nu}{\rm GHz}\right)^{-2}\ \left(\frac{\theta_{\rm maj}\, \theta_{\rm min}}{\rm arcsec^2}\right)^{-1}
\end{equation}
Here $\nu$ is the frequency of the observation, $S_{\nu}$ is the flux density at that frequency,  $\theta_{ \rm maj}$ is the FWHM major axis of the source and $\theta_{\rm min}$ is the FWHM minor axis.
In this case, we estimated the diameter of V3890 Sgr from 3$\sigma$ contour levels, as listed in Table \ref{tab:radio}. 

\begin{figure}
    \centering
    \includegraphics[width=80mm]{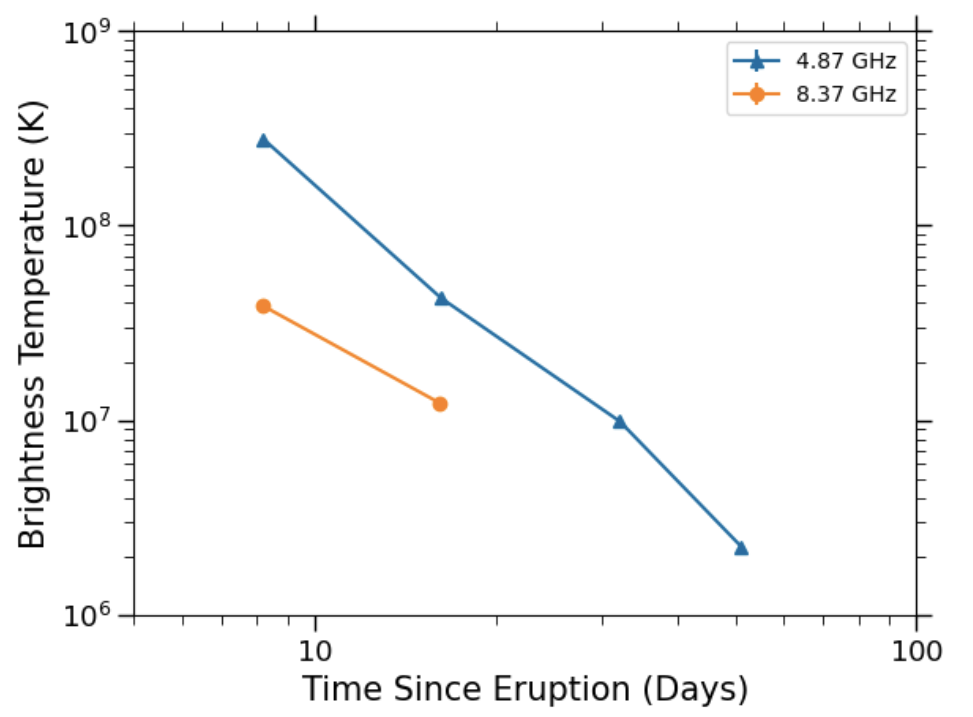}
    \caption{ Brightness temperature measurements as a function of time following V3890~Sgr's 2019 eruption, measured from our VLBA images in $4.87$ GHz (plotted as blue triangles) and $8.37$ GHz (plotted as orange circles). }
    \label{fig:bright}
\end{figure}

Figure~\ref{fig:bright} plots the calculated brightness temperature as a function of time following the 2019 eruption.  The $4.87$ GHz brightness temperature starts at $2.8 \times 10^8$ K on day 8 and declines to a value of  $2.2 \times 10^6$ K by day 51. The $8.37$ GHz brightness temperature starts lower at a value of  $3.9 \times 10^7$ K and declines to $1.2 \times 10^7$ K on day 16. At all times the brightness temperature is high, staying well above a thermal expectation of $\lesssim 5 \times 10^4$ K \citep{Chomiuk+21radio}. 
These $T_{\rm B}$ estimates are consistent with the conclusion in the previous subsection,
that the vast majority of the flux is synchrotron emission.


\cite{Nyamai+23} estimated the brightness temperature of V3890~Sgr from VLA observations of the integrated flux, 
by assuming 
a spherically symmetric shell expanding at a constant velocity of $4200$ km s$^{-1}$ at a distance of 9 kpc.
The flux densities used were from the $1.28$ GHz VLA observations. 
Our brightness temperature measurements are similar to \cite{Nyamai+23} on day 8 ($2.8 \times 10^8$K vs their $\sim 2.5 \times 10^8$ K). However, our brightness temperature decreases much faster (51 days after eruption we have a value $2.2 \times 10^6$ K  vs their $2 \times 10^7$ K). This difference can partially be explained by the negative spectral index and the difference in frequencies used to estimate $T_{\rm B}$. Our initial velocity measurement is also faster than the modeled velocity used in \citet{Nyamai+23} and decelerates faster. Our VLBA estimates of $T_{\rm B}$ are made directly from images, 
and are therefore more reliable than those estimated from integrated fluxes 
relying on assumptions about the emitting region size.


We also estimate the brightness temperature of the diffuse flux detected by the EVN+e-MERLIN on day 48 (panel (h) of Figure~\ref{fig:bigcomp}). The emission is patchy, and peaks at 0.27 mJy beam$^{-1}$, which corresponds to a peak brightness temperature of $1.3 \times 10^5$ K. Such a high surface brightness makes it unlikely to be thermal emission, as might be expected if circumstellar material is photoionized by the nova; in this case, we would expect $T_{\rm B} \approx 10^4$ K. The relatively high brightness temperature of the diffuse emission and spectral index of the second radio bump (\S \ref{sec:alpha}) make it likely that the diffuse emission is synchrotron in origin.

\subsection{Magnetic Field Strength}\label{sec:bfield}

The magnetic field strength in V3890~Sgr's synchrotron-emitting nova shell  was calculated using the revised equipartition formula from \cite{Beck_Krause05}:
\begin{equation} \label{eq:B_field}
    B_{\rm eq} = \left(\frac{4\pi (1 - 2\alpha) (K_0 + 1) I_{\nu} E_{\rm p}^{1+2\alpha} (\nu/2c_1)^{-\alpha}}{(-2\alpha-1)c_2lc_4}\right)^{\frac{1}{3-\alpha}}
\end{equation}
Here $K_0$ is the proton-to-electron number density ratio amongst the relativistic particles, $\alpha$ is the radio spectral index defined as $S_{\nu} \propto \nu^{\alpha}$, $I_{\nu}$ is the synchrotron intensity at frequency $\nu$, $E_{\rm p}$ is the proton rest energy ($938.26$ MeV), and $l$ is the path length through the synchrotron emitting region.
$c_1 = 6.264 \times 10^{18}$ erg$^{-2}$s$^{-1}$G$^{-1}$ and  $c_4 = (2/3)^{(\alpha+1)/2}$, as defined in \cite{Beck_Krause05} in Appendix A. 
$c_2$ is defined in \cite{Beck_Krause05}: for an $\alpha=-0.7$, $c_2 = 1.016 \times 10^{-23}$erg G$^{-1}$ sterad $^{-1}$.

The spectral index of V3890~Sgr is relatively flat compared to other astrophysical synchrotron sources. On Day 16, we estimate $\alpha = -0.25$ from our VLBA images. Even at late times, the spectral index converges to $\alpha \approx -0.4$ \citep{Nyamai+23}.
However, Equation \ref{eq:B_field} is only valid for $\alpha$ values steeper than $-0.5$, so we used $\alpha=-0.7$ for the magnetic field estimates, as is typical for other synchrotron-emitting astrophysical shocks (e.g., \citealt{Weiler+02,Green19}). 
$I_{\nu}$ was found by dividing the measured flux density by the angular area of the nova, measured from our VLBA images. 
We set $K_0$ to be 40 and 100; $K_0=100$ comes from measurements of local Galactic cosmic rays and $K_0=40$ is expected for strong shocks \citep{Beck_Krause05}. The path length $l$ takes a distance of 6.8 kpc and the estimates of the  minor axis angular size from Table \ref{tab:radio}.


\begin{table*}
    \centering
    \renewcommand{\arraystretch}{1.1}
    \caption{Properties Derived from VLBA Radio Images of V3890~Sgr \label{tab:Bfield}}
    \begin{tabular}{|l|l|l|l|l|l|l|l|l|l|l|}
    \hline
    \hline
        Day & Freq & $l$\tablenotemark{a} & B$_{K_0 = 40}$\tablenotemark{b} & B$_{K_0 = 100}$\tablenotemark{b}& $v_{\rm s}$ &  $\epsilon_{\rm B}(K_0 = 40)$\tablenotemark{c} & $\epsilon_{\rm B}(K_0 = 100)$\tablenotemark{c}  \\ 
       & (GHz) & (cm) & (G) & (G) & (km s$^{-1}$) & \\
        \hline
        8.1 &8.37 & $8.0 \times 10^{14}$& $0.13$& $0.17$& 5940\tablenotemark{d} & $2.9 \times 10^{-4}$ & $5.0 \times 10^{-4}$\\ \hline        
         16.0 &8.37& $1.4 \times 10^{15}$& $0.08$& $0.11$& 3970\tablenotemark{e} & $6.7 \times 10^{-4}$ & $1.3 \times 10^{-3}$\\ \hline
        32.0 &4.87  & $2.2 \times 10^{15}$& $0.05$& $0.06$& 3130\tablenotemark{f} & $1.4 \times 10^{-3}$ & $2.0 \times 10^{-3}$ \\ \hline
        51.0 &4.87 & $3.2 \times 10^{15}$& $0.03$& $0.04$& 3390\tablenotemark{g}& $1.1 \times 10^{-3}$ & $1.9 \times 10^{-3}$ \\
        \hline
    \end{tabular}
\tablenotetext{a}{Diameter of the radio emission, assuming a distance of 6.8 kpc.}    
\tablenotetext{b}{Magnetic field estimated using $\alpha = -0.7$.}
\tablenotetext{c}{$\epsilon_{\rm B}$ estimated assuming $v_{\rm s}$ from this Table and $n_{\rm e}$ from Table \ref{tab:Xray}.}
\tablenotetext{d}{Velocity between day 0--8.1, representing the geometric mean of the major and minor axes.}
\tablenotetext{e}{Velocity between day 8.1--16.0.}
\tablenotetext{f}{Velocity between day 16.0--32.0.}
\tablenotetext{g}{Velocity between day 32.0--51.0.}
\end{table*}


The resulting magnetic field estimates are listed in Table \ref{tab:Bfield}.
On day 8 the magnetic field is $0.13$ G (for $K_0=40$) and diminishes to $0.03$ G by day 51. For $K_0=100$ on day 8 the magnetic field is slightly higher at $0.17$ G and declines to $0.04$ G by day 51. 

These magnetic field strengths in V3890~Sgr are comparable with other novae that have been imaged with the VLBA.
In V1535 Sco,  \citet{Linford+17} estimate magnetic field strengths of $0.13$ G for $K_0=40$ and $0.17$ G for $K_0=100$ about 7.7 days after eruption using a path length of $\sim 10^{13}$ cm, comparable to our day 8.1 estimates in V3890~Sgr. 
In imaging of V1674 Her, M.\ Williams et al.\ in prep. estimate the magnetic field on day 20 to be $0.09$ G for $K_0=40$ and $0.11$ G for $K_0=100$, very similar to what we find for V3890~Sgr 16 days after eruption. 
RS~Oph was also found to have similar magnetic field values, of $0.09$ G for $K_0=40$ and $0.11$ G for $K_0=100$ about 21 days after eruption \citep{Rupen+08}. 

We also estimate the magnetic field pressure, given by $P_{\rm B}=  \frac{B^2}{8\pi}$.
The magnetic field pressures on day 8.1 are $7 \times 10^{-4}$ dyn cm$^{-2}$ for $K_0=40$ and $0.001$ dyn cm$^{-2}$ for $K_0=100$.  On day 16 the magnetic pressure has decreased to $3 \times 10^{-4}$ dyn cm$^{-2}$ for $K_0=40$ and $5 \times 10^{-4}$ dyn cm$^{-2}$ for $K_0=100$. 
By day 51 the magnetic pressure has further declined to $4 \times 10^{-5}$ dyn cm$^{-2}$ for $K_0=40$ and $6 \times 10^{-5}$ dyn cm$^{-2}$ for $K_0=100$. 

These magnetic field determinations will be used to estimate the magnetic pressure fraction
($\epsilon_{\rm B}$) in \S \ref{sec:epsb}. 


 \begin{figure}
    \centering
    \includegraphics[width=80mm]{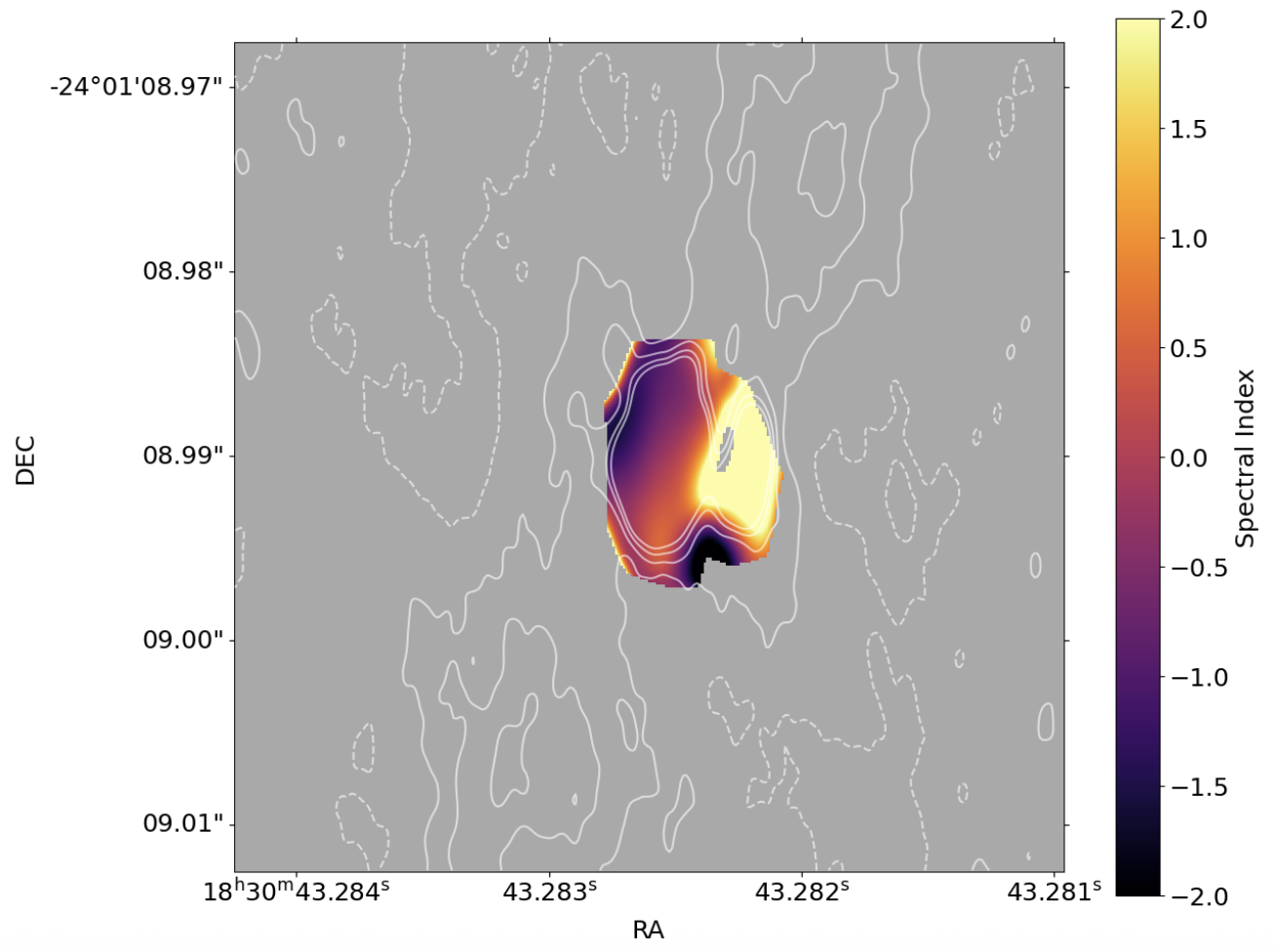}
    \caption{Spectral index map of the September 4 observation, 8.1 days after eruption, using $4.87$ GHz and $8.37$ GHz. The September 4 $8.37$ GHz image contours are overlaid on the SIM.}
    \label{fig:SIM1}

\end{figure}
\begin{figure}
    \centering
    \includegraphics[width=80mm]{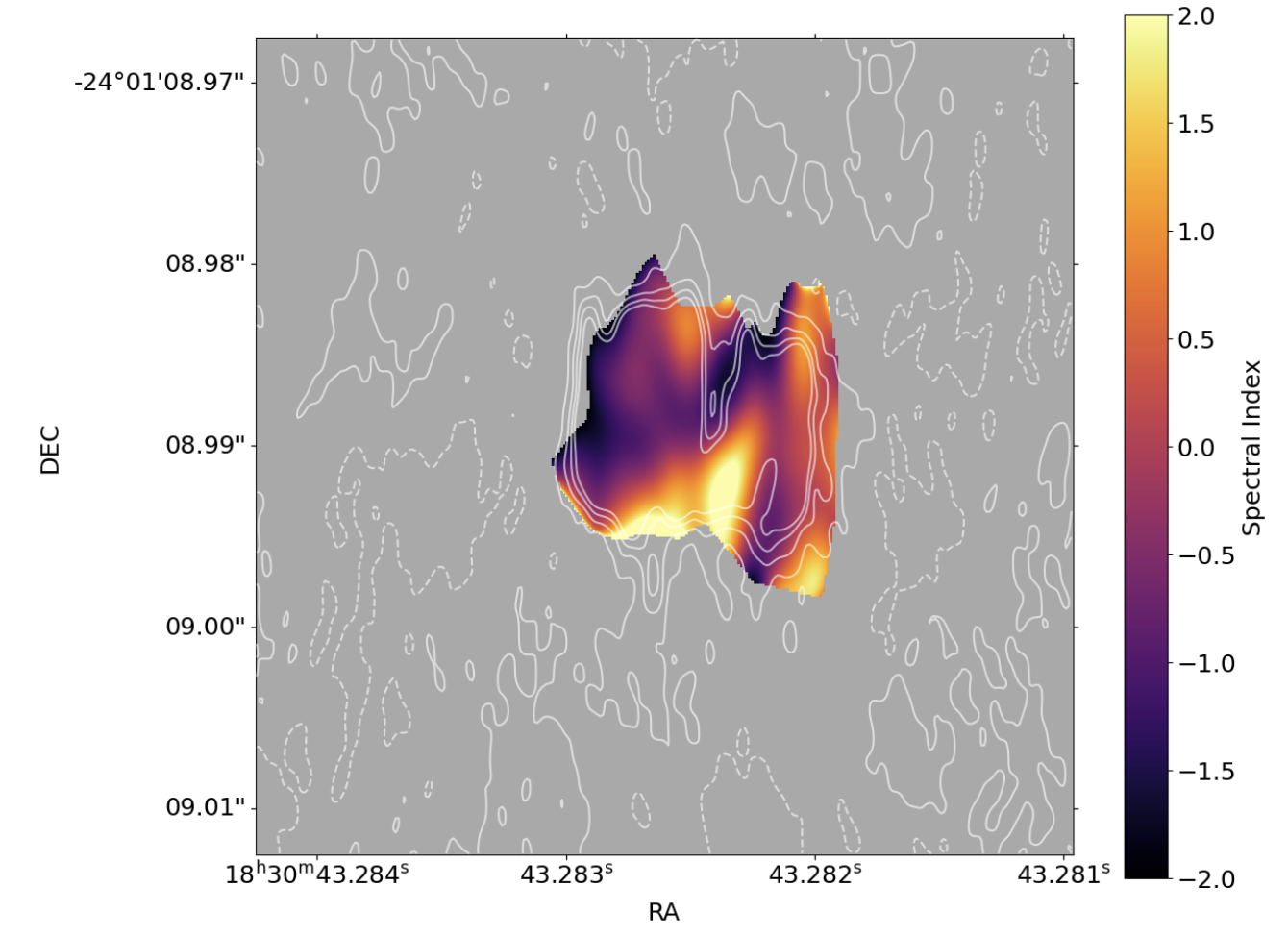}
    \caption{Spectral index map of the September 12 observation, 16 days after eruption, using $4.87$ GHz and $8.37$ GHz. The  September 12 $8.37$ GHz image contours are overlaid on the SIM.}
    \label{fig:SIM2}
\end{figure}

\subsection{Spectral Index Maps}\label{sec:alpha}

We combine our 4.87 GHz and 8.37 GHz VLBA images to measure the spectral index $\alpha$ across V3890~Sgr on day 8 and 16 of eruption.
The spectral index maps (SIMs) use images of the nova made in \textsc{difmap}. 
 
Each epoch had identical map and cell sizes. 
The $4.87$ GHz UV data were plotted in \textsc{radpl}, and the data on the shortest baselines were cut so that the data began at the same minimum UV distance as the corresponding $8.37$ GHz data. The $8.37$ GHz images were also tapered to have the same beam size as the $4.87$ GHz images. This way the UV ranges sampled by the two frequencies are brought into alignment.
The fits files were then loaded back into \textsc{AIPS}  to make the spectral index maps. Using \textsc{blank} the nova was outlined 
and regions without significant nova emission were blanked out in both frequencies. Then an image cube was made using \textsc{fqub} which was then transposed  using \textsc{trans}. 
The SIM was made using \textsc{spixr}.

The SIMs are shown in Figures~\ref{fig:SIM1} and \ref{fig:SIM2}. 
The more yellow regions correspond to a positive spectral index, brighter at higher frequencies. Contours of the $8.37$ GHz image flux (identical to those in panels (a) and (c) of Figure \ref{fig:bigcomp}) are added to further show how the spectral index is changing over the source. 


From the SIM on day 8 (Figure~\ref{fig:SIM1}), we see that the spectral index changes over the source, with the eastern region having a spectral index of around $-0.5$ to $0.5$. The western component is more positive with values of $\alpha \approx 2$. This could be explained by the eastern region being optically thin synchrotron emission and the western region still being optically thick, free-free absorbed by intervening CSM (see
Figure~2 of \citealt{Munari+22} for a nice illustration). The hole located just to the west of image center is due to a small region with negative flux density in the 4.87 GHz image.  
At this same location
there is significant positive flux from the source in the 8.37 GHz image, $\sim 0.0015$Jy. 
When the two images are overlaid, this resulted in a non physical spectral index.
The values on the edge of this hole should also be disregarded. Here we have reduced the scale to only allow for a range of $\alpha = -2 $ to $2$.  Due to noise and the differences between the apparent structures in the 
$4.87$ GHz and $8.37$ GHz images, the spectral index calculated at the edges of the SIMs should also be disregarded.

In the day 16 SIM the spectral index in the eastern region becomes more negative with values ranging from $-1$ to $0.5$. The western region has become flatter with values ranging from $0.5$ to $2$. 
The negative spectral index seen in the eastern region is as expected for synchrotron emission from shocks originating from ejecta hitting up against strong CSM \citep{Linford+17}. This is also seen in RS~Oph's early radio emission \citep{Eyres+09}. 


To attempt to constrain the spectral index of the diffuse emission in the day 48 EVN+e-MERLIN image, we considered the VLA observations published in \cite{Nyamai+23}. High frequency ($\gtrsim$30 GHz) observations obtained in the VLA's A configuration should be able to resolve this diffuse emission. Unfortunately, while the VLA observations obtained on day 45 and 51 were in A configuration, they were limited to $<$8 GHz, where resolution is poorer. The VLA observation from day 67 did include higher frequency data, but were obtained in A-to-D configuration move time, which made extracting a high-resolution image difficult, with limited baselines probing the small angular scales. We conclude that there is no bright flux at 30 GHz resolvable on $\sim$0.06 arcsec scales,  and the VLA data are all consistent with point-like emission. This, along with the fact that the integrated radio spectrum during the second bump maintains a negative spectral index ($\alpha = -0.4$; \citealt{Nyamai+23}), leads us to conclude that it is likely that the diffuse emission is brighter at lower frequencies, as expected for synchrotron emission.

\section{{\it Fermi} GeV $\gamma$-ray Analysis } \label{sec:fermi}

\subsection{LAT Light Curve}\label{sec:fermi_lc}

\begin{figure}
    \centering
    \includegraphics[]{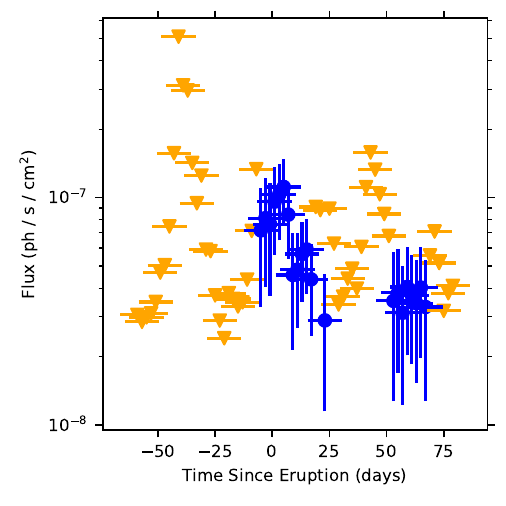}
    \caption{$\gamma$-ray light curve for V3890~Sgr obtained using {\it Fermi}/LAT data using a bin width of 15 days. Blue points indicate bins with a test statistic of 4 or greater, corresponding to at least a $2 \sigma$ detection, while orange triangles show 
95~per~cent upper limits on the $\gamma$-ray flux. }
    \label{fig:LAT_LC}

\end{figure}

V3890~Sgr's $>$100 MeV $\gamma$-ray light curve, as measured by {\it Fermi}/LAT, is shown in Figure~\ref{fig:LAT_LC} with time bins that are 15 days wide and sampling every 2 days. The displayed light curve spans 2019 June 29 to November 15 with the maximum flux being 2019 Sep 2 (5 days after start of eruption/optical peak). To convert between photon and energy fluxes, we assume a standard spectral model based on V906 Car (currently the highest S/N $\gamma$-ray detection of a nova; \citealt{Aydi+20}), which has been found to be reasonable for most novae \citep{Franckowiak+18}. For our energy band, this translates to a conversion factor of 1 photon cm$^{-2}$ s$^{-1} = 1.2794 \times 10^{-3}$ erg cm$^{-2}$ s$^{-1}$ (the same one used by \citealt{Craig+26}). Given the low S/N of the V3890~Sgr detections, the available data are not able to provide better constraints on the $\gamma$-ray spectrum of this nova. A maximum flux value of $1.43 \times 10^{-10}$ erg s$^{-1}$ cm$^{-2}$ with an error of $4.6 \times 10^{-11}$ erg s$^{-1}$ cm$^{-2}$ were found. A narrower bin, beginning two days before the optical peak to ensure that we capture the start of the $\gamma$-ray emission, and ending at the observed maximum in the light curve, with free spectral parameters, yields a somewhat larger flux of  $2.4 \times 10^{-10}$ erg s$^{-1}$ cm$^{-2}$, with a relatively large uncertainty of 1.5 $\times 10^{-10}$ erg s$^{-1}$ cm$^{-2}$. We take this to be the maximum $\gamma$-ray flux during the eruption to estimate the system's $\gamma$-ray luminosity.

At a distance of 6.8 kpc, this {\it Fermi}/LAT flux corresponds to a peak $>$100 MeV luminosity of $(1.3\pm0.8) \times 10^{36}$ erg s$^{-1}$, or about 1 per cent of the white-dwarf Eddington luminosity. This is more luminous than V745 Sco ($<5.2 \times 10^{35}$ erg s$^{-1}$ at 8.2 kpc; \citealt{Franckowiak+18, Molina+24}) and
V407 Cyg ($6.6 \times 10^{35}$ erg s$^{-1}$ at 3.5 kpc; \citealt{Abdo+10, Chomiuk+21}). V3890~Sgr is significantly less $\gamma$-ray luminous than 
RS~Oph at peak ($6.7 \times 10^{36}$ erg s$^{-1}$ at 2.7 kpc; \citealt{Acciari+22,Cheung+22}). V3890~Sgr's $\gamma$-ray luminosity ranks in the top $\sim$quarter of the classical nova sample of \citet{Craig+26}, which is focused on novae with main-sequence companions but provides an overview of the state-of-the-art of the population of $\gamma$-ray detected novae. 

The light curve in Figure~\ref{fig:LAT_LC} indicates that V3890~Sgr had a peak in the $\gamma$-rays shortly after optical maximum, and emitted significantly until 23 days after optical peak. Interestingly, there is a marginal detection (with a detection defined here as a test statistic of at least 4) around day 60, after a series of non-detections. The timing of this $\gamma$-ray reappearance is strikingly similar to the timing of the second radio bump (day $\sim$40--80; Figure~\ref{fig:light} and \citealt{Nyamai+23}). Possible explanations for this resurgence of $\gamma$-rays and radio emission are considered in \S \ref{sec:late_halo}.

\subsection{LAT Test Statistic}

\begin{figure}
    \centering
    \includegraphics[]{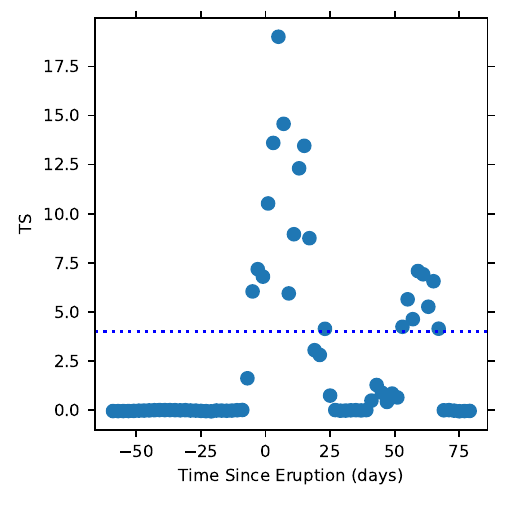}
    \caption{{\it Fermi}/LAT test statistic on V3890~Sgr as a function of time, with windows of 15 days. Time shown is relative to the optical peak. The horizontal dotted line displays the $TS = 4$ significance cutoff adopted for the light curve in Figure~\ref{fig:LAT_LC}.}
    \label{fig:TEST}

\end{figure}

A plot displaying the test statistic throughout our {\it Fermi}/LAT  light curve is shown in Figure~\ref{fig:TEST}. Each point matches a point in the light curve, so the sampling is again a series of bins with a total width of 15 days, sampled every 2 days. A horizontal line displays the threshold for a significant detection in the light curve, set to the 2-$\sigma$ level. The highest test statistic recovered in the light curve is 19.0, found on day 5 (also the day with the maximum observed flux), which corresponds to a $4.3\sigma$ detection. We note that a larger test statistic can be found with a longer time window, exceeding a $5 \sigma$ detection. In the secondary peak around day 60, there are a series of several bins with test statistics between 4 and 7, which meet our cutoff for significance in the light curve.

\begin{figure}
    \centering
    \includegraphics[width=\columnwidth]{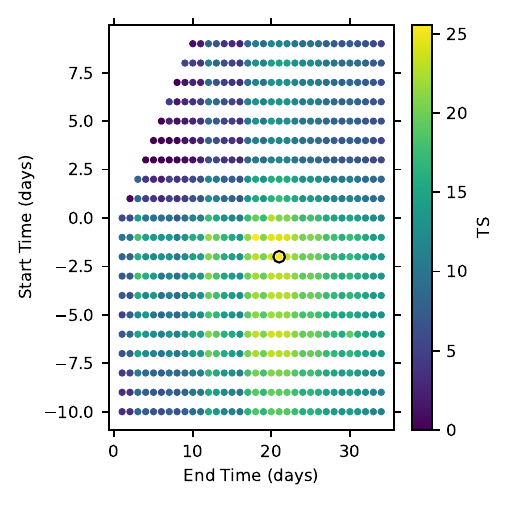}
    \caption{Grid of TS values assuming different windows of data selection for V3890~Sgr. The circled point corresponds to the maximum recovered TS value of 25.5, and a duration of 23 days. The upper left corner is blank because in this region the end times would be before the start times, and therefore there are no data.}
    \label{fig:TSGrid}
\end{figure}

Measuring the duration of the $\gamma$-ray emission is tricky in light curves with large, overlapping time windows. Following the methods of \cite{Franckowiak+18}, we find the test statistic across a range of start and end times. This includes start times at 1 day intervals ranging from 10 days before the optical peak until 10 days after the peak, and end times at 1 day intervals ranging from 1 to 35 days following optical peak. The exceptions are any start and end time combinations that lead to the start time being the same or after the end time. From this analysis, the maximum TS value is obtained starting 2 days before the optical peak, and continuing until 21 days after the peak, corresponding to a $\gamma$-ray duration of 23 days. This solution also corresponds to our maximum TS value of 25.5, a $5 \sigma$ detection. The assumed spectral model is again a power law with an exponential cutoff (equation \ref{eqn:powerlawexp}, but here the power law index and energy cutoff are allowed to vary. The resulting grid of TS values is displayed in Figure~\ref{fig:TSGrid}.

\begin{figure}
    \centering
    \includegraphics[]{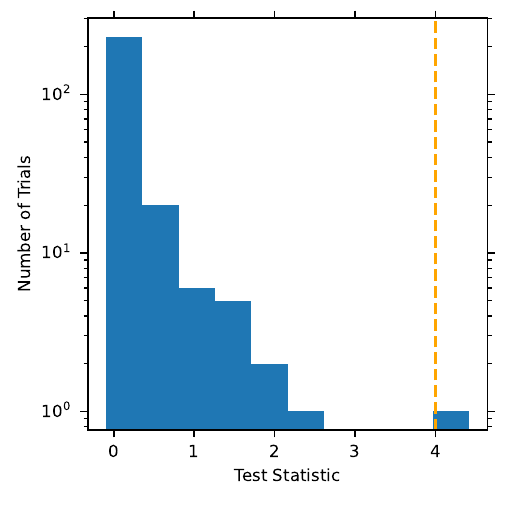}
    \caption{Histogram showing the obtained {\it Fermi}/LAT test statistic values from background sampled regions. The vertical orange line displays the cutoff for a significant detection. Only $0.38$~per~cent of our background samples would meet this criterion, and none match the significance of the detections during the eruption of V3890~Sgr.}
    \label{fig:SIG}
\end{figure}

As the secondary peak around day 60 is only a marginal detection, with the significance peaking at 2.5$\sigma$ in our light curve, we investigate the rate of false positives at this significance level. The goal is to eliminate the possibility that we are getting false detections due to a poor fit to the background or similar modelling difficulty. In order to make the fairest comparison with our generated light curve, we extend the start time of our light curve to 600 days before the optical peak for this nova. Data more than 60 days before the eruption certainly contain no $\gamma$-ray emission from the nova eruption itself, so we repeat our analysis at these times to search for false positives. The distribution of test statistics recovered from this data set is shown in Figure~\ref{fig:SIG}. The vertical line displays the same cutoff used in our light curve. Out of 266 samples, we find only one that exceeds our significance cutoff. We note that these trials are not independent, as the sampling follows our light curve procedure and therefore nearby points contain some data overlap. However, this low false detection rate leads us to have confidence that the secondary $\gamma$-ray peak in V3890~Sgr is real.

\section{Modelling the Shock in V3890~Sgr} \label{sec:model}

\begin{table*}
    \centering
    \renewcommand{\arraystretch}{1.1}
    \caption{Properties Derived from X-ray Spectra of V3890~Sgr \label{tab:Xray}}
    \begin{tabular}{|l|l|l|l|l|l|l|l|l|l|l|}
    \hline
    \hline
        Day & kT$_{\rm hot}$ & Norm$_{\rm hot}$ & kT$_{\rm cool}$ & Norm$_{\rm cool}$ & $n_{\rm H}$(hot) & $\dot{M_{\rm w}}$ \\ 
       & (keV) & (cm$^{-5}$) & (keV) & (cm$^{-5}$) &  ($10^6$ cm$^{-3}$)&   (M$_{\odot}$ yr$^{-1}$)\\
        \hline
        7.7& $2.8^{+1.1}_{-0.7}$ & $0.064^{+0.014}_{-0.010}$& -- & -- & $11.1 \pm 1.1$& $2.2\times 10^{-7}$\\ \hline        
         15.5 &$2.1^{+0.3}_{-0.2}$ & $0.041^{+0.003}_{-0.005}$ & $0.76^{+0.20}_{-0.15}$& $0.017^{+0.006}_{-0.006}$ &$4.2\pm 0.2$ & $2.3 \times 10^{-7}$\\ \hline
        31.9& $1.5^{+5.9}_{-0.3}$  & $0.017^{+0.006}_{-0.012}$ & $0.62^{+0.37}_{-0.29}$& $0.011^{+0.007}_{-0.005}$ & $1.3 \pm 0.3$  & $1.9\times 10^{-7}$\\ \hline
        51.4& $<0.9$ & $0.008^{+0.001}_{-0.001}$ & -- & -- & $0.49 \pm 0.03$ & $1.6\times 10^{-7}$\\ \hline
    \end{tabular}
\end{table*}

\subsection{Density of the Shocked Gas} \label{sec:density}

The density of the gas being shocked, $\rho$, is a key parameter in determining shock signatures.  In V3890~Sgr, $\rho$ is the density of the pre-existing CSM, expelled from the red giant as a wind. Here we assume a spherical wind of constant mass-loss rate $\dot{M_{\rm w}}$ and velocity $v_{\rm w}$, which yields a CSM density profile of:
\begin{equation} \label{eq:mdot}
\rho = \frac{\dot{M_{\rm w}}}{4 \pi v_{\rm w} r^2}
\end{equation}
While the CSM in V3890~Sgr is likely more complex than a simple spherical wind, our VLBI images show that, at least for the first 32 days, the radio-emitting shock is likely to be interacting with the lower density spherical component (as opposed to the EDE). This shock component, with relatively low absorption (as opposed to the shocks in the orbital plane, which are likely dominating the $\gamma$-rays; \S \ref{sec:simul}), is also most likely to produce the X-rays detectable by \textit{Swift}/XRT.

We therefore estimate $\rho$ from the hard X-ray emission observed by \textit{Swift}/XRT \citep{Page+20};  APEC normalization factors and temperatures from these observations are listed in Table \ref{tab:Xray}. While the X-ray plasma temperatures listed in Table \ref{tab:Xray} are rather low, similarly modest X-ray temperatures have been measured in other symbiotic recurrent novae at \~1 week after optical peak (e.g., RS Oph and V745 Sco; \citealt{Orio+23,Delgado_Hernanz19}). It is possible that the hot X-ray plasma is mixing with cooler gas from behind the shock reducing the temperature. 

The APEC normalization factor constrains the number density of electrons and ions as:
\begin{equation} \label{eq:norm}
    {\rm Norm} = \frac{10^{-14}}{4 \pi D^2} \int n_{\rm e} n_{\rm H} dV 
\end{equation}
We solve for $n_H$ assuming that $n_{\rm e}=n_{\rm H}$, $D = 6.8$ kpc, and $V$ is the volume of the hot shocked region. 
We assume the hot APEC component is filling the volume between the forward shock and the contact discontinuity, take the radius of the forward shock from our VLBA images, and estimate the radius of the contact discontinuity ($R_{\rm CD}$) from the formalism described in \cite{Tang&Chevalier17} and \cite{Nyamai+23}: $R_{\rm CD} \approx 0.35\, R_{\rm FS}$ over the time-scale of our VLBA observations. The resulting densities of the hot shocked gas are listed in Table \ref{tab:Xray}.

These $n_{\rm H}$ estimates can be used to estimate $\rho$, the density of the gas being shocked, assuming a compression ratio of 4 as expected for a strong shock. 
We estimate the mass-loss rate of the wind $\dot{M_{\rm w}}$ from Equation \ref{eq:mdot}, assuming  $v_{\rm w} = 10$ km s$^{-1}$ as expected for a red giant, and the radii from our VLBA images, assuming $D = 6.8$ kpc. These $\dot{M_{\rm w}}$ estimates are also given in Table \ref{tab:Xray}.

We find that $\dot{M_{\rm w}} = 2 \times 10^{-7}$ M$_{\odot}$ yr$^{-1}$ in V3890~Sgr, consistently across our four imaging epochs as expected if the CSM truly is distributed as a wind. To sanity check this estimate of $\dot{M_{\rm w}}$, we searched for radio observations of V3890~Sgr in quiescence, when free-free thermal emission is expected to dominate, driven by the CSM ionized by the accreting white dwarf \citep{Seaquist&Taylor90}. While there are no targeted observations in the VLA archive, we obtained limits on the radio flux from the VLA Sky Survey (VLASS; \citealt{Lacy+20}). VLASS Epoch 1.1 observed V3890~Sgr on 2018 Feb 17, and reveals a low-significance peak ($2.9 \sigma$) near the position of V3890~Sgr. We are not confident that this is a detection, and therefore take a 3$\sigma$ upper limit of $<$0.59 mJy at 3.0 GHz. Using Equation 2 of \citet{Seaquist&Taylor90} and assuming $\alpha = 0.6$ in quiescence \citep{Wright&Barlow75}, we estimate an upper limit on $\dot{M_{\rm w}}$ of $< 5 \times 10^{-7}$ M$_{\odot}$ yr$^{-1}$, consistent with our X-ray estimates.

 $\dot{M_{\rm w}} = 2 \times 10^{-7}$ M$_{\odot}$ yr$^{-1}$ is typical for a symbiotic star, whose $\dot{M_{\rm w}}$ span $10^{-8}-10^{-5}$  M$_{\odot}$ yr$^{-1}$ \citep{Seaquist&Taylor90, Seaquist+93}. However, it is significantly higher than other symbiotic recurrent novae, like V745 Sco ($\dot{M_{\rm w}} = 2 \times 10^{-8}$ M$_{\odot}$ yr$^{-1}$; \citealt{Orlando+17, Molina+24}), and T CrB ($\dot{M_{\rm w}} = 2 \times 10^{-9}$ M$_{\odot}$ yr$^{-1}$; \citealt{Orlando+25}). The symbiotic recurrent nova RS~Oph also has a lower $\dot{M_{\rm w}}  \approx 6 \times 10^{-8}$ M$_{\odot}$ yr$^{-1}$ in the spherical CSM component \citep{Orlando+25}, but $\dot{M_{\rm EDE}} \approx 2 \times 10^{-7}$ M$_{\odot}$ yr$^{-1}$ in the EDE (for comparable $v_{\rm w} = 10$ km s$^{-1}$; \citealt{Lico+24}),  implying CSM that is more strongly confined to the orbital plane than in V3890~Sgr.
 Our estimate of $\dot{M_{\rm w}}$ is also significantly higher than the mass-loss rate estimated from modelling V3890~Sgr's radio light curve in \cite{Nyamai+23}: 
 $\dot{M_{\rm w}} \approx 10^{-8}$ M$_{\odot}$ yr$^{-1}$ (but we note their estimate depends crucially on the poorly known efficiencies of amplifying magnetic fields and accelerating relativistic electrons, and that it is possible to get good fits to the radio light curve with $\dot{M_{\rm w}} = 2 \times 10^{-7}$ M$_{\odot}$ yr$^{-1}$).
 The relatively dense environment of V3890~Sgr, and that this CSM appears relatively spherically distributed (as opposed to confined in an EDE)
 likely explains why its radio flux is so bright and radio images reach good S/N, even at a relatively large distance of 6.8 kpc. Its 5 GHz spectral radio luminosity at $d=6.8$ kpc is $2.9 \times 10^{21}$ erg s$^{-1}$ Hz$^{-1}$, a factor of $\sim 4-5$ greater than RS~Oph (\citealt{Eyres+09}, Molina et al. 2026, in preparation) and V745 Sco \citep{Molina+24}. 

\subsection{Efficiency of Magnetic Field Amplification and Lepton Acceleration} \label{sec:epsb}

We can use the magnetic field estimates from \S \ref{sec:bfield} to constrain the efficiency with which the shock  accelerates leptons ($\epsilon_e$) and the fraction of the post-shock pressure in magnetic fields ($\epsilon_{\rm B}$). We remind the reader that $B$ was estimated assuming equipartition with relativistic electrons, such that implicitly $\epsilon_{\rm B} = \epsilon_{\rm e}$, where $\epsilon_{\rm e}$ is the fraction of the post-shock pressure in relativistic electrons, $\epsilon_{\rm B}$ is defined such that $\frac{B^2}{8\pi} = \epsilon_{\rm B}\,\rho\,v_{\rm s}^2$, and $\rho$ is the density of gas being shocked.  These $\epsilon$ factors are commonly used in modelling radio emission from diverse astrophysical transients 
(e.g. \citealt{Chevalier82, Chandra&Frail12, Margutti&Chornock21, Bietenholz+21}). Table \ref{tab:Bfield} lists $B$ and $v_{\rm s}$ estimates from our VLBI images, and we estimate $\rho$ from the spherical CSM component $\dot{M_{\rm w}}$ estimated in the previous sub-section. The resulting values of $\epsilon_{\rm B}$ are listed in Table \ref{tab:Bfield}.

Our estimates of $\epsilon_{\rm B}$ and $\epsilon_{\rm e}$ range from $3 \times 10^{-4}$ to $2 \times 10^{-3}$, depending on the epoch and the assumed value of $K_0$. These equipartition parameters are smaller than what is often taken or inferred in studies of supernovae and $\gamma$-ray bursts \citep{Chevalier&Fransson06,Chomiuk+16,Chandra&Frail12}, which might be the result of the relatively low shock speeds in novae. For example, if the magnetic field is amplified via the non-resonant streaming instability \citep[][]{Bell04}, then $\epsilon_{\rm B}$ scales as $v_{\rm s}\epsilon_{\rm p}$, where $\epsilon_{\rm p}$ is the proton acceleration efficiency \citep[see, e.g.][for a detailed implementation]{Diesing+23}. Moreover, estimates of $\epsilon_{\rm e}$ in supernova remnants, which have shock velocities similar to those seen in novae, find similarly low efficiencies \citep{Morlino&Caprioli12, Sarbadhicary+17}.  However,  surprisingly, the efficiencies seem to increase between day 8 and 32 in V3890~Sgr, counter to expectations for a decelerating shock. This apparent increase might be attributed to shock expansion increasing the fraction of the emitting volume that is optically thin to radio (see Sections \ref{sec:simul} and \ref{sec:late_halo} for a more detailed discussion).


\subsection{Lessons from Simultaneous Modelling of Radio and $\gamma$-Ray Emission}\label{sec:simul}

V3890~Sgr achieved a peak $\gamma$-ray luminosity of $L_{\gamma} \approx1.3\times10^{36}$ erg s$^{-1}$ around the time of optical peak, assuming a distance of 6.8 kpc. The maximum possible $L_\gamma$ is the calorimetric limit, in which accelerated protons lose all of their energy in proton-proton collisions such that $L_{\gamma} = \kappa L_{\rm p}$, where $\kappa \approx 0.1$ is the fraction of the proton energy that goes into $\gamma-$rays, and $L_{\rm p}$ is the cosmic ray (proton) luminosity 
(e.g. \citealt{Metzger+15}). Taking $L_{\rm p} = \epsilon_{\rm p}L_{\rm s} \approx \epsilon_{\rm p}M_{\rm  ej}v_{\rm s}^2/t_{\rm age}$, where $L_s$ is the shock luminosity, $M_{\rm ej}$ is the nova ejecta mass, and $t_{\rm age}$ is the nova age, we approximate:
\begin{equation}
\begin{split}
    L_{\gamma} \approx 2.3\times10^{35}\text{ erg s}^{-1}  \\ \times \bigg(\frac{\epsilon_{\rm p}}{10^{-2}}\bigg)\bigg(\frac{M_{\rm ej}}{10^{-6}M_\odot}\bigg)\bigg(\frac{v_{\rm s}}{1000\text{ km s}^{-1}}\bigg)^2\bigg(\frac{t_{\rm age}}{\text{day}}\bigg)^{-1}.
\end{split}
\end{equation}
Thus, for $M_{\rm ej} \approx 10^{-6}M_\odot$ \citep{Nyamai+23} and $v_{\rm s} \approx 5000$ km s$^{-1}$, we obtain the observed $L_\gamma \approx 1.3\times10^{36}$ erg s$^{-1}$ at $t_{\rm age} = 4$ days. 

Of course, even higher values of $L_\gamma$ may be achievable at earlier times (the peak date of the $\gamma$-rays is uncertain) or with less conservative assumptions about the ejecta mass and acceleration efficiency. However, the fact remains that, to produce sufficient $\gamma$-ray flux, the density of the shocked gas must be very large, such that the pion production time-scale, $\tau_{\rm pp}$, is comparable to (or less than) $t_{\rm age}$. Taking $\tau_{\rm pp} \approx 260 \text{ days} \times(10^8\text{ cm}^{-3}/n_{\rm H})$ \citep{Diesing+23}, we obtain a density, $n_{\rm H} \gtrsim 6.5\times10^9$ cm$^{-3}$, at $t_{\rm age} = 5$ days. Not only is this density much larger than that inferred in Section \ref{sec:density}, it is almost certainly opaque to radio due to the large free-free optical depth. Note that a comparable density must be maintained until at least $t_{\rm age} \gtrsim 20$ days---well after the radio detection---since the $\gamma$-ray emission remains bright until that time.

In short, consistent with previous analysis in the context of classical novae 
(e.g. \citealt{Vlasov+16}), the $\gamma$-ray emission cannot arise from the same region as the radio. The most natural explanation is CSM characterized by an EDE. In this picture, a spherical outflow interacts with an aspherical environment: the interaction between the outflow and the dense equatorial medium produces a large $\gamma$-ray luminosity (but free-free absorption occludes any radio emission), while the interaction between the outflow and the lower-density red giant wind in the polar regions is responsible for the radio observations. This picture is consistent with the bipolar structure observed with VLBA at intermediate times (16 days $\lesssim t_{\rm age} \lesssim$ 32 days).

This picture is also supported by a comparison of radio and $\gamma$-ray luminosities of novae with evolved companions. While the peak 5 GHz radio luminosity scales with the density of the spherical wind component (\S \ref{sec:density}), the $\gamma$-ray luminosity does not.
 V407 Cyg had denser spherical CSM than V3890~Sgr ($\dot{M_{\rm w}} =  10^{-6}$ M$_{\odot}$ yr$^{-1}$; \citealt{Chomiuk+12}) but less luminous $\gamma$-rays (\S \ref{sec:fermi_lc}), while RS~Oph is characterized by lower $\dot{M_{\rm w}}$ and yet has more luminous $\gamma$-rays. We therefore predict that, of these three novae, RS~Oph had the densest EDE and V407 Cyg had the least dense. This is consistent with expectations for mass transfer by wind Roche lobe overflow, where a larger fraction of the CSM is expected to be confined in the EDE for symbiotic binaries with smaller separations \citep{Mohamed&Podsiadlowski12}.






\subsection{Origin of the Late-Time Diffuse Radio Emission}
\label{sec:late_halo}


The late-time radio excess observed in V3890~Sgr around $t \simeq 45$--$60$ days
cannot be explained by direct emission from the forward shock traced by our VLBI images.
At these epochs, the angular extent of the diffuse radio emission detected by the EVN+e-MERLIN
($\sim 0.1^{\prime\prime}$, corresponding to physical scales $\sim 10^{16}$~cm at $d = 6.8$~kpc)
is far larger than can be reached by ballistic nova ejecta expanding at the observed velocities
($\lesssim 5000$~km~s$^{-1}$; \S \ref{sec:dist}).
This disfavors an origin in freshly shocked ejecta or a simple density enhancement encountered by the forward shock.

Instead, we propose that the late-time radio emission arises from a {\it synchrotron halo}:
non-thermal particles escaping from the $\gamma$-ray--producing shock and diffusing upstream, to interact with pre-existing CSM at a relatively large radius.
In this picture, we focus on the shock in the equatorial region which, due to the large inferred post-shock density, is likely radiative; \S \ref{sec:simul}. Relativistic electrons and positrons---produced in proton-proton interactions within the dense, radiative, and relatively unmagnetized shell---leak out of the acceleration region,
propagate into the unshocked CSM, and radiate synchrotron emission over much larger spatial scales.  
In Appendix~\ref{app:halo}, we present a simple analytic model for this synchrotron halo,
demonstrating that the observed radio luminosity, spectral properties,
and spatial extent can be reproduced with physically reasonable parameters.

In this framework, the forward shock continues to accelerate hadrons and leptons at late times, particularly as indicated by the $\gamma$-ray re-brightening, but a fraction of the non-thermal particle population escapes upstream.
Such escape is expected once the shock weakens and the surrounding medium becomes less confining, particularly for high-energy protons and secondary leptons.
Moreover, if the observed $\gamma$-rays are produced predominantly through hadronic interactions,
then charged pion decay naturally injects relativistic electron-positron pairs with Lorentz factors
$\gamma \sim 10^2$--$10^4$, comparable to those required to produce GHz synchrotron emission
in magnetic fields of order $10^{-4}$--$10^{-2}$~G.

As these particles stream or diffuse into the unshocked red giant wind,
they encounter pre-existing magnetic fields carried by the wind itself.
For plausible red giant surface fields of order $\sim$G and flux-freezing in the outflow,
the magnetic field strength at radii $\sim 10^{15}$--$10^{16}$~cm is expected to be
$B \sim 10^{-3}$--$10^{-2}$~G, comparable to or slightly below the magnetic fields
we infer in the shocked region at earlier times (\S\ref{sec:bfield}).
Additional amplification may occur if escaping non-thermal ions drive plasma instabilities
in the upstream medium (e.g. \citealt{Bell04}) or if the latter gas was shocked and compressed by a previous recurrent nova eruption (\S\ref{subsec:mfa}).

Because the synchrotron cooling time of relativistic leptons in such fields
is much longer than both the wind expansion time and the particle transport time,
the emitting particles are firmly in the slow-cooling regime (\S\ref{subsec:slowcool}).
As a result, the radio luminosity reflects the cumulative injection and transport
of non-thermal particles rather than local radiative losses.
This naturally produces a low-surface-brightness, extended radio halo
whose characteristic size is set by particle propagation rather than by ejecta dynamics.


A natural question raised by this scenario is why the diffuse radio halo (and corresponding bump in the radio light curve)
appears contemporaneously with the second $\gamma$-ray peak, rather than
being dominated by particles accelerated during the earlier, more luminous
$\gamma$-ray phase near optical maximum.
Those early shocks undoubtedly produced large numbers of non-thermal
particles, which in principle could also escape upstream and propagate
to large radii.

If pairs were only produced during the first $\gamma$-ray peak, when the shock was still compact, then the delayed rise-time of the radio halo could in principle be attributed to the finite time required for the pairs to diffuse upstream into the red giant wind.  However, it would then be a coincidence that the radio rebrightening occurred roughly contemporaneous with the second $\gamma$-ray peak.  If the latter indicates ongoing shock interaction continuing to larger radii, the pairs released from this later shock phase may dominate the observed emission.
In particular, particles injected at early times escape at relatively small radii,
$r_0 \ll 10^{15}$~cm, and subsequently suffer substantial adiabatic energy
losses as they propagate outward through the red giant wind.
By the time such particles reach radii $\sim 10^{16}$~cm, their Lorentz
factors have been reduced by factors of $\sim r_0/r$, shifting their
synchrotron emission to frequencies well below the GHz band and rendering their contribution to the observed radio halo negligible.

The second $\gamma$-ray peak could mark the interaction of
the nova ejecta with a denser circumstellar structure at
$r \sim 10^{15}$~cm, such as a secondary density enhancement in the red giant wind or EDE inferred from the radio light curve. Shocks formed at these larger radii inject non-thermal particles with a
much larger effective escape radius $r_0$.
Because the subsequent adiabatic degradation from $r_0 \sim 10^{15}$~cm
to $r \sim 10^{16}$~cm is modest, these particles retain sufficiently high
Lorentz factors to radiate efficiently at GHz frequencies.
As a result, the late-time shocks dominate the observable synchrotron
halo, even if the total non-thermal energy injected at earlier times was
larger.  


\section{Conclusion} \label{sec:concl}

We present VLBI radio imaging and {\it Fermi}/LAT GeV $\gamma$-ray monitoring of the 2019 eruption of the symbiotic recurrent nova V3890~Sgr. VLBI imaging spans 8--51 days following eruption, concurrent with the detected $\gamma$-rays. We conclude:
\begin{itemize}
    \item The VLBI images start out asymmetric on day 8, with the eastern component brighter than the western component. However, the western component brightens with time, and by day 32, the radio emitting shell is quite circularly symmetric. We explain this evolution as bipolar expansion, with a density enhancement in the orbital plane oriented north--south and the western component absorbed by the EDE at early times (as in RS~Oph; \citealt{Munari+22, Lico+24}). In the EVN+e-MERLIN image from day 48, the shell appears brighter 
in the north--south direction, implying that at late times the shocks interacting with the denser EDE become visible at radio frequencies.
    \item We measure the expansion of the radio-emitting shell and compare it with radial velocities from optical spectroscopy. We thereby estimate an expansion parallax distance to V3890~Sgr of 6.8 kpc (\S \ref{sec:dist}). The radio images suggest 
significant deceleration of the shock between the first imaging epoch on day 8 and later epochs, from $\sim$6000 km s$^{-1}$ to 3100 km s$^{-1}$.
    \item By comparing the flux recovered from our VLBI imaging with integrated flux from the VLA \citep{Nyamai+23}, we conclude that the VLBA is detecting most ($>$80 per cent) of the radio flux on days 8--32, confirming that the radio emission is characterized by high brightness temperatures only achievable with non-thermal emission ($>10^7$ K). 
    \item On day 51, the VLBA resolves out $\sim$ a third of the integrated flux, but the EVN+e-MERLIN recovers this missing flux and finds it is distributed in a diffuse halo $\gtrsim 10^{16}$ cm across with peak $T_{\rm B} \approx 10^5$ K. The appearance of this radio halo is coincident with a second peak in the VLA radio light curve (Figure~\ref{fig:light}) and in the $\gamma$-ray light curve (Figure~\ref{fig:LAT_LC}). The radio spectrum of this second peak and the brightness temperature of the diffuse flux significantly in excess of $10^4$ K lead us to conclude that this halo is powered by synchrotron emission.
    \item We find the $>$100 MeV $\gamma$-rays peak at a flux of $(1.9 \pm 1.2) \times 10^{-7}$ phot s$^{-1}$ cm$^{-2}$ ($L_{\gamma} = 1.3 \times 10^{36}$ erg s$^{-1}$ at 6.8 kpc) around the time of optical maximum. The $\gamma$-rays remain detectable for 23 days, and then reappear at 2.5$\sigma$ significance around day 60, contemporaneous with the second peak in the radio light curve (\S \ref{sec:fermi_lc}).
    \item By modelling the simultaneous $\gamma$-ray and radio emission from V3890~Sgr, we conclude that radio emission and $\gamma$-rays must be originating from different portions of the shock: the high $\gamma$-ray luminosity requires a very dense medium for efficient pion production, while such a dense medium would free-free absorb radio emission. It is likely that the radio-emitting shock marks the ejecta's interaction with a spherical wind-like component of the CSM, while the $\gamma$-rays originate from interaction with an over-density of CSM in the orbital plane (\S \ref{sec:simul}).
    \item We estimate the density of the CSM from X-ray observations of the hot shocked gas \citep{Page+20}, finding that the CSM can be parameterized as a spherical wind with  $\dot{M_{\rm w}} = 2 \times 10^{-7}$ M$_{\odot}$ yr$^{-1}$ (assuming $v_{\rm w} = 10$ km s$^{-1}$), plus denser material in the orbital plane. This is significantly denser than the CSM estimated in other symbiotic recurrent novae (\S \ref{sec:density}).
    \item By combining our velocity, magnetic field, and CSM density measurements, we can estimate the fraction of the post-shock pressure in magnetic fields, and find $\epsilon_{\rm B} \approx$ few $\times 10^{-4}$ to few $\times 10^{-3}$ (\S \ref{sec:epsb}). We assume equipartition with relativistic electrons, so similar values of $\epsilon_{\rm e}$ are implied.
    \item The $\sim 10^{16}$ cm synchrotron halo and late-time radio bump can be explained by relativistic leptons that escape the shocked shell (most of secondary origin) and interact with CSM upstream of the shock (\S \ref{sec:late_halo}).
\end{itemize}

\section*{Acknowledgements}

Support for this work was provided by the NSF through the Grote Reber Fellowship Program administered by Associated Universities, Inc./National Radio Astronomy Observatory. Nova science at Michigan State is supported by NSF grant AST-2107070 and NASA grants 80NSSC23K0497, 80NSSC25K7334, and  80NSSC23K1247.  RD and BDM were supported in part by NASA (grant number 80NSSC22K0807, 80NSSC24K0408), the Simons Foundation (grant PG013158), and Columbia University through the Research Stabilization Fund.  The Flatiron Institute is supported by the Simons Foundation. KLP acknowledges support from the UK Space Agency. JLS acknowledges support from NASA grants 80NSSC25K0622 and 80NSSC25K7082.

The National Radio Astronomy Observatory and Green Bank Observatory are facilities of the U.S. National Science Foundation operated under cooperative agreement by Associated Universities, Inc. This work made use of the Swinburne University of Technology software correlator \citep{Deller+11}, developed as part of the Australian Major National Research Facilities Programme and operated under license. The European VLBI Network is a joint facility of independent European, African, Asian, and North American radio astronomy institutes. Scientific results from data presented in this publication are derived from the following EVN project code(s): RY008. e-MERLIN is a National Facility operated by the University of Manchester at Jodrell Bank Observatory on behalf of STFC. The research leading to these results has received funding from the European Commission Horizon 2020 Research and Innovation Programme under grant agreement No. 730562 (RadioNet). e-VLBI research infrastructure in Europe is supported by the European Union's Seventh Framework Programme (FP7/2007-2013) under grant agreement number RI-261525 NEXPReS.

\section*{Data Availability}
The radio images presented in this paper are available for download in fits format from \href{https://doi.org/10.5281/zenodo.18988180}{Zenodo}.

\appendix
\section{A Synchrotron Halo Model for the Late-Time Radio Emission}
\label{app:halo}

In this Appendix we outline a simple analytic model for the diffuse radio halo proposed in
\S~\ref{sec:late_halo}.
The goal is not to provide a detailed numerical treatment,
but rather to demonstrate that the observed radio luminosity,
characteristic size, and spectral properties follow naturally from
non-thermal particles escaping the $\gamma$-ray--producing shock
into the red giant wind.

\subsection{Injection and Transport of Escaping Pairs}

The late-time $\gamma$-ray luminosity is
$L_\gamma \sim 10^{35.5}$~erg~s$^{-1}$
(Figure~\ref{fig:LAT_LC}).
If the $\gamma$-rays originate from hadronic interactions,
a comparable power is injected into relativistic $e^\pm$ pairs through charged pion decay.
We therefore parametrize the power injected into escaping pairs as
\begin{equation}
L_{\pm} = \xi_{\pm} L_\gamma,
\end{equation}
with $\xi_{\pm}\sim \mathcal{O}(1)$.

Pairs escape the dense acceleration region at a characteristic radius
\begin{equation}
r_0 \sim 10^{15}\ \mathrm{cm},
\end{equation}
motivated by the VLBI size at $t\sim 50$ days (Table~\ref{tab:radio}) and the timing of the second radio peak (Figure \ref{fig:light}).  We take the injected pairs to have characteristic Lorentz factors
$\gamma_{\pm,0} \sim 10^{2}$--$10^{4}$, in the range expected if the observed
$E_{\gamma} \sim 1$--$10$~GeV emission arises from neutral pion decay
following inelastic $pp$ collisions.
The corresponding parent proton energies
($E_{\rm p} \sim 10$--$100$~GeV) imply secondary $e^\pm$ from charged pion
decay with energies $E_{\pm} \sim 0.1$--$10$~GeV, consistent with this
range of Lorentz factors.

After escape, particles stream or diffuse outward with an effective radial speed $v_{\pm}$.
To populate the halo out to $\sim 10^{16}$~cm within $\sim 50$ days requires
$v_{\pm}\gtrsim 0.1c$.
In steady state, conservation of particle flux gives a pair density profile
\begin{equation}
n_{\pm}(r) =
\frac{\dot{N}_{\pm}}{4\pi r^2 v_{\pm}}
=
\frac{L_{\pm}}{4\pi r^2 v_{\pm}\gamma_{\pm,0} m_{\rm e} c^2}.
\end{equation}

\subsection{Red Giant Wind Density and Magnetic Field}

We assume a steady red giant wind with mass-loss rate $\dot{M}_{\rm w}$
and velocity $v_{\rm w}$, yielding a density profile
\begin{equation}
\rho(r) = \frac{\dot{M}_{\rm w}}{4\pi v_{\rm w} r^2}.
\end{equation}
We adopt $\dot{M}_{\rm w} \simeq 2\times10^{-7}\,M_\odot\,\mathrm{yr^{-1}}$
and $v_{\rm w} \simeq 10$~km~s$^{-1}$, consistent with the X-ray constraints
(Table~\ref{tab:Xray}).

We assume the wind carries a magnetic field that is approximately frozen into the flow.
Beyond the Alfv\'en radius, a convenient scaling is
\begin{equation}
B(r) = B_\ast\left(\frac{R_\ast}{r}\right),
\end{equation}
where $B_\ast$ is the red-giant surface field at radius $R_\ast$.
For $B_\ast \sim 1$--$10$~G and $R_\ast \sim 1$~AU,
this yields $B \sim 10^{-3}$--$10^{-2}$~G at $r \sim 10^{15}$--$10^{16}$~cm,
comparable to the magnetic field strengths inferred in the shocked region
(Table~\ref{tab:Bfield}).

\subsection{Magnetic Field Amplification}
\label{subsec:mfa}
The magnetic field calculated in the previous section assumes that the red giant wind has reached $r \sim10^{16}$ cm since the previous (1990) nova eruption. Depending on the wind velocity, this may not be the case. However, we argue that $B \sim 10^{-3} - 10^{-2}$ is still a reasonable estimate. Not only can the post-shock region of the previous nova outburst contribute to the magnetic field at large radii, but escaping particles (ions and pairs) drive magnetic field amplification via streaming instabilities \citep[e.g.,][]{Weibel59,Skilling75,Bell04}. Assuming energy equipartition between the escaping pairs and the amplified field, we obtain,
\begin{equation}
    \frac{B^2(r)}{8\pi} \sim u_{\rm \pm}(r)\simeq \frac{L_\pm}{4\pi r^2 v_\pm}.
\end{equation}
Taking $v_\pm = 0.1c$, we find $B \sim 10^{-3}$ G for $L_\pm = L_\gamma = 10^{35}$ erg s$^{-1}$ and $r=10^{16}$ cm.

\subsection{Slow cooling, characteristic frequency, and adiabatic losses}
\label{subsec:slowcool}
At the field strengths relevant here, synchrotron cooling is extremely slow compared to
the transport/expansion time-scale.
We therefore assume that pairs cool predominantly adiabatically as they propagate outward, so that
\begin{equation}
\gamma_{\pm}(r) \simeq \gamma_{\pm,0}\left(\frac{r_0}{r}\right).
\end{equation}

The characteristic synchrotron frequency at radius $r$ is
\begin{equation}
\begin{aligned}
\nu_{\mathrm{syn}}(r) \simeq\;&
6\ \mathrm{GHz}
\left(\frac{B_\ast}{1\,\mathrm{G}}\right)
\left(\frac{R_\ast}{1\,\mathrm{AU}}\right) \\
&\times
\left(\frac{\gamma_{\pm,0}}{10^4}\right)^2
\left(\frac{r_0}{10^{16}\,\mathrm{cm}}\right)^2
\left(\frac{r}{10^{16}\,\mathrm{cm}}\right)^{-3}.
\end{aligned}
\label{eq:nusyn}
\end{equation}
so that GHz emission naturally arises at $r\sim 10^{16}$~cm for fiducial parameters.

The synchrotron cooling time is
\begin{equation}
\begin{aligned}
t_{\mathrm{syn}}(r) \simeq\;&
1.1\times10^{4}\ \mathrm{yr}
\left(\frac{B_\ast}{1\,\mathrm{G}}\right)^{-2}
\left(\frac{R_\ast}{1\,\mathrm{AU}}\right)^{-2} \\
&\times
\left(\frac{\gamma_{\pm,0}}{10^4}\right)^{-1}
\left(\frac{r_0}{10^{16}\,\mathrm{cm}}\right)^{-1}
\left(\frac{r}{10^{16}\,\mathrm{cm}}\right)^3.
\end{aligned}
\end{equation}
while the transport/adiabatic time-scale is $t_{\mathrm{ad}}(r)\sim r/v_\pm$.
For the parameter space of interest, $t_{\mathrm{ad}}\ll t_{\mathrm{syn}}$ at all relevant radii,
confirming that the pairs are in the slow-cooling regime.

\subsection{Radio luminosity and predicted flux density}
\label{app:halo_lum}

A useful way to estimate the synchrotron output in the slow-cooling regime is to consider the
energy contained in pairs within a logarithmic radial interval at radius $r$.
The residence time in a shell of width $d\ln r$ is $\sim r/v_\pm$, so the energy content is
\begin{equation}
\frac{dE_{\pm}}{d\ln r}(r)\ \simeq\ \left(\frac{r}{v_\pm}\right)\dot{N}_{\pm}\,\gamma_{\pm}(r)\,m_{\rm e} c^2,
\qquad
\dot{N}_{\pm}=\frac{L_{\pm}}{\gamma_{\pm,0}m_{\rm e}c^2}.
\end{equation}
Using $\gamma_{\pm}(r)=\gamma_{\pm,0}(r_0/r)$, the dependence on $r$ and $\gamma_{\pm,0}$ cancels, giving
\begin{equation}
\begin{aligned}
\frac{dE_{\pm}}{d\ln r} \simeq\;&
\frac{L_{\pm} r_0}{v_\pm} \\
\simeq\;&
3.3\times10^{39}
\left(\frac{L_{\pm}}{10^{35}\ \mathrm{erg\,s^{-1}}}\right)
\left(\frac{r_0}{10^{16}\ \mathrm{cm}}\right) \\
&\times
\left(\frac{v_\pm}{0.1c}\right)^{-1}
\ \mathrm{erg}.
\end{aligned}
\end{equation}

In the slow-cooling regime, the synchrotron power radiated from this interval is
\begin{equation}
\frac{dL_{\mathrm{syn}}}{d\ln r}\ \simeq\ \frac{1}{t_{\mathrm{syn}}(r)}\frac{dE_{\pm}}{d\ln r}.
\end{equation}
A convenient estimate of the spectral luminosity near the characteristic frequency emitted at radius $r$ is then
\begin{equation}
L_\nu(r)\ \sim\ \frac{1}{\nu_{\mathrm{syn}}(r)}\frac{dL_{\mathrm{syn}}}{d\ln r}
\ \sim\ \frac{1}{t_{\mathrm{syn}}(r)\,\nu_{\mathrm{syn}}(r)}\frac{dE_{\pm}}{d\ln r}.
\end{equation}
Substituting the expressions above and evaluating at the radius contributing to GHz emission
(i.e. $r\sim 10^{16}$~cm for fiducial parameters) yields
\begin{equation}
\begin{aligned}
L_\nu \simeq\;&
1.5\times10^{19}
\left(\frac{L_{\pm}}{10^{35}\ \mathrm{erg\,s^{-1}}}\right)
\left(\frac{B_\ast}{1\ \mathrm{G}}\right)
\left(\frac{R_\ast}{1\ \mathrm{AU}}\right) \\
&\times
\left(\frac{\gamma_{\pm,0}}{10^4}\right)^{-1}
\left(\frac{r_0}{10^{16}\ \mathrm{cm}}\right)
\left(\frac{v_\pm}{0.1c}\right)^{-1}
\ \mathrm{erg\,s^{-1}\,Hz^{-1}}.
\end{aligned}
\end{equation}

The corresponding flux density is
\begin{equation}
\begin{aligned}
F_\nu \simeq\;&
0.28\
\left(\frac{L_{\pm}}{10^{35}\ \mathrm{erg\,s^{-1}}}\right)
\left(\frac{B_\ast}{1\ \mathrm{G}}\right)
\left(\frac{R_\ast}{1\ \mathrm{AU}}\right) \\
&\times
\left(\frac{\gamma_{\pm,0}}{10^4}\right)^{-1}
\left(\frac{r_0}{10^{16}\ \mathrm{cm}}\right)
\left(\frac{v_\pm}{0.1c}\right)^{-1} \\
&\times
\left(\frac{d}{6.8\ \mathrm{kpc}}\right)^{-2}
\ \mathrm{mJy}.
\end{aligned}
\label{eq:Fnu}
\end{equation}

For $\xi_\pm\sim 1$ and $L_\gamma \sim 10^{35.5}$~erg~s$^{-1}$ (Figure~\ref{fig:LAT_LC}),
this predicts $F_\nu \sim 0.1$--$1$ mJy for plausible values of $(B_\ast,R_\ast,\gamma_{\pm,0},v_\pm)$.
This is of the same order as the observed diffuse component dominating the second radio bump
in the integrated light curve (Figure~\ref{fig:light}), up to order-unity factors associated with
the breadth of the synchrotron spectrum and the mapping between observing frequency and emission radius.

Finally, because $\nu_{\mathrm{syn}}\propto r^{-3}$ under the scalings adopted above,
a broad range of radii contributes across a decade in observing frequency.
This naturally produces a shallow, approximately power-law radio spectrum and
a low-surface-brightness halo whose characteristic extent is set by particle transport.

Although non-thermal particles are produced throughout the eruption,
the contribution of a given population to the observed GHz radio halo
depends sensitively on the radius $r_0$ at which the particles escape
the dense shock region.
For particles injected at early times ($r_0 \ll 10^{15}$~cm), adiabatic
losses reduce their Lorentz factors as $\gamma_\pm(r)\propto r_0/r$,
so that by the time they reach radii $\sim 10^{16}$~cm their characteristic
synchrotron frequencies lie well below the GHz band.
In contrast, particles injected at later times, when the shock forms at
$r_0 \sim 10^{15}$~cm, experience only modest adiabatic degradation and
retain sufficient energy to emit efficiently at radio frequencies.

This effect is explicitly reflected in the scalings derived below:
while the characteristic synchrotron frequency depends on both $r_0$ and
$r$ (equation~\eqref{eq:nusyn}), the normalization of the radio luminosity or flux scales linearly with $r_0$ (equation~\eqref{eq:Fnu}).
As a result, late-time injection at larger radii naturally dominates the
observable synchrotron halo, even if earlier shocks were more luminous
in $\gamma$-rays.

\newpage

\bibliography{radionovae}

@string{june = {June}}

@ARTICLE{Delgado_Hernanz19,
       author = {{Delgado}, L. and {Hernanz}, M.},
        title = "{Early multiwavelength analysis of the recurrent nova V745 Sco}",
      journal = {\mnras},
         year = 2019,
        month = dec,
       volume = {490},
       number = {3},
        pages = {3691-3704},
          doi = {10.1093/mnras/stz2765},
archivePrefix = {arXiv},
       eprint = {1910.08940},
 primaryClass = {astro-ph.HE},
       adsurl = {https://ui.adsabs.harvard.edu/abs/2019MNRAS.490.3691D}
}

@ARTICLE{Orio+23,
       author = {{Orio}, Marina and {Gendreau}, Keith and {Giese}, Morgan and {Luna}, Gerardo Juan M. and {Magdolen}, Jozef and {Strohmayer}, Tod E. and {Zhang}, Andy E. and {Altamirano}, Diego and {Dobrotka}, Andrej and {Enoto}, Teruaki and {Ferrara}, Elizabeth C. and {Ignace}, Richard and {Heinz}, Sebastian and {Markwardt}, Craig and {Nichols}, Joy S. and {Parker}, Michael L. and {Pasham}, Dheeraj R. and {Pei}, Songpeng and {Pradhan}, Pragati and {Remillard}, Ron and {Steiner}, James F. and {Tombesi}, Francesco},
        title = "{The RS Oph Outburst of 2021 Monitored in X-Rays with NICER}",
      journal = {\apj},
         year = 2023,
        month = sep,
       volume = {955},
       number = {1},
          eid = {37},
        pages = {37},
          doi = {10.3847/1538-4357/ace9bd},
archivePrefix = {arXiv},
       eprint = {2307.11485},
 primaryClass = {astro-ph.HE},
       adsurl = {https://ui.adsabs.harvard.edu/abs/2023ApJ...955...37O}
}

@article{1998PASP..110....3G,
 adsurl = {https://ui.adsabs.harvard.edu/abs/1998PASP..110....3G},
 author = {{Gehrz}, Robert D. and {Truran}, James W. and {Williams}, Robert E. and {Starrfield}, Sumner},
 doi = {10.1086/316107},
 journal = {\pasp},
 month = {January},
 number = {743},
 pages = {3-26},
 title = {{Nucleosynthesis in Classical Novae and Its Contribution to the Interstellar Medium}},
 volume = {110},
 year = {1998}
}

@article{2014A&A...563L...9D,
 adsurl = {https://ui.adsabs.harvard.edu/abs/2014A&A...563L...9D},
 archiveprefix = {arXiv},
 author = {{Darnley}, M.~J. and {Williams}, S.~C. and {Bode}, M.~F. and {Henze}, M. and {Ness}, J.-U. and {Shafter}, A.~W. and {Hornoch}, K. and {Votruba}, V.},
 doi = {10.1051/0004-6361/201423411},
 eid = {L9},
 eprint = {1401.2905},
 journal = {\aap},
 month = {March},
 pages = {L9},
 primaryclass = {astro-ph.SR},
 title = {{A remarkable recurrent nova in M 31: The optical observations}},
 volume = {563},
 year = {2014}
}

@article{2020A&A...638A.130G,
 adsurl = {https://ui.adsabs.harvard.edu/abs/2020A&A...638A.130G},
 archiveprefix = {arXiv},
 author = {{Giroletti}, M. and {Munari}, U. and {K{\"o}rding}, E. and {Mioduszewski}, A. and {Sokoloski}, J. and {Cheung}, C.~C. and {Corbel}, S. and {Schinzel}, F. and {Sokolovsky}, K. and {O'Brien}, T.~J.},
 doi = {10.1051/0004-6361/202038142},
 eid = {A130},
 eprint = {2005.06473},
 journal = {\aap},
 month = {June},
 pages = {A130},
 primaryclass = {astro-ph.SR},
 title = {{Very long baseline interferometry imaging of the advancing ejecta in the first gamma-ray nova V407 Cygni}},
 volume = {638},
 year = {2020}
}

@inproceedings{2021gacv.workE..44D,
 adsurl = {https://ui.adsabs.harvard.edu/abs/2021gacv.workE..44D},
 archiveprefix = {arXiv},
 author = {{Darnley}, M.~J.},
 booktitle = {The Golden Age of Cataclysmic Variables and Related Objects V},
 doi = {10.22323/1.368.0044},
 eid = {44},
 eprint = {1912.13209},
 month = {February},
 pages = {44},
 primaryclass = {astro-ph.SR},
 title = {{Accrete, Accrete, Accrete{\textellipsis} Bang! (and repeat): The remarkable Recurrent Novae}},
 volume = {2-7},
 year = {2021}
}

@article{2025ApJ...994..229B,
 adsurl = {https://ui.adsabs.harvard.edu/abs/2025ApJ...994..229B},
 archiveprefix = {arXiv},
 author = {{Basu}, Judhajeet and {Anupama}, G.~C. and {Ness}, Jan-Uwe and {Singh}, Kulinder Pal and {Barway}, Sudhanshu and {Chamoli}, Shatakshi},
 doi = {10.3847/1538-4357/ae13e1},
 eid = {229},
 eprint = {2510.11231},
 journal = {\apj},
 month = {December},
 number = {2},
 pages = {229},
 primaryclass = {astro-ph.HE},
 title = {{Survival of the Accretion Disk in LMC Recurrent Nova 1968-12a: UV--X-Ray Case Study of the 2024 Eruption}},
 volume = {994},
 year = {2025}
}

@article{Abdo+10,
 adsurl = {http://adsabs.harvard.edu/abs/2010Sci...329..817A},
 archiveprefix = {arXiv},
 author = {{Abdo}, A.~A. and {Ackermann}, M. and {Ajello}, M. and others},
 doi = {10.1126/science.1192537},
 eprint = {1008.3912},
 journal = {Science},
 month = {August},
 pages = {817-821},
 primaryclass = {astro-ph.HE},
 title = {{Gamma-Ray Emission Concurrent with the Nova in the Symbiotic Binary V407 Cygni}},
 volume = {329},
 year = {2010}
}

@article{Abdollahi_etal_2022,
 adsurl = {https://ui.adsabs.harvard.edu/abs/2022ApJS..260...53A},
 archiveprefix = {arXiv},
 author = {{Abdollahi}, S. and {Acero}, F. and {Baldini}, L. and {Ballet}, J. and {Bastieri}, D. and {Bellazzini}, R. and {Berenji}, B. and {Berretta}, A. and {Bissaldi}, E. and {Blandford}, R.~D. and {Bloom}, E. and {Bonino}, R. and {Brill}, A. and {Britto}, R.~J. and {Bruel}, P. and {Burnett}, T.~H. and {Buson}, S. and {Cameron}, R.~A. and {Caputo}, R. and {Caraveo}, P.~A. and {Castro}, D. and {Chaty}, S. and {Cheung}, C.~C. and {Chiaro}, G. and {Cibrario}, N. and {Ciprini}, S. and {Coronado-Bl{\'a}zquez}, J. and {Crnogorcevic}, M. and {Cutini}, S. and {D'Ammando}, F. and {De Gaetano}, S. and {Digel}, S.~W. and {Di Lalla}, N. and {Dirirsa}, F. and {Di Venere}, L. and {Dom{\'\i}nguez}, A. and {Fallah Ramazani}, V. and {Fegan}, S.~J. and {Ferrara}, E.~C. and {Fiori}, A. and {Fleischhack}, H. and {Franckowiak}, A. and {Fukazawa}, Y. and {Funk}, S. and {Fusco}, P. and {Galanti}, G. and {Gammaldi}, V. and {Gargano}, F. and {Garrappa}, S. and {Gasparrini}, D. and {Giacchino}, F. and {Giglietto}, N. and {Giordano}, F. and {Giroletti}, M. and {Glanzman}, T. and {Green}, D. and {Grenier}, I.~A. and {Grondin}, M. -H. and {Guillemot}, L. and {Guiriec}, S. and {Gustafsson}, M. and {Harding}, A.~K. and {Hays}, E. and {Hewitt}, J.~W. and {Horan}, D. and {Hou}, X. and {J{\'o}hannesson}, G. and {Karwin}, C. and {Kayanoki}, T. and {Kerr}, M. and {Kuss}, M. and {Landriu}, D. and {Larsson}, S. and {Latronico}, L. and {Lemoine-Goumard}, M. and {Li}, J. and {Liodakis}, I. and {Longo}, F. and {Loparco}, F. and {Lott}, B. and {Lubrano}, P. and {Maldera}, S. and {Malyshev}, D. and {Manfreda}, A. and {Mart{\'\i}-Devesa}, G. and {Mazziotta}, M.~N. and {Mereu}, I. and {Meyer}, M. and {Michelson}, P.~F. and {Mirabal}, N. and {Mitthumsiri}, W. and {Mizuno}, T. and {Moiseev}, A.~A. and {Monzani}, M.~E. and {Morselli}, A. and {Moskalenko}, I.~V. and {Negro}, M. and {Nuss}, E. and {Omodei}, N. and {Orienti}, M. and {Orlando}, E. and {Paneque}, D. and {Pei}, Z. and {Perkins}, J.~S. and {Persic}, M. and {Pesce-Rollins}, M. and {Petrosian}, V. and {Pillera}, R. and {Poon}, H. and {Porter}, T.~A. and {Principe}, G. and {Rain{\`o}}, S. and {Rando}, R. and {Rani}, B. and {Razzano}, M. and {Razzaque}, S. and {Reimer}, A. and {Reimer}, O. and {Reposeur}, T. and {S{\'a}nchez-Conde}, M. and {Saz Parkinson}, P.~M. and {Scotton}, L. and {Serini}, D. and {Sgr{\`o}}, C. and {Siskind}, E.~J. and {Smith}, D.~A. and {Spandre}, G. and {Spinelli}, P. and {Sueoka}, K. and {Suson}, D.~J. and {Tajima}, H. and {Tak}, D. and {Thayer}, J.~B. and {Thompson}, D.~J. and {Torres}, D.~F. and {Troja}, E. and {Valverde}, J. and {Wood}, K. and {Zaharijas}, G.},
 doi = {10.3847/1538-4365/ac6751},
 eid = {53},
 eprint = {2201.11184},
 journal = {\apjs},
 month = {June},
 number = {2},
 pages = {53},
 primaryclass = {astro-ph.HE},
 title = {{Incremental Fermi Large Area Telescope Fourth Source Catalog}},
 volume = {260},
 year = {2022}
}

@article{Acciari+22,
 adsurl = {https://ui.adsabs.harvard.edu/abs/2022NatAs...6..689A},
 archiveprefix = {arXiv},
 author = {{Acciari}, V.~A. and {Ansoldi}, S. and {Antonelli}, L.~A. and {Arbet Engels}, A. and {Artero}, M. and {Asano}, K. and {Baack}, D. and {Babi{\'c}}, A. and {Baquero}, A. and {Barres de Almeida}, U. and {Barrio}, J.~A. and {Batkovi{\'c}}, I. and {Becerra Gonz{\'a}lez}, J. and {Bednarek}, W. and {Bellizzi}, L. and {Bernardini}, E. and {Bernardos}, M. and {Berti}, A. and {Besenrieder}, J. and {Bhattacharyya}, W. and {Bigongiari}, C. and {Biland}, A. and {Blanch}, O. and {B{\"o}kenkamp}, H. and {Bonnoli}, G. and {Bo{\v{s}}njak}, {\v{Z}}. and {Busetto}, G. and {Carosi}, R. and {Ceribella}, G. and {Cerruti}, M. and {Chai}, Y. and {Chilingarian}, A. and {Cikota}, S. and {Colak}, S.~M. and {Colombo}, E. and {Contreras}, J.~L. and {Cortina}, J. and {Covino}, S. and {D'Amico}, G. and {D'Elia}, V. and {Da Vela}, P. and {Dazzi}, F. and {De Angelis}, A. and {De Lotto}, B. and {Del Popolo}, A. and {Delfino}, M. and {Delgado}, J. and {Delgado Mendez}, C. and {Depaoli}, D. and {Di Pierro}, F. and {Di Venere}, L. and {Do Souto Espi{\~n}eira}, E. and {Prester}, D. Dominis and {Donini}, A. and {Dorner}, D. and {Doro}, M. and {Elsaesser}, D. and {Fallah Ramazani}, V. and {Fari{\~n}a Alonso}, L. and {Fattorini}, A. and {Fonseca}, M.~V. and {Font}, L. and {Fruck}, C. and {Fukami}, S. and {Fukazawa}, Y. and {Garc{\'\i}a L{\'o}pez}, R.~J. and {Garczarczyk}, M. and {Gasparyan}, S. and {Gaug}, M. and {Giglietto}, N. and {Giordano}, F. and {Gliwny}, P. and {Godinovi{\'c}}, N. and {Green}, J.~G. and {Green}, D. and {Hadasch}, D. and {Hahn}, A. and {Hassan}, T. and {Heckmann}, L. and {Herrera}, J. and {Hoang}, J. and {Hrupec}, D. and {H{\"u}tten}, M. and {Inada}, T. and {Ishio}, K. and {Iwamura}, Y. and {Jim{\'e}nez Mart{\'\i}nez}, I. and {Jormanainen}, J. and {Jouvin}, L. and {Kerszberg}, D. and {Kobayashi}, Y. and {Kubo}, H. and {Kushida}, J. and {Lamastra}, A. and {Lelas}, D. and {Leone}, F. and {Lindfors}, E. and {Linhoff}, L. and {Lombardi}, S. and {Longo}, F. and {L{\'o}pez-Coto}, R. and {L{\'o}pez-Moya}, M. and {L{\'o}pez-Oramas}, A. and {Loporchio}, S. and {Machado de Oliveira Fraga}, B. and {Maggio}, C. and {Majumdar}, P. and {Makariev}, M. and {Mallamaci}, M. and {Maneva}, G. and {Manganaro}, M. and {Mannheim}, K. and {Maraschi}, L. and {Mariotti}, M. and {Mart{\'\i}nez}, M. and {Mas Aguilar}, A. and {Mazin}, D. and {Menchiari}, S. and {Mender}, S. and {Mi{\'c}anovi{\'c}}, S. and {Miceli}, D. and {Miener}, T. and {Miranda}, J.~M. and {Mirzoyan}, R. and {Molina}, E. and {Moralejo}, A. and {Morcuende}, D. and {Moreno}, V. and {Moretti}, E. and {Nakamori}, T. and {Nava}, L. and {Neustroev}, V. and {Nievas Rosillo}, M. and {Nigro}, C. and {Nilsson}, K. and {Nishijima}, K. and {Noda}, K. and {Nozaki}, S. and {Ohtani}, Y. and {Oka}, T. and {Otero-Santos}, J. and {Paiano}, S. and {Palatiello}, M. and {Paneque}, D. and {Paoletti}, R. and {Paredes}, J.~M. and {Pavleti{\'c}}, L. and {Pe{\~n}il}, P. and {Persic}, M. and {Pihet}, M. and {Prada Moroni}, P.~G. and {Prandini}, E. and {Priyadarshi}, C. and {Puljak}, I. and {Rhode}, W. and {Rib{\'o}}, M. and {Rico}, J. and {Righi}, C. and {Rugliancich}, A. and {Sahakyan}, N. and {Saito}, T. and {Sakurai}, S. and {Satalecka}, K. and {Saturni}, F.~G. and {Schleicher}, B. and {Schmidt}, K. and {Schweizer}, T. and {Sitarek}, J. and {{\v{S}}nidari{\'c}}, I. and {Sobczynska}, D. and {Spolon}, A. and {Stamerra}, A. and {Stri{\v{s}}kovi{\'c}}, J. and {Strom}, D. and {Strzys}, M. and {Suda}, Y. and {Suri{\'c}}, T. and {Takahashi}, M. and {Takeishi}, R. and {Tavecchio}, F. and {Temnikov}, P. and {Terzi{\'c}}, T. and {Teshima}, M. and {Tosti}, L. and {Truzzi}, S. and {Tutone}, A. and {Ubach}, S. and {van Scherpenberg}, J. and {Vanzo}, G. and {Vazquez Acosta}, M. and {Ventura}, S. and {Verguilov}, V. and {Vigorito}, C.~F. and {Vitale}, V. and {Vovk}, I. and {Will}, M. and {Wunderlich}, C. and {Yamamoto}, T. and {Zari{\'c}}, D. and {Ambrosino}, F.},
 doi = {10.1038/s41550-022-01640-z},
 eprint = {2202.07681},
 journal = {Nature Astronomy},
 month = {April},
 pages = {689-697},
 primaryclass = {astro-ph.HE},
 title = {{Proton acceleration in thermonuclear nova explosions revealed by gamma rays}},
 volume = {6},
 year = {2022}
}

@article{Ackermann+14,
 adsurl = {https://ui.adsabs.harvard.edu/abs/2014Sci...345..554A},
 archiveprefix = {arXiv},
 author = {{Ackermann}, M. and {Ajello}, M. and {Albert}, A. and {Baldini}, L. and {Ballet}, J. and {Barbiellini}, G. and {Bastieri}, D. and {Bellazzini}, R. and {Bissaldi}, E. and {Blandford}, R.~D. and {Bloom}, E.~D. and {Bottacini}, E. and {Brandt}, T.~J. and {Bregeon}, J. and {Bruel}, P. and {Buehler}, R. and {Buson}, S. and {Caliandro}, G.~A. and {Cameron}, R.~A. and {Caragiulo}, M. and {Caraveo}, P.~A. and {Cavazzuti}, E. and {Charles}, E. and {Chekhtman}, A. and {Cheung}, C.~C. and {Chiang}, J. and {Chiaro}, G. and {Ciprini}, S. and {Claus}, R. and {Cohen-Tanugi}, J. and {Conrad}, J. and {Corbel}, S. and {D'Ammando}, F. and {de Angelis}, A. and {den Hartog}, P.~R. and {de Palma}, F. and {Dermer}, C.~D. and {Desiante}, R. and {Digel}, S.~W. and {Di Venere}, L. and {do Couto e Silva}, E. and {Donato}, D. and {Drell}, P.~S. and {Drlica-Wagner}, A. and {Favuzzi}, C. and {Ferrara}, E.~C. and {Focke}, W.~B. and {Franckowiak}, A. and {Fuhrmann}, L. and {Fukazawa}, Y. and {Fusco}, P. and {Gargano}, F. and {Gasparrini}, D. and {Germani}, S. and {Giglietto}, N. and {Giordano}, F. and {Giroletti}, M. and {Glanzman}, T. and {Godfrey}, G. and {Grenier}, I.~A. and {Grove}, J.~E. and {Guiriec}, S. and {Hadasch}, D. and {Harding}, A.~K. and {Hayashida}, M. and {Hays}, E. and {Hewitt}, J.~W. and {Hill}, A.~B. and {Hou}, X. and {Jean}, P. and {Jogler}, T. and {J{\'o}hannesson}, G. and {Johnson}, A.~S. and {Johnson}, W.~N. and {Kerr}, M. and {Kn{\"o}dlseder}, J. and {Kuss}, M. and {Larsson}, S. and {Latronico}, L. and {Lemoine-Goumard}, M. and {Longo}, F. and {Loparco}, F. and {Lott}, B. and {Lovellette}, M.~N. and {Lubrano}, P. and {Manfreda}, A. and {Martin}, P. and {Massaro}, F. and {Mayer}, M. and {Mazziotta}, M.~N. and {McEnery}, J.~E. and {Michelson}, P.~F. and {Mitthumsiri}, W. and {Mizuno}, T. and {Monzani}, M.~E. and {Morselli}, A. and {Moskalenko}, I.~V. and {Murgia}, S. and {Nemmen}, R. and {Nuss}, E. and {Ohsugi}, T. and {Omodei}, N. and {Orienti}, M. and {Orlando}, E. and {Ormes}, J.~F. and {Paneque}, D. and {Panetta}, J.~H. and {Perkins}, J.~S. and {Pesce-Rollins}, M. and {Piron}, F. and {Pivato}, G. and {Porter}, T.~A. and {Rain{\`o}}, S. and {Rando}, R. and {Razzano}, M. and {Razzaque}, S. and {Reimer}, A. and {Reimer}, O. and {Reposeur}, T. and {Saz Parkinson}, P.~M. and {Schaal}, M. and {Schulz}, A. and {Sgr{\`o}}, C. and {Siskind}, E.~J. and {Spandre}, G. and {Spinelli}, P. and {Stawarz}, {\L}. and {Suson}, D.~J. and {Takahashi}, H. and {Tanaka}, T. and {Thayer}, J.~G. and {Thayer}, J.~B. and {Thompson}, D.~J. and {Tibaldo}, L. and {Tinivella}, M. and {Torres}, D.~F. and {Tosti}, G. and {Troja}, E. and {Uchiyama}, Y. and {Vianello}, G. and {Winer}, B.~L. and {Wolff}, M.~T. and {Wood}, D.~L. and {Wood}, K.~S. and {Wood}, M. and {Charbonnel}, S. and {Corbet}, R.~H.~D. and {De Gennaro Aquino}, I. and {Edlin}, J.~P. and {Mason}, E. and {Schwarz}, G.~J. and {Shore}, S.~N. and {Starrfield}, S. and {Teyssier}, F. and {Fermi-LAT Collaboration}},
 doi = {10.1126/science.1253947},
 eprint = {1408.0735},
 journal = {Science},
 month = {August},
 number = {6196},
 pages = {554-558},
 primaryclass = {astro-ph.HE},
 title = {{Fermi establishes classical novae as a distinct class of gamma-ray sources}},
 volume = {345},
 year = {2014}
}

@article{Anupama+13,
 adsurl = {http://adsabs.harvard.edu/abs/2013A%26A...559A.121A},
 author = {{Anupama}, G.~C. and {Kamath}, U.~S. and {Ramaprakash}, A.~N. and 
{Kantharia}, N.~G. and {Hegde}, M. and {Mohan}, V. and {Kulkarni}, M. and 
{Bode}, M.~F. and {Eyres}, S.~P.~S. and {Evans}, A. and {O'Brien}, T.~J.},
 doi = {10.1051/0004-6361/201321262},
 eid = {A121},
 journal = {\aap},
 month = {November},
 pages = {A121},
 title = {{The 2010 outburst and pre-outburst optical spectrum of the recurrent nova U Scorpii}},
 volume = {559},
 year = {2013}
}

@article{Anupama_Sethi94,
 adsurl = {https://ui.adsabs.harvard.edu/abs/1994MNRAS.269..105A},
 author = {{Anupama}, G.~C. and {Sethi}, S.},
 doi = {10.1093/mnras/269.1.105},
 journal = {\mnras},
 month = {July},
 pages = {105},
 title = {{Spectroscopy of the Recurrent Nova V3890-SAGITTARII 18-DAYS after the 1990 Outburst}},
 volume = {269},
 year = {1994}
}

@article{Aydi+20,
 adsurl = {https://ui.adsabs.harvard.edu/abs/2020NatAs...4..776A},
 archiveprefix = {arXiv},
 author = {{Aydi}, Elias and {Sokolovsky}, Kirill V. and {Chomiuk}, Laura and {Steinberg}, Elad and {Li}, Kwan Lok and {Vurm}, Indrek and {Metzger}, Brian D. and {Strader}, Jay and {Mukai}, Koji and {Pejcha}, Ond{\v{r}}ej and {Shen}, Ken J. and {Wade}, Gregg A. and {Kuschnig}, Rainer and {Moffat}, Anthony F.~J. and {Pablo}, Herbert and {Pigulski}, Andrzej and {Popowicz}, Adam and {Weiss}, Werner and {Zwintz}, Konstanze and {Izzo}, Luca and {Pollard}, Karen R. and {Handler}, Gerald and {Ryder}, Stuart D. and {Filipovi{\'c}}, Miroslav D. and {Alsaberi}, Rami Z.~E. and {Manojlovi{\'c}}, Perica and {Lopes de Oliveira}, Raimundo and {Walter}, Frederick M. and {Vallely}, Patrick J. and {Buckley}, David A.~H. and {Brown}, Michael J.~I. and {Harvey}, Eamonn J. and {Kawash}, Adam and {Kniazev}, Alexei and {Kochanek}, Christopher S. and {Linford}, Justin and {Mikolajewska}, Joanna and {Molaro}, Paolo and {Orio}, Marina and {Page}, Kim L. and {Shappee}, Benjamin J. and {Sokoloski}, Jennifer L.},
 doi = {10.1038/s41550-020-1070-y},
 eprint = {2004.05562},
 journal = {Nature Astronomy},
 month = {April},
 pages = {776-780},
 primaryclass = {astro-ph.HE},
 title = {{Direct evidence for shock-powered optical emission in a nova}},
 volume = {4},
 year = {2020}
}

@article{Beck_Krause05,
 adsurl = {https://ui.adsabs.harvard.edu/abs/2005AN....326..414B},
 archiveprefix = {arXiv},
 author = {{Beck}, R. and {Krause}, M.},
 doi = {10.1002/asna.200510366},
 eprint = {astro-ph/0507367},
 journal = {Astronomische Nachrichten},
 month = {July},
 number = {6},
 pages = {414-427},
 primaryclass = {astro-ph},
 title = {{Revised equipartition and minimum energy formula for magnetic field strength estimates from radio synchrotron observations}},
 volume = {326},
 year = {2005}
}

@article{Bell04,
 adsurl = {https://ui.adsabs.harvard.edu/abs/2004MNRAS.353..550B},
 author = {{Bell}, A.~R.},
 doi = {10.1111/j.1365-2966.2004.08097.x},
 journal = {\mnras},
 month = {September},
 number = {2},
 pages = {550-558},
 title = {{Turbulent amplification of magnetic field and diffusive shock acceleration of cosmic rays}},
 volume = {353},
 year = {2004}
}

@article{Bietenholz+21,
 adsurl = {https://ui.adsabs.harvard.edu/abs/2021ApJ...908...75B},
 archiveprefix = {arXiv},
 author = {{Bietenholz}, M.~F. and {Bartel}, N. and {Argo}, M. and {Dua}, R. and {Ryder}, S. and {Soderberg}, A.},
 doi = {10.3847/1538-4357/abccd9},
 eid = {75},
 eprint = {2011.11737},
 journal = {\apj},
 month = {February},
 number = {1},
 pages = {75},
 primaryclass = {astro-ph.HE},
 title = {{The Radio Luminosity-risetime Function of Core-collapse Supernovae}},
 volume = {908},
 year = {2021}
}

@book{Bode&Evans08,
 address = {Cambridge},
 adsurl = {http://cdsads.u-strasbg.fr/abs/2008clno.book.....B},
 author = {{Bode}, M.~F. and {Evans}, A.},
 booktitle = {Classical Novae, 2nd Edition.~Edited by M.F.~Bode and A.~Evans.~Cambridge Astrophysics Series, No.~43, Cambridge: Cambridge University Press, 2008.},
 doi = {10.1017/CBO9780511536168},
 month = {April},
 publisher = {Cambridge University Press},
 title = {{Classical Novae}},
 year = {2008}
}

@inproceedings{Booth+14,
 adsurl = {https://ui.adsabs.harvard.edu/abs/2014IAUS..296..382B},
 author = {{Booth}, Richard A. and {Mohamed}, Shazrene and {Podsiadlowski}, Philipp},
 booktitle = {Supernova Environmental Impacts},
 doi = {10.1017/S1743921313009939},
 editor = {{Ray}, Alak and {McCray}, Richard A.},
 month = {January},
 pages = {382-383},
 series = {IAU Symposium},
 title = {{Simulations of RS Oph and the CSM in Type Ia Supernovae}},
 volume = {296},
 year = {2014}
}

@article{Buckley+90,
 adsurl = {https://ui.adsabs.harvard.edu/abs/1990IAUC.5019....1B},
 author = {{Buckley}, D.~A.~H. and {Wargau}, W.~F. and {Soltynski}, M.~G. and {Shao}, C.-Y. and {Hazen}, M.~L.},
 journal = {\iaucirc},
 month = {May},
 pages = {1},
 title = {{V3890 Sagittarii}},
 volume = {5019},
 year = {1990}
}

@article{Buson+19,
 adsurl = {https://ui.adsabs.harvard.edu/abs/2019ATel13114....1B},
 author = {{Buson}, S. and {Jean}, P. and {Cheung}, C.~C.},
 journal = {The Astronomer's Telegram},
 month = {September},
 pages = {1},
 title = {{Fermi-LAT Gamma-ray Detection of Symbiotic Recurrent Nova V3890 Sgr}},
 volume = {13114},
 year = {2019}
}

@article{Chandra&Frail12,
 adsurl = {https://ui.adsabs.harvard.edu/abs/2012ApJ...746..156C},
 archiveprefix = {arXiv},
 author = {{Chandra}, Poonam and {Frail}, Dale A.},
 doi = {10.1088/0004-637X/746/2/156},
 eid = {156},
 eprint = {1110.4124},
 journal = {\apj},
 month = {February},
 number = {2},
 pages = {156},
 primaryclass = {astro-ph.CO},
 title = {{A Radio-selected Sample of Gamma-Ray Burst Afterglows}},
 volume = {746},
 year = {2012}
}

@article{Chen+16,
 adsurl = {https://ui.adsabs.harvard.edu/abs/2016MNRAS.458.2916C},
 archiveprefix = {arXiv},
 author = {{Chen}, Hai-Liang and {Woods}, T.~E. and {Yungelson}, L.~R. and {Gilfanov}, M. and {Han}, Zhanwen},
 doi = {10.1093/mnras/stw458},
 eprint = {1602.07849},
 journal = {\mnras},
 month = {May},
 number = {3},
 pages = {2916-2927},
 primaryclass = {astro-ph.SR},
 title = {{Modelling nova populations in galaxies}},
 volume = {458},
 year = {2016}
}

@ARTICLE{Cheung+22,
       author = {{Cheung}, C.~C. and {Johnson}, T.~J. and {Jean}, P. and {Kerr}, M. and {Page}, K.~L. and {Osborne}, J.~P. and {Beardmore}, A.~P. and {Sokolovsky}, K.~V. and {Teyssier}, F. and {Ciprini}, S. and {Mart{\'\i}-Devesa}, G. and {Mereu}, I. and {Razzaque}, S. and {Wood}, K.~S. and {Shore}, S.~N. and {Korotkiy}, S. and {Levina}, A. and {Blumenzweig}, A.},
        title = "{Fermi LAT Gamma-ray Detection of the Recurrent Nova RS Ophiuchi during its 2021 Outburst}",
      journal = {\apj},
         year = 2022,
        month = aug,
       volume = {935},
       number = {1},
          eid = {44},
        pages = {44},
          doi = {10.3847/1538-4357/ac7eb7},
archivePrefix = {arXiv},
       eprint = {2207.02921},
 primaryClass = {astro-ph.HE},
       adsurl = {https://ui.adsabs.harvard.edu/abs/2022ApJ...935...44C}
}

@article{Chevalier&Fransson06,
 adsurl = {https://ui.adsabs.harvard.edu/abs/2006ApJ...651..381C},
 archiveprefix = {arXiv},
 author = {{Chevalier}, Roger A. and {Fransson}, Claes},
 doi = {10.1086/507606},
 eprint = {astro-ph/0607196},
 journal = {\apj},
 month = {November},
 number = {1},
 pages = {381-391},
 primaryclass = {astro-ph},
 title = {{Circumstellar Emission from Type Ib and Ic Supernovae}},
 volume = {651},
 year = {2006}
}

@article{Chevalier82,
 adsurl = {https://ui.adsabs.harvard.edu/abs/1982ApJ...259..302C},
 author = {{Chevalier}, R.~A.},
 doi = {10.1086/160167},
 journal = {\apj},
 month = {August},
 pages = {302-310},
 title = {{The radio and X-ray emission from type II supernovae.}},
 volume = {259},
 year = {1982}
}

@article{Chomiuk+12,
 adsurl = {http://adsabs.harvard.edu/abs/2012ApJ...761..173C},
 archiveprefix = {arXiv},
 author = {{Chomiuk}, L. and {Krauss}, M.~I. and {Rupen}, M.~P. and {Nelson}, T. and 
{Roy}, N. and {Sokoloski}, J.~L. and {Mukai}, K. and {Munari}, U. and 
{Mioduszewski}, A. and {Weston}, J. and {O'Brien}, T.~J. and 
{Eyres}, S.~P.~S. and {Bode}, M.~F.},
 doi = {10.1088/0004-637X/761/2/173},
 eid = {173},
 eprint = {1210.6029},
 journal = {\apj},
 month = {December},
 pages = {173},
 primaryclass = {astro-ph.HE},
 title = {{The Radio Light Curve of the Gamma-Ray Nova in V407 Cyg: Thermal Emission from the Ionized Symbiotic Envelope, Devoured from within by the Nova Blast}},
 volume = {761},
 year = {2012}
}

@article{Chomiuk+14,
 adsurl = {http://adsabs.harvard.edu/abs/2014Natur.514..339C},
 archiveprefix = {arXiv},
 author = {{Chomiuk}, L. and {Linford}, J.~D. and {Yang}, J. and {O'Brien}, T.~J. and 
{Paragi}, Z. and {Mioduszewski}, A.~J. and {Beswick}, R.~J. and 
{Cheung}, C.~C. and {Mukai}, K. and {Nelson}, T. and {Ribeiro}, V.~A.~R.~M. and 
{Rupen}, M.~P. and {Sokoloski}, J.~L. and {Weston}, J. and {Zheng}, Y. and 
{Bode}, M.~F. and {Eyres}, S. and {Roy}, N. and {Taylor}, G.~B.},
 doi = {10.1038/nature13773},
 eprint = {1410.3473},
 journal = {\nat},
 month = {October},
 pages = {339-342},
 primaryclass = {astro-ph.HE},
 title = {{Binary orbits as the driver of {$\gamma$}-ray emission and mass ejection in classical novae}},
 volume = {514},
 year = {2014}
}

@article{Chomiuk+16,
 adsurl = {https://ui.adsabs.harvard.edu/abs/2016ApJ...821..119C},
 archiveprefix = {arXiv},
 author = {{Chomiuk}, Laura and {Soderberg}, Alicia M. and {Chevalier}, Roger A. and {Bruzewski}, Seth and {Foley}, Ryan J. and {Parrent}, Jerod and {Strader}, Jay and {Badenes}, Carles and {Fransson}, Claes and {Kamble}, Atish and {Margutti}, Raffaella and {Rupen}, Michael P. and {Simon}, Joshua D.},
 doi = {10.3847/0004-637X/821/2/119},
 eid = {119},
 eprint = {1510.07662},
 journal = {\apj},
 month = {April},
 number = {2},
 pages = {119},
 primaryclass = {astro-ph.HE},
 title = {{A Deep Search for Prompt Radio Emission from Thermonuclear Supernovae with the Very Large Array}},
 volume = {821},
 year = {2016}
}

@article{Chomiuk+21,
 adsurl = {https://ui.adsabs.harvard.edu/abs/2021ARA&A..59..391C},
 archiveprefix = {arXiv},
 author = {{Chomiuk}, Laura and {Metzger}, Brian D. and {Shen}, Ken J.},
 doi = {10.1146/annurev-astro-112420-114502},
 eprint = {2011.08751},
 journal = {\araa},
 month = {September},
 pages = {391-444},
 primaryclass = {astro-ph.HE},
 title = {{New Insights into Classical Novae}},
 volume = {59},
 year = {2021}
}

@article{Chomiuk+21radio,
 adsurl = {https://ui.adsabs.harvard.edu/abs/2021ApJS..257...49C},
 archiveprefix = {arXiv},
 author = {{Chomiuk}, Laura and {Linford}, Justin D. and {Aydi}, Elias and {Bannister}, Keith W. and {Krauss}, Miriam I. and {Mioduszewski}, Amy J. and {Mukai}, Koji and {Nelson}, Thomas J. and {Rupen}, Michael P. and {Ryder}, Stuart D. and {Sokoloski}, Jennifer L. and {Sokolovsky}, Kirill V. and {Strader}, Jay and {Filipovi{\'c}}, Miroslav D. and {Finzell}, Tom and {Kawash}, Adam and {Kool}, Erik C. and {Metzger}, Brian D. and {Nyamai}, Miriam M. and {Ribeiro}, Val{\'e}rio A.~R.~M. and {Roy}, Nirupam and {Urquhart}, Ryan and {Weston}, Jennifer},
 doi = {10.3847/1538-4365/ac24ab},
 eid = {49},
 eprint = {2107.06251},
 journal = {\apjs},
 month = {December},
 number = {2},
 pages = {49},
 primaryclass = {astro-ph.HE},
 title = {{Classical Novae at Radio Wavelengths}},
 volume = {257},
 year = {2021}
}

@article{Craig+26,
 adsurl = {https://ui.adsabs.harvard.edu/abs/2025MNRAS.tmp.2143C},
 archiveprefix = {arXiv},
 author = {{Craig}, Peter and {Aydi}, Elias and {Chomiuk}, Laura and {Stone}, Ashley and {Strader}, Jay and {Chong}, Atticus and {Li}, Kwan-Lok and {Fan}, Jhih-Ling and {Bahramian}, Arash and {Buckley}, David A.~H. and {Izzo}, Luca and {Kawash}, Adam and {Metzger}, Brian D. and {Mukai}, Koji and {Linford}, Justin D. and {Orio}, Marina and {Sokoloski}, J.~L. and {Sokolovsky}, Kirill V. and {Tremou}, Evangelia and {Walter}, Frederick M. and {Fl{\'o}}, Joan Guarro and {Boussin}, Christophe and {Charbonnel}, St{\'e}phane and {Garde}, Olivier and {Belyakov}, Konstantin and {Monard}, Libert A.~G. and {Hambsch}, Franz-Josef and {Thomas}, Neil},
 doi = {10.1093/mnras/staf2270},
 eprint = {2508.15900},
 journal = {\mnras},
 month = {December},
 primaryclass = {astro-ph.HE},
 title = {{What determines the {\ensuremath{\gamma}}-ray luminosities of classical novae?}},
 year = {2025}
}

@article{Deller+11,
 adsurl = {https://ui.adsabs.harvard.edu/abs/2011PASP..123..275D},
 archiveprefix = {arXiv},
 author = {{Deller}, A.~T. and {Brisken}, W.~F. and {Phillips}, C.~J. and {Morgan}, J. and {Alef}, W. and {Cappallo}, R. and {Middelberg}, E. and {Romney}, J. and {Rottmann}, H. and {Tingay}, S.~J. and {Wayth}, R.},
 doi = {10.1086/658907},
 eprint = {1101.0885},
 journal = {\pasp},
 month = {March},
 number = {901},
 pages = {275},
 primaryclass = {astro-ph.IM},
 title = {{DiFX-2: A More Flexible, Efficient, Robust, and Powerful Software Correlator}},
 volume = {123},
 year = {2011}
}

@article{Diesing+23,
 adsurl = {https://ui.adsabs.harvard.edu/abs/2023ApJ...947...70D},
 archiveprefix = {arXiv},
 author = {{Diesing}, Rebecca and {Metzger}, Brian D. and {Aydi}, Elias and {Chomiuk}, Laura and {Vurm}, Indrek and {Gupta}, Siddhartha and {Caprioli}, Damiano},
 doi = {10.3847/1538-4357/acc105},
 eid = {70},
 eprint = {2211.02059},
 journal = {\apj},
 month = {April},
 number = {2},
 pages = {70},
 primaryclass = {astro-ph.HE},
 title = {{Evidence for Multiple Shocks from the {\ensuremath{\gamma}}-Ray Emission of RS Ophiuchi}},
 volume = {947},
 year = {2023}
}

@article{Evans+22,
 adsurl = {https://ui.adsabs.harvard.edu/abs/2022MNRAS.517.6077E},
 archiveprefix = {arXiv},
 author = {{Evans}, A. and {Geballe}, T.~R. and {Woodward}, C.~E. and {Banerjee}, D.~P.~K. and {Gehrz}, R.~D. and {Starrfield}, S. and {Shahbandeh}, M.},
 doi = {10.1093/mnras/stac2363},
 eprint = {2208.09356},
 journal = {\mnras},
 month = {December},
 number = {4},
 pages = {6077-6090},
 primaryclass = {astro-ph.SR},
 title = {{Infrared spectroscopy of the 2019 eruption of the recurrent nova V3890 Sgr: Separation into equatorial and polar winds revealed}},
 volume = {517},
 year = {2022}
}

@article{Eyres+09,
 adsurl = {https://ui.adsabs.harvard.edu/abs/2009MNRAS.395.1533E},
 archiveprefix = {arXiv},
 author = {{Eyres}, S.~P.~S. and {O'Brien}, T.~J. and {Beswick}, R. and {Muxlow}, T.~W.~B. and {Anupama}, G.~C. and {Kantharia}, N.~G. and {Bode}, M.~F. and {Gawro{\'n}ski}, M.~P. and {Feiler}, R. and {Evans}, A. and {Rushton}, M.~T. and {Davis}, R.~J. and {Prabhu}, T. and {Porcas}, R. and {Hassall}, B.~J.~M.},
 doi = {10.1111/j.1365-2966.2009.14633.x},
 eprint = {0902.2265},
 journal = {\mnras},
 month = {May},
 number = {3},
 pages = {1533-1540},
 primaryclass = {astro-ph.SR},
 title = {{Double radio peak and non-thermal collimated ejecta in RS Ophiuchi following the 2006 outburst}},
 volume = {395},
 year = {2009}
}

@article{Finzell+18,
 adsurl = {https://ui.adsabs.harvard.edu/abs/2018ApJ...852..108F},
 archiveprefix = {arXiv},
 author = {{Finzell}, Thomas and {Chomiuk}, Laura and {Metzger}, Brian D. and {Walter}, Frederick M. and {Linford}, Justin D. and {Mukai}, Koji and {Nelson}, Thomas and {Weston}, Jennifer H.~S. and {Zheng}, Yong and {Sokoloski}, Jennifer L. and {Mioduszewski}, Amy and {Rupen}, Michael P. and {Dong}, Subo and {Starrfield}, Sumner and {Cheung}, C.~C. and {Woodward}, Charles E. and {Taylor}, Gregory B. and {Bohlsen}, Terry and {Buil}, Christian and {Prieto}, Jose and {Wagner}, R. Mark and {Bensby}, Thomas and {Bond}, I.~A. and {Sumi}, T. and {Bennett}, D.~P. and {Abe}, F. and {Koshimoto}, N. and {Suzuki}, D. and {Tristram}, P.~J. and {Christie}, Grant W. and {Natusch}, Tim and {McCormick}, Jennie and {Yee}, Jennifer and {Gould}, Andy},
 doi = {10.3847/1538-4357/aaa12a},
 eid = {108},
 eprint = {1701.03094},
 journal = {\apj},
 month = {January},
 number = {2},
 pages = {108},
 primaryclass = {astro-ph.SR},
 title = {{A Detailed Observational Analysis of V1324 Sco, the Most Gamma-Ray-luminous Classical Nova to Date}},
 volume = {852},
 year = {2018}
}

@article{Franckowiak+18,
 adsurl = {https://ui.adsabs.harvard.edu/abs/2018A&A...609A.120F},
 archiveprefix = {arXiv},
 author = {{Franckowiak}, A. and {Jean}, P. and {Wood}, M. and {Cheung}, C.~C. and {Buson}, S.},
 doi = {10.1051/0004-6361/201731516},
 eid = {A120},
 eprint = {1710.04736},
 journal = {\aap},
 month = {February},
 pages = {A120},
 primaryclass = {astro-ph.HE},
 title = {{Search for gamma-ray emission from Galactic novae with the Fermi -LAT}},
 volume = {609},
 year = {2018}
}

@article{Gaia16,
 adsurl = {https://ui.adsabs.harvard.edu/abs/2016A&A...595A...1G},
 archiveprefix = {arXiv},
 author = {{Gaia Collaboration} and {Prusti}, T. and {de Bruijne}, J.~H.~J. and {Brown}, A.~G.~A. and {Vallenari}, A. and {Babusiaux}, C. and {Bailer-Jones}, C.~A.~L. and {Bastian}, U. and {Biermann}, M. and {Evans}, D.~W. and et al.},
 doi = {10.1051/0004-6361/201629272},
 eid = {A1},
 eprint = {1609.04153},
 journal = {\aap},
 month = {November},
 pages = {A1},
 primaryclass = {astro-ph.IM},
 title = {{The Gaia mission}},
 volume = {595},
 year = {2016}
}

@article{Gaia23,
 adsurl = {https://ui.adsabs.harvard.edu/abs/2023A&A...674A...1G},
 archiveprefix = {arXiv},
 author = {{Gaia Collaboration} and {Vallenari}, A. and {Brown}, A.~G.~A. and {Prusti}, T. and {de Bruijne}, J.~H.~J. and {Arenou}, F. and {Babusiaux}, C. and {Biermann}, M. and {Creevey}, O.~L. and {Ducourant}, C. and {Evans}, D.~W. and {Eyer}, L. and {Guerra}, R. and {Hutton}, A. and {Jordi}, C. and {Klioner}, S.~A. and {Lammers}, U.~L. and {Lindegren}, L. and {Luri}, X. and {Mignard}, F. and {Panem}, C. and {Pourbaix}, D. and {Randich}, S. and {Sartoretti}, P. and {Soubiran}, C. and {Tanga}, P. and {Walton}, N.~A. and {Bailer-Jones}, C.~A.~L. and {Bastian}, U. and {Drimmel}, R. and {Jansen}, F. and {Katz}, D. and {Lattanzi}, M.~G. and {van Leeuwen}, F. and {Bakker}, J. and {Cacciari}, C. and {Casta{\~n}eda}, J. and {De Angeli}, F. and {Fabricius}, C. and {Fouesneau}, M. and {Fr{\'e}mat}, Y. and {Galluccio}, L. and {Guerrier}, A. and {Heiter}, U. and {Masana}, E. and {Messineo}, R. and {Mowlavi}, N. and {Nicolas}, C. and {Nienartowicz}, K. and {Pailler}, F. and {Panuzzo}, P. and {Riclet}, F. and {Roux}, W. and {Seabroke}, G.~M. and {Sordo}, R. and {Th{\'e}venin}, F. and {Gracia-Abril}, G. and {Portell}, J. and {Teyssier}, D. and {Altmann}, M. and {Andrae}, R. and {Audard}, M. and {Bellas-Velidis}, I. and {Benson}, K. and {Berthier}, J. and {Blomme}, R. and {Burgess}, P.~W. and {Busonero}, D. and {Busso}, G. and {C{\'a}novas}, H. and {Carry}, B. and {Cellino}, A. and {Cheek}, N. and {Clementini}, G. and {Damerdji}, Y. and {Davidson}, M. and {de Teodoro}, P. and {Nu{\~n}ez Campos}, M. and {Delchambre}, L. and {Dell'Oro}, A. and {Esquej}, P. and {Fern{\'a}ndez-Hern{\'a}ndez}, J. and {Fraile}, E. and {Garabato}, D. and {Garc{\'\i}a-Lario}, P. and {Gosset}, E. and {Haigron}, R. and {Halbwachs}, J. -L. and {Hambly}, N.~C. and {Harrison}, D.~L. and {Hern{\'a}ndez}, J. and {Hestroffer}, D. and {Hodgkin}, S.~T. and {Holl}, B. and {Jan{\ss}en}, K. and {Jevardat de Fombelle}, G. and {Jordan}, S. and {Krone-Martins}, A. and {Lanzafame}, A.~C. and {L{\"o}ffler}, W. and {Marchal}, O. and {Marrese}, P.~M. and {Moitinho}, A. and {Muinonen}, K. and {Osborne}, P. and {Pancino}, E. and {Pauwels}, T. and {Recio-Blanco}, A. and {Reyl{\'e}}, C. and {Riello}, M. and {Rimoldini}, L. and {Roegiers}, T. and {Rybizki}, J. and {Sarro}, L.~M. and {Siopis}, C. and {Smith}, M. and {Sozzetti}, A. and {Utrilla}, E. and {van Leeuwen}, M. and {Abbas}, U. and {{\'A}brah{\'a}m}, P. and {Abreu Aramburu}, A. and {Aerts}, C. and {Aguado}, J.~J. and {Ajaj}, M. and {Aldea-Montero}, F. and {Altavilla}, G. and {{\'A}lvarez}, M.~A. and {Alves}, J. and {Anders}, F. and {Anderson}, R.~I. and {Anglada Varela}, E. and {Antoja}, T. and {Baines}, D. and {Baker}, S.~G. and {Balaguer-N{\'u}{\~n}ez}, L. and {Balbinot}, E. and {Balog}, Z. and {Barache}, C. and {Barbato}, D. and {Barros}, M. and {Barstow}, M.~A. and {Bartolom{\'e}}, S. and {Bassilana}, J. -L. and {Bauchet}, N. and {Becciani}, U. and {Bellazzini}, M. and {Berihuete}, A. and {Bernet}, M. and {Bertone}, S. and {Bianchi}, L. and {Binnenfeld}, A. and {Blanco-Cuaresma}, S. and {Blazere}, A. and {Boch}, T. and {Bombrun}, A. and {Bossini}, D. and {Bouquillon}, S. and {Bragaglia}, A. and {Bramante}, L. and {Breedt}, E. and {Bressan}, A. and {Brouillet}, N. and {Brugaletta}, E. and {Bucciarelli}, B. and {Burlacu}, A. and {Butkevich}, A.~G. and {Buzzi}, R. and {Caffau}, E. and {Cancelliere}, R. and {Cantat-Gaudin}, T. and {Carballo}, R. and {Carlucci}, T. and {Carnerero}, M.~I. and {Carrasco}, J.~M. and {Casamiquela}, L. and {Castellani}, M. and {Castro-Ginard}, A. and {Chaoul}, L. and {Charlot}, P. and {Chemin}, L. and {Chiaramida}, V. and {Chiavassa}, A. and {Chornay}, N. and {Comoretto}, G. and {Contursi}, G. and {Cooper}, W.~J. and {Cornez}, T. and {Cowell}, S. and {Crifo}, F. and {Cropper}, M. and {Crosta}, M. and {Crowley}, C. and {Dafonte}, C. and {Dapergolas}, A. and {David}, M. and {David}, P. and {de Laverny}, P. and {De Luise}, F. and {De March}, R.},
 doi = {10.1051/0004-6361/202243940},
 eid = {A1},
 eprint = {2208.00211},
 journal = {\aap},
 month = {June},
 pages = {A1},
 primaryclass = {astro-ph.GA},
 title = {{Gaia Data Release 3. Summary of the content and survey properties}},
 volume = {674},
 year = {2023}
}

@article{Gallagher&Starrfield78,
 adsurl = {https://ui.adsabs.harvard.edu/abs/1978ARA&A..16..171G},
 author = {{Gallagher}, J.~S. and {Starrfield}, S.},
 doi = {10.1146/annurev.aa.16.090178.001131},
 journal = {\araa},
 month = {January},
 pages = {171-214},
 title = {{Theory and observations of classical novae.}},
 volume = {16},
 year = {1978}
}

@article{Gordon+21,
 adsurl = {https://ui.adsabs.harvard.edu/abs/2021ApJ...910..134G},
 archiveprefix = {arXiv},
 author = {{Gordon}, A.~C. and {Aydi}, E. and {Page}, K.~L. and {Li}, Kwan-Lok and {Chomiuk}, L. and {Sokolovsky}, K.~V. and {Mukai}, K. and {Seitz}, J.},
 doi = {10.3847/1538-4357/abe547},
 eid = {134},
 eprint = {2010.15930},
 journal = {\apj},
 month = {April},
 number = {2},
 pages = {134},
 primaryclass = {astro-ph.HE},
 title = {{Surveying the X-Ray Behavior of Novae as They Emit {\ensuremath{\gamma}}-Rays}},
 volume = {910},
 year = {2021}
}

@article{Green19,
 adsurl = {https://ui.adsabs.harvard.edu/abs/2019JApA...40...36G},
 archiveprefix = {arXiv},
 author = {{Green}, D.~A.},
 doi = {10.1007/s12036-019-9601-6},
 eid = {36},
 eprint = {1907.02638},
 journal = {Journal of Astrophysics and Astronomy},
 month = {August},
 number = {4},
 pages = {36},
 primaryclass = {astro-ph.GA},
 title = {{A revised catalogue of 294 Galactic supernova remnants}},
 volume = {40},
 year = {2019}
}

@inbook{Greisen03,
 adsurl = {https://ui.adsabs.harvard.edu/abs/2003ASSL..285..109G},
 author = {{Greisen}, E.~W.},
 booktitle = {Information Handling in Astronomy - Historical Vistas},
 publisher = {Springer, Dordrecht},
 doi = {10.1007/0-306-48080-8\_7},
 editor = {{Heck}, Andr{\'e}},
 pages = {109},
 title = {{AIPS, the VLA, and the VLBA}},
 volume = {285},
 year = {2003}
}

@article{Harrison+93,
 adsurl = {https://ui.adsabs.harvard.edu/abs/1993AJ....105..320H},
 author = {{Harrison}, T.~E. and {Johnson}, J.~J. and {Spyromilio}, J.},
 doi = {10.1086/116429},
 journal = {\aj},
 month = {January},
 pages = {320},
 title = {{Infrared Observations of the Recurrent Novae V745 SCO and V3890 SGR}},
 volume = {105},
 year = {1993}
}

@article{Kaminsky+22,
 adsurl = {https://ui.adsabs.harvard.edu/abs/2022MNRAS.517.6064K},
 archiveprefix = {arXiv},
 author = {{Kaminsky}, B. and {Evans}, A. and {Pavlenko}, Ya V. and {Woodward}, C.~E. and {Banerjee}, D.~P.~K. and {Gehrz}, R.~D. and {Walter}, F. and {Starrfield}, S. and {Ilyin}, I. and {Strassmeier}, K.~G. and {Wagner}, R.~M.},
 doi = {10.1093/mnras/stac2199},
 eprint = {2207.14721},
 journal = {\mnras},
 month = {December},
 number = {4},
 pages = {6064-6076},
 primaryclass = {astro-ph.SR},
 title = {{The recurrent nova V3890 Sgr: a near-infrared and optical study of the red giant component and its environment}},
 volume = {517},
 year = {2022}
}

@article{Kemp+21,
 adsurl = {https://ui.adsabs.harvard.edu/abs/2021MNRAS.504.6117K},
 archiveprefix = {arXiv},
 author = {{Kemp}, Alex J. and {Karakas}, Amanda I. and {Casey}, Andrew R. and {Izzard}, Robert G. and {Ruiter}, Ashley J. and {Agrawal}, Poojan and {Broekgaarden}, Floor S. and {Temmink}, Karel D.},
 doi = {10.1093/mnras/stab1160},
 eprint = {2104.10870},
 journal = {\mnras},
 month = {July},
 number = {4},
 pages = {6117-6143},
 primaryclass = {astro-ph.SR},
 title = {{Population synthesis of accreting white dwarfs: rates and evolutionary pathways of H and He novae}},
 volume = {504},
 year = {2021}
}

@book{Kenyon86,
 adsurl = {https://ui.adsabs.harvard.edu/abs/1986syst.book.....K},
 author = {{Kenyon}, S.~J.},
 publisher = {Cambridge University Press},
 title = {{The symbiotic stars}},
 year = {1986}
}

@article{Lacy+20,
 adsurl = {https://ui.adsabs.harvard.edu/abs/2020PASP..132c5001L},
 archiveprefix = {arXiv},
 author = {{Lacy}, M. and {Baum}, S.~A. and {Chandler}, C.~J. and {Chatterjee}, S. and {Clarke}, T.~E. and {Deustua}, S. and {English}, J. and {Farnes}, J. and {Gaensler}, B.~M. and {Gugliucci}, N. and {Hallinan}, G. and {Kent}, B.~R. and {Kimball}, A. and {Law}, C.~J. and {Lazio}, T.~J.~W. and {Marvil}, J. and {Mao}, S.~A. and {Medlin}, D. and {Mooley}, K. and {Murphy}, E.~J. and {Myers}, S. and {Osten}, R. and {Richards}, G.~T. and {Rosolowsky}, E. and {Rudnick}, L. and {Schinzel}, F. and {Sivakoff}, G.~R. and {Sjouwerman}, L.~O. and {Taylor}, R. and {White}, R.~L. and {Wrobel}, J. and {Andernach}, H. and {Beasley}, A.~J. and {Berger}, E. and {Bhatnager}, S. and {Birkinshaw}, M. and {Bower}, G.~C. and {Brandt}, W.~N. and {Brown}, S. and {Burke-Spolaor}, S. and {Butler}, B.~J. and {Comerford}, J. and {Demorest}, P.~B. and {Fu}, H. and {Giacintucci}, S. and {Golap}, K. and {G{\"u}th}, T. and {Hales}, C.~A. and {Hiriart}, R. and {Hodge}, J. and {Horesh}, A. and {Ivezi{\'c}}, {\v{Z}}. and {Jarvis}, M.~J. and {Kamble}, A. and {Kassim}, N. and {Liu}, X. and {Loinard}, L. and {Lyons}, D.~K. and {Masters}, J. and {Mezcua}, M. and {Moellenbrock}, G.~A. and {Mroczkowski}, T. and {Nyland}, K. and {O'Dea}, C.~P. and {O'Sullivan}, S.~P. and {Peters}, W.~M. and {Radford}, K. and {Rao}, U. and {Robnett}, J. and {Salcido}, J. and {Shen}, Y. and {Sobotka}, A. and {Witz}, S. and {Vaccari}, M. and {van Weeren}, R.~J. and {Vargas}, A. and {Williams}, P.~K.~G. and {Yoon}, I.},
 doi = {10.1088/1538-3873/ab63eb},
 eid = {035001},
 eprint = {1907.01981},
 journal = {\pasp},
 month = {March},
 number = {1009},
 pages = {035001},
 primaryclass = {astro-ph.IM},
 title = {{The Karl G. Jansky Very Large Array Sky Survey (VLASS). Science Case and Survey Design}},
 volume = {132},
 year = {2020}
}

@article{Lico+24,
 adsurl = {https://ui.adsabs.harvard.edu/abs/2024A&A...692A.107L},
 archiveprefix = {arXiv},
 author = {{Lico}, R. and {Giroletti}, M. and {Munari}, U. and {O'Brien}, T.~J. and {Marcote}, B. and {Williams}, D.~R.~A. and {Yang}, J. and {Veres}, P. and {Woudt}, P.},
 doi = {10.1051/0004-6361/202451364},
 eid = {A107},
 eprint = {2407.05794},
 journal = {\aap},
 month = {December},
 pages = {A107},
 primaryclass = {astro-ph.HE},
 title = {{High-resolution imaging of the evolving bipolar outflows in symbiotic novae: The case of the RS Ophiuchi 2021 nova outburst}},
 volume = {692},
 year = {2024}
}

@article{Linford+17,
 adsurl = {https://ui.adsabs.harvard.edu/abs/2017ApJ...842...73L},
 archiveprefix = {arXiv},
 author = {{Linford}, J.~D. and {Chomiuk}, L. and {Nelson}, T. and {Finzell}, T. and {Walter}, F.~M. and {Sokoloski}, J.~L. and {Mukai}, K. and {Mioduszewski}, A.~J. and {van der Horst}, A.~J. and {Weston}, J.~H.~S. and {Rupen}, M.~P.},
 doi = {10.3847/1538-4357/aa7512},
 eid = {73},
 eprint = {1703.03333},
 journal = {\apj},
 month = {June},
 number = {2},
 pages = {73},
 primaryclass = {astro-ph.HE},
 title = {{The Peculiar Multiwavelength Evolution Of V1535 Sco}},
 volume = {842},
 year = {2017}
}

@article{Margutti&Chornock21,
 adsurl = {https://ui.adsabs.harvard.edu/abs/2021ARA&A..59..155M},
 archiveprefix = {arXiv},
 author = {{Margutti}, Raffaella and {Chornock}, Ryan},
 doi = {10.1146/annurev-astro-112420-030742},
 eprint = {2012.04810},
 journal = {\araa},
 month = {September},
 pages = {155-202},
 primaryclass = {astro-ph.HE},
 title = {{First Multimessenger Observations of a Neutron Star Merger}},
 volume = {59},
 year = {2021}
}

@article{Metzger+15,
 adsurl = {https://ui.adsabs.harvard.edu/abs/2015MNRAS.450.2739M},
 archiveprefix = {arXiv},
 author = {{Metzger}, B.~D. and {Finzell}, T. and {Vurm}, I. and {Hasco{\"e}t}, R. and {Beloborodov}, A.~M. and {Chomiuk}, L.},
 doi = {10.1093/mnras/stv742},
 eprint = {1501.05308},
 journal = {\mnras},
 month = {July},
 number = {3},
 pages = {2739-2748},
 primaryclass = {astro-ph.HE},
 title = {{Gamma-ray novae as probes of relativistic particle acceleration at non-relativistic shocks}},
 volume = {450},
 year = {2015}
}

@article{Metzger+25,
 adsurl = {https://ui.adsabs.harvard.edu/abs/2025ApJ...988..211M},
 archiveprefix = {arXiv},
 author = {{Metzger}, Brian D. and {Lancaster}, Lachlan and {Diesing}, Rebecca},
 doi = {10.3847/1538-4357/ade711},
 eid = {211},
 eprint = {2505.08907},
 journal = {\apj},
 month = {August},
 number = {2},
 pages = {211},
 primaryclass = {astro-ph.HE},
 title = {{Suppression of Shock X-Ray Emission in Novae from Turbulent Mixing with Cool Gas}},
 volume = {988},
 year = {2025}
}

@article{Mikolajewska+21,
 adsurl = {https://ui.adsabs.harvard.edu/abs/2021MNRAS.504.2122M},
 archiveprefix = {arXiv},
 author = {{Miko{\l}ajewska}, J. and {I{\l}kiewicz}, K. and {Ga{\l}an}, C. and {Monard}, B. and {Otulakowska-Hypka}, M. and {Shara}, M.~M. and {Udalski}, A.},
 doi = {10.1093/mnras/stab1058},
 eprint = {2104.06218},
 journal = {\mnras},
 month = {June},
 number = {2},
 pages = {2122-2132},
 primaryclass = {astro-ph.SR},
 title = {{The symbiotic recurrent nova V3890 Sgr: binary parameters and pre-outburst activity}},
 volume = {504},
 year = {2021}
}

@article{Miller91,
 adsurl = {https://ui.adsabs.harvard.edu/abs/1991JAVSO..20..182M},
 author = {{Miller}, La Tanya P.},
 journal = {\jaavso},
 month = {October},
 number = {2},
 pages = {182-184},
 title = {{V3890 Sagittarii Brightness Update}},
 volume = {20},
 year = {1991}
}

@article{Mitrani+25,
 adsurl = {https://ui.adsabs.harvard.edu/abs/2025ApJ...989..166M},
 archiveprefix = {arXiv},
 author = {{Mitrani}, Sharon and {Behar}, Ehud and {Orio}, Marina and {Worley}, Jack},
 doi = {10.3847/1538-4357/adf1a3},
 eid = {166},
 eprint = {2507.02465},
 journal = {\apj},
 month = {August},
 number = {2},
 pages = {166},
 primaryclass = {astro-ph.HE},
 title = {{X-Ray Observations of Nova Sco 2023: Spectroscopic Evidence of Charge Exchange}},
 volume = {989},
 year = {2025}
}

@article{Mohamed&Podsiadlowski12,
 adsurl = {https://ui.adsabs.harvard.edu/abs/2012BaltA..21...88M},
 author = {{Mohamed}, S. and {Podsiadlowski}, Ph.},
 doi = {10.1515/astro-2017-0362},
 journal = {Baltic Astronomy},
 month = {January},
 pages = {88-96},
 title = {{Mass Transfer in Mira-type Binaries}},
 volume = {21},
 year = {2012}
}

@inproceedings{Mohamed+2013,
 adsurl = {https://ui.adsabs.harvard.edu/abs/2013IAUS..281..195M},
 author = {{Mohamed}, S. and {Booth}, R. and {Podsiadlowski}, Ph.},
 booktitle = {Binary Paths to Type Ia Supernovae Explosions},
 doi = {10.1017/S1743921312014998},
 editor = {{Di Stefano}, Rosanne and {Orio}, Marina and {Moe}, Maxwell},
 month = {January},
 pages = {195-198},
 series = {IAU Symposium},
 title = {{The Asymmetric Outflow of RS Ophiuchi}},
 volume = {281},
 year = {2013}
}

@article{Molina+24,
 adsurl = {https://ui.adsabs.harvard.edu/abs/2024MNRAS.534.1227M},
 author = {{Molina}, Isabella and {Chomiuk}, Laura and {Linford}, Justin D. and {Aydi}, Elias and {Mioduszewski}, Amy J. and {Mukai}, Koji and {Sokolovsky}, Kirill V. and {Strader}, Jay and {Craig}, Peter and {Dong}, Dillon and {Harris}, Chelsea E. and {Nyamai}, Miriam M. and {Rupen}, Michael P. and {Sokoloski}, Jennifer L. and {Walter}, Frederick M. and {Weston}, Jennifer H.~S. and {Williams}, Montana N.},
 doi = {10.1093/mnras/stae2093},
 journal = {\mnras},
 month = {October},
 number = {2},
 pages = {1227-1246},
 title = {{The symbiotic recurrent nova V745 Sco at radio wavelengths}},
 volume = {534},
 year = {2024}
}

@article{Morlino&Caprioli12,
 adsurl = {https://ui.adsabs.harvard.edu/abs/2012A&A...538A..81M},
 archiveprefix = {arXiv},
 author = {{Morlino}, G. and {Caprioli}, D.},
 doi = {10.1051/0004-6361/201117855},
 eid = {A81},
 eprint = {1105.6342},
 journal = {\aap},
 month = {February},
 pages = {A81},
 primaryclass = {astro-ph.HE},
 title = {{Strong evidence for hadron acceleration in Tycho's supernova remnant}},
 volume = {538},
 year = {2012}
}

@article{Munari+22,
 adsurl = {https://ui.adsabs.harvard.edu/abs/2022A&A...666L...6M},
 archiveprefix = {arXiv},
 author = {{Munari}, U. and {Giroletti}, M. and {Marcote}, B. and {O'Brien}, T.~J. and {Veres}, P. and {Yang}, J. and {Williams}, D.~R.~A. and {Woudt}, P.},
 doi = {10.1051/0004-6361/202244821},
 eid = {L6},
 eprint = {2209.12794},
 journal = {\aap},
 month = {October},
 pages = {L6},
 primaryclass = {astro-ph.SR},
 title = {{Radio interferometric imaging of RS Oph bipolar ejecta for the 2021 nova outburst}},
 volume = {666},
 year = {2022}
}

@article{Nelson+12,
 adsurl = {http://adsabs.harvard.edu/abs/2012ApJ...748...43N},
 archiveprefix = {arXiv},
 author = {{Nelson}, T. and {Donato}, D. and {Mukai}, K. and {Sokoloski}, J. and 
{Chomiuk}, L.},
 doi = {10.1088/0004-637X/748/1/43},
 eid = {43},
 eprint = {1201.5643},
 journal = {\apj},
 month = {March},
 pages = {43},
 primaryclass = {astro-ph.SR},
 title = {{X-Ray Emission from an Asymmetric Blast Wave and a Massive White Dwarf in the Gamma-Ray Emitting Nova V407 Cyg}},
 volume = {748},
 year = {2012}
}

@article{Nelson+19,
 adsurl = {https://ui.adsabs.harvard.edu/abs/2019ApJ...872...86N},
 archiveprefix = {arXiv},
 author = {{Nelson}, Thomas and {Mukai}, Koji and {Li}, Kwan-Lok and {Vurm}, Indrek and {Metzger}, Brian D. and {Chomiuk}, Laura and {Sokoloski}, J.~L. and {Linford}, Justin D. and {Bohlsen}, Terry and {Luckas}, Paul},
 doi = {10.3847/1538-4357/aafb6d},
 eid = {86},
 eprint = {1901.00030},
 journal = {\apj},
 month = {February},
 number = {1},
 pages = {86},
 primaryclass = {astro-ph.HE},
 title = {{NuSTAR Detection of X-Rays Concurrent with Gamma-Rays in the Nova V5855 Sgr}},
 volume = {872},
 year = {2019}
}

@article{Ness+22,
 adsurl = {https://ui.adsabs.harvard.edu/abs/2022A&A...658A.169N},
 archiveprefix = {arXiv},
 author = {{Ness}, J. -U. and {Beardmore}, A.~P. and {Bezak}, P. and {Dobrotka}, A. and {Drake}, J.~J. and {Vander Meulen}, B. and {Osborne}, J.~P. and {Orio}, M. and {Page}, K.~L. and {Pinto}, C. and {Singh}, K.~P. and {Starrfield}, S.},
 doi = {10.1051/0004-6361/202142037},
 eid = {A169},
 eprint = {2111.00331},
 journal = {\aap},
 month = {February},
 pages = {A169},
 primaryclass = {astro-ph.HE},
 title = {{The super-soft source phase of the recurrent nova V3890 Sgr}},
 volume = {658},
 year = {2022}
}

@article{Nyamai+19,
 adsurl = {https://ui.adsabs.harvard.edu/abs/2019ATel13089....1N},
 author = {{Nyamai}, M.~M. and {Woudt}, P.~A. and {Ribeiro}, V.~A.~R.~M. and {Chomiuk}, L.},
 journal = {The Astronomer's Telegram},
 month = {September},
 pages = {1},
 title = {{Radio detection of the recurrent nova V3890 Sgr with MeerKAT at 1.28 GHz}},
 volume = {13089},
 year = {2019}
}

@article{Nyamai+23,
 adsurl = {https://ui.adsabs.harvard.edu/abs/2023MNRAS.523.1661N},
 archiveprefix = {arXiv},
 author = {{Nyamai}, Miriam M. and {Linford}, Justin D. and {Allison}, James R. and {Chomiuk}, Laura and {Woudt}, Patrick A. and {Ribeiro}, Val{\'e}rio A.~R.~M. and {Sarbadhicary}, Sumit K.},
 doi = {10.1093/mnras/stad1534},
 eprint = {2301.09116},
 journal = {\mnras},
 month = {August},
 number = {2},
 pages = {1661-1675},
 primaryclass = {astro-ph.HE},
 title = {{Synchrotron emission from double-peaked radio light curves of the symbiotic recurrent nova V3890 Sagitarii}},
 volume = {523},
 year = {2023}
}

@article{Orio+20,
 adsurl = {https://ui.adsabs.harvard.edu/abs/2020ApJ...895...80O},
 archiveprefix = {arXiv},
 author = {{Orio}, M. and {Drake}, J.~J. and {Ness}, J. -U. and {Behar}, E. and {Luna}, G.~J.~M. and {Darnley}, M.~J. and {Gallagher}, J. and {Gehrz}, R.~D. and {Kuin}, N.~P.~M. and {Mikolajewska}, J. and {Ospina}, N. and {Page}, K.~L. and {Poggiani}, R. and {Starrfield}, S. and {Williams}, R. and {Woodward}, C.~E.},
 doi = {10.3847/1538-4357/ab8c4d},
 eid = {80},
 eprint = {2004.11263},
 journal = {\apj},
 month = {June},
 number = {2},
 pages = {80},
 primaryclass = {astro-ph.HE},
 title = {{Chandra High Energy Transmission Gratings Spectra of V3890 Sgr}},
 volume = {895},
 year = {2020}
}

@article{Orlando+09,
 adsurl = {https://ui.adsabs.harvard.edu/abs/2009A&A...493.1049O},
 archiveprefix = {arXiv},
 author = {{Orlando}, S. and {Drake}, J.~J. and {Laming}, J.~M.},
 doi = {10.1051/0004-6361:200810109},
 eprint = {0811.3941},
 journal = {\aap},
 month = {January},
 number = {3},
 pages = {1049-1059},
 primaryclass = {astro-ph},
 title = {{Three-dimensional modeling of the asymmetric blast wave from the 2006 outburst of RS Ophiuchi: Early X-ray emission}},
 volume = {493},
 year = {2009}
}

@article{Orlando+17,
 adsurl = {https://ui.adsabs.harvard.edu/abs/2017MNRAS.464.5003O},
 archiveprefix = {arXiv},
 author = {{Orlando}, Salvatore and {Drake}, Jeremy J. and {Miceli}, Marco},
 doi = {10.1093/mnras/stw2718},
 eprint = {1610.05692},
 journal = {\mnras},
 month = {February},
 number = {4},
 pages = {5003-5017},
 primaryclass = {astro-ph.HE},
 title = {{Origin of asymmetries in X-ray emission lines from the blast wave of the 2014 outburst of nova V745 Sco}},
 volume = {464},
 year = {2017}
}

@article{Orlando+25,
 adsurl = {https://ui.adsabs.harvard.edu/abs/2025A&A...704A.144O},
 archiveprefix = {arXiv},
 author = {{Orlando}, S. and {Chomiuk}, L. and {Drake}, J.~J. and {Miceli}, M. and {Bocchino}, F. and {Petruk}, O.},
 doi = {10.1051/0004-6361/202556617},
 eid = {A144},
 eprint = {2507.20334},
 journal = {\aap},
 month = {December},
 pages = {A144},
 primaryclass = {astro-ph.HE},
 title = {{Predicting the X-ray signatures of the imminent T Coronae Borealis outburst through 3D hydrodynamic modeling}},
 volume = {704},
 year = {2025}
}

@article{Ozdonmez+18,
 adsurl = {https://ui.adsabs.harvard.edu/abs/2018MNRAS.476.4162O},
 archiveprefix = {arXiv},
 author = {{{\"O}zd{\"o}nmez}, Aykut and {Ege}, Erg{\"u}n and {G{\"u}ver}, Tolga and {Ak}, Tansel},
 doi = {10.1093/mnras/sty432},
 eprint = {1802.05725},
 journal = {\mnras},
 month = {May},
 number = {3},
 pages = {4162-4186},
 primaryclass = {astro-ph.SR},
 title = {{A new catalogue of Galactic novae: investigation of the MMRD relation and spatial distribution}},
 volume = {476},
 year = {2018}
}

@article{Page+15,
 adsurl = {https://ui.adsabs.harvard.edu/abs/2015MNRAS.454.3108P},
 archiveprefix = {arXiv},
 author = {{Page}, K.~L. and {Osborne}, J.~P. and {Kuin}, N.~P.~M. and {Henze}, M. and {Walter}, F.~M. and {Beardmore}, A.~P. and {Bode}, M.~F. and {Darnley}, M.~J. and {Delgado}, L. and {Drake}, J.~J. and {Hernanz}, M. and {Mukai}, K. and {Nelson}, T. and {Ness}, J. -U. and {Schwarz}, G.~J. and {Shore}, S.~N. and {Starrfield}, S. and {Woodward}, C.~E.},
 doi = {10.1093/mnras/stv2144},
 eprint = {1509.04004},
 journal = {\mnras},
 month = {December},
 number = {3},
 pages = {3108-3120},
 primaryclass = {astro-ph.SR},
 title = {{Swift detection of the super-swift switch-on of the super-soft phase in nova V745 Sco (2014)}},
 volume = {454},
 year = {2015}
}

@article{Page+19a,
 adsurl = {https://ui.adsabs.harvard.edu/abs/2019ATel13084....1P},
 author = {{Page}, K.~L. and {Beardmore}, A.~P. and {Osborne}, J.~P. and {Orio}, M. and {Sokolovsky}, K.~V. and {Darnley}, M.~J.},
 journal = {The Astronomer's Telegram},
 month = {September},
 pages = {1},
 title = {{Swift detection of super-soft X-ray emission from V3890 Sgr}},
 volume = {13084},
 year = {2019}
}

@article{Page+20,
 adsurl = {https://ui.adsabs.harvard.edu/abs/2020MNRAS.499.4814P},
 archiveprefix = {arXiv},
 author = {{Page}, K.~L. and {Kuin}, N.~P.~M. and {Beardmore}, A.~P. and {Walter}, F.~M. and {Osborne}, J.~P. and {Markwardt}, C.~B. and {Ness}, J. -U. and {Orio}, M. and {Sokolovsky}, K.~V.},
 doi = {10.1093/mnras/staa3083},
 eprint = {2010.01001},
 journal = {\mnras},
 month = {December},
 number = {4},
 pages = {4814-4831},
 primaryclass = {astro-ph.HE},
 title = {{The 2019 eruption of recurrent nova V3890 Sgr: observations by Swift, NICER, and SMARTS}},
 volume = {499},
 year = {2020}
}

@article{Polisensky+19,
 adsurl = {https://ui.adsabs.harvard.edu/abs/2019ATel13185....1P},
 author = {{Polisensky}, E. and {Linford}, J.~D. and {Giacintucci}, S. and {Clarke}, T. and {Kassim}, N. and {Sokolovsky}, K.~V. and {van der Horst}, A.~J. and {Rupen}, M. and {Chomiuk}, L. and {Sokoloski}, J.~L. and {Mukai}, K. and {Mioduszewski}, A. and {Aydi}, E. and {Barrett}, P. and {Babul}, A. and {Kawash}, A.},
 journal = {The Astronomer's Telegram},
 month = {October},
 pages = {1},
 title = {{VLITE/VLA Detection of Nova V3890 Sgr Reveals Sub-GHz Turnover}},
 volume = {13185},
 year = {2019}
}

@article{Rupen+08,
 adsurl = {https://ui.adsabs.harvard.edu/abs/2008ApJ...688..559R},
 archiveprefix = {arXiv},
 author = {{Rupen}, Michael P. and {Mioduszewski}, Amy J. and {Sokoloski}, Jennifer L.},
 doi = {10.1086/525555},
 eprint = {0711.1142},
 journal = {\apj},
 month = {November},
 number = {1},
 pages = {559-567},
 primaryclass = {astro-ph},
 title = {{An Expanding Shell and Synchrotron Jet in RS Ophiuchi}},
 volume = {688},
 year = {2008}
}

@article{Sarbadhicary+17,
 adsurl = {https://ui.adsabs.harvard.edu/abs/2017MNRAS.464.2326S},
 archiveprefix = {arXiv},
 author = {{Sarbadhicary}, Sumit K. and {Badenes}, Carles and {Chomiuk}, Laura and {Caprioli}, Damiano and {Huizenga}, Daniel},
 doi = {10.1093/mnras/stw2566},
 eprint = {1605.04923},
 journal = {\mnras},
 month = {January},
 number = {2},
 pages = {2326-2340},
 primaryclass = {astro-ph.HE},
 title = {{Supernova remnants in the Local Group - I. A model for the radio luminosity function and visibility times of supernova remnants}},
 volume = {464},
 year = {2017}
}

@article{Schaefer10,
 adsurl = {http://adsabs.harvard.edu/abs/2010ApJS..187..275S},
 archiveprefix = {arXiv},
 author = {{Schaefer}, B.~E.},
 doi = {10.1088/0067-0049/187/2/275},
 eprint = {0912.4426},
 journal = {\apjs},
 month = {April},
 pages = {275-373},
 primaryclass = {astro-ph.SR},
 title = {{Comprehensive Photometric Histories of All Known Galactic Recurrent Novae}},
 volume = {187},
 year = {2010}
}

@article{Seaquist&Taylor90,
 adsurl = {https://ui.adsabs.harvard.edu/abs/1990ApJ...349..313S},
 author = {{Seaquist}, E.~R. and {Taylor}, A.~R.},
 doi = {10.1086/168315},
 journal = {\apj},
 month = {January},
 pages = {313},
 title = {{The Collective Radio Properties of Symbiotic Stars}},
 volume = {349},
 year = {1990}
}

@article{Seaquist+93,
 adsurl = {https://ui.adsabs.harvard.edu/abs/1993ApJ...410..260S},
 author = {{Seaquist}, E.~R. and {Krogulec}, M. and {Taylor}, A.~R.},
 doi = {10.1086/172742},
 journal = {\apj},
 month = {June},
 pages = {260},
 title = {{A Highly Sensitive Radio Survey of Symbiotic Stars at 3.6 Centimeters}},
 volume = {410},
 year = {1993}
}

@inproceedings{Shepherd_Difmap_1997,
 adsurl = {https://ui.adsabs.harvard.edu/abs/1997ASPC..125...77S},
 author = {{Shepherd}, M.~C.},
 booktitle = {Astronomical Data Analysis Software and Systems VI},
 editor = {{Hunt}, Gareth and {Payne}, Harry},
 month = {January},
 pages = {77},
 series = {Astronomical Society of the Pacific Conference Series},
 title = {{Difmap: an Interactive Program for Synthesis Imaging}},
 volume = {125},
 year = {1997}
}

@article{Sokoloski+06,
 adsurl = {http://adsabs.harvard.edu/abs/2006Natur.442..276S},
 author = {{Sokoloski}, J.~L. and {Luna}, G.~J.~M. and {Mukai}, K. and 
{Kenyon}, S.~J.},
 doi = {10.1038/nature04893},
 eprint = {arXiv:astro-ph/0605326},
 journal = {\nat},
 month = {July},
 pages = {276-278},
 title = {{An X-ray-emitting blast wave from the recurrent nova RS Ophiuchi}},
 volume = {442},
 year = {2006}
}

@article{Sokolovsky+19,
 adsurl = {https://ui.adsabs.harvard.edu/abs/2019ATel13050....1S},
 author = {{Sokolovsky}, K.~V. and {Orio}, M. and {Page}, K.~L. and {Beardmore}, A. and {Osborne}, J.~P. and {Kuin}, P. and {Leahy-McGregor}, J. and {Aydi}, E. and {Chomiuk}, L. and {Kawash}, A. and {Strader}, J. and {Linford}, J.~D. and {Rupen}, M.},
 journal = {The Astronomer's Telegram},
 month = {August},
 pages = {1},
 title = {{Swift X-ray detection during the optical peak of the recurrent nova V3890 Sgr}},
 volume = {13050},
 year = {2019}
}

@article{Starrfield+16,
 adsurl = {https://ui.adsabs.harvard.edu/abs/2016PASP..128e1001S},
 archiveprefix = {arXiv},
 author = {{Starrfield}, S. and {Iliadis}, C. and {Hix}, W.~R.},
 doi = {10.1088/1538-3873/128/963/051001},
 eprint = {1605.04294},
 journal = {\pasp},
 month = {May},
 number = {963},
 pages = {051001},
 primaryclass = {astro-ph.SR},
 title = {{The Thermonuclear Runaway and the Classical Nova Outburst}},
 volume = {128},
 year = {2016}
}

@article{Steinberg&Metzger18,
 adsurl = {https://ui.adsabs.harvard.edu/abs/2018MNRAS.479..687S},
 archiveprefix = {arXiv},
 author = {{Steinberg}, Elad and {Metzger}, Brian D.},
 doi = {10.1093/mnras/sty1641},
 eprint = {1805.03223},
 journal = {\mnras},
 month = {September},
 number = {1},
 pages = {687-702},
 primaryclass = {astro-ph.HE},
 title = {{The multidimensional structure of radiative shocks: suppressed thermal X-rays and relativistic ion acceleration}},
 volume = {479},
 year = {2018}
}

@article{Strader+19,
 adsurl = {https://ui.adsabs.harvard.edu/abs/2019ATel13047....1S},
 author = {{Strader}, J. and {Chomiuk}, L. and {Aydi}, E. and {Kawash}, A. and {Miller}, J. and {Sokolovsky}, K.~V. and {Swihart}, S. and {Stanek}, K. and {Kochanek}, C. and {Shappee}, B.},
 journal = {The Astronomer's Telegram},
 month = {August},
 pages = {1},
 title = {{SOAR spectroscopic confirmation of a new eruption of the recurrent nova V3890 Sgr}},
 volume = {13047},
 year = {2019}
}

@inbook{Synthesis_1999,
 adsurl = {https://ui.adsabs.harvard.edu/abs/1999ASPC..180.....T},
 author = {J.M. Wrobel and R.C. Walker},
 booktitle = {Synthesis Imaging in Radio Astronomy II},
 chapter = {Lecture 9: Sensitivity},
 month = {January},
 pages = {171--185},
 publisher = {Astronomical Society of the Pacific Conference Series},
 series = {Astronomical Society of the Pacific Conference Series},
 title = {{Synthesis Imaging in Radio Astronomy II}},
 volume = {180},
 year = {1999}
}

@article{Tang&Chevalier17,
 adsurl = {https://ui.adsabs.harvard.edu/abs/2017MNRAS.465.3793T},
 archiveprefix = {arXiv},
 author = {{Tang}, Xiaping and {Chevalier}, Roger A.},
 doi = {10.1093/mnras/stw2978},
 eprint = {1607.06391},
 journal = {\mnras},
 month = {March},
 number = {4},
 pages = {3793-3802},
 primaryclass = {astro-ph.HE},
 title = {{Shock evolution in non-radiative supernova remnants}},
 volume = {465},
 year = {2017}
}

@article{Vlasov+16,
 adsurl = {https://ui.adsabs.harvard.edu/abs/2016MNRAS.463..394V},
 archiveprefix = {arXiv},
 author = {{Vlasov}, Andrey and {Vurm}, Indrek and {Metzger}, Brian D.},
 doi = {10.1093/mnras/stw1949},
 eprint = {1603.05194},
 journal = {\mnras},
 month = {November},
 number = {1},
 pages = {394-412},
 primaryclass = {astro-ph.HE},
 title = {{Shocks in nova outflows - II. Synchrotron radio emission}},
 volume = {463},
 year = {2016}
}

@article{Walder+08,
 adsurl = {https://ui.adsabs.harvard.edu/abs/2008A&A...484L...9W},
 archiveprefix = {arXiv},
 author = {{Walder}, R. and {Folini}, D. and {Shore}, S.~N.},
 doi = {10.1051/0004-6361:200809703},
 eprint = {0804.2628},
 journal = {\aap},
 month = {June},
 number = {1},
 pages = {L9-L12},
 primaryclass = {astro-ph},
 title = {{3D simulations of RS Ophiuchi: from accretion to nova blast}},
 volume = {484},
 year = {2008}
}

@article{Walter+12,
 adsurl = {http://adsabs.harvard.edu/abs/2012PASP..124.1057W},
 archiveprefix = {arXiv},
 author = {{Walter}, F.~M. and {Battisti}, A. and {Towers}, S.~E. and {Bond}, H.~E. and 
{Stringfellow}, G.~S.},
 doi = {10.1086/668404},
 eprint = {1209.1583},
 journal = {\pasp},
 month = {October},
 pages = {1057-1072},
 primaryclass = {astro-ph.SR},
 title = {{The Stony Brook/SMARTS Atlas of (mostly) Southern Novae}},
 volume = {124},
 year = {2012}
}

@article{Weiler+02,
 adsurl = {https://ui.adsabs.harvard.edu/abs/2002ARA&A..40..387W},
 author = {{Weiler}, Kurt W. and {Panagia}, Nino and {Montes}, Marcos J. and {Sramek}, Richard A.},
 doi = {10.1146/annurev.astro.40.060401.093744},
 journal = {\araa},
 month = {January},
 pages = {387-438},
 title = {{Radio Emission from Supernovae and Gamma-Ray Bursters}},
 volume = {40},
 year = {2002}
}

@article{Wenzel90,
 adsurl = {https://ui.adsabs.harvard.edu/abs/1990IBVS.3517....1W},
 author = {{Wenzel}, W.},
 journal = {Information Bulletin on Variable Stars},
 month = {September},
 pages = {1},
 title = {{On the Amplitude of the Recurrent Nova V3890 Sagittarii}},
 volume = {3517},
 year = {1990}
}

@article{Weston+16a,
 adsurl = {http://adsabs.harvard.edu/abs/2016MNRAS.457..887W},
 archiveprefix = {arXiv},
 author = {{Weston}, J.~H.~S. and {Sokoloski}, J.~L. and {Metzger}, B.~D. and 
{Zheng}, Y. and {Chomiuk}, L. and {Krauss}, M.~I. and {Linford}, J.~D. and 
{Nelson}, T. and {Mioduszewski}, A.~J. and {Rupen}, M.~P. and 
{Finzell}, T. and {Mukai}, K.},
 doi = {10.1093/mnras/stv3019},
 eprint = {1505.05879},
 journal = {\mnras},
 month = {March},
 pages = {887-901},
 primaryclass = {astro-ph.SR},
 title = {{Non-thermal radio emission from colliding flows in classical nova V1723 Aql}},
 volume = {457},
 year = {2016}
}

@article{Wright&Barlow75,
 adsurl = {https://ui.adsabs.harvard.edu/abs/1975MNRAS.170...41W},
 author = {{Wright}, A.~E. and {Barlow}, M.~J.},
 doi = {10.1093/mnras/170.1.41},
 journal = {\mnras},
 month = {January},
 pages = {41-51},
 title = {{The radio and infrared spectrum of early type stars undergoing mass loss.}},
 volume = {170},
 year = {1975}
}

@article{Yaron+05,
 adsurl = {http://adsabs.harvard.edu/abs/2005ApJ...623..398Y},
 author = {{Yaron}, O. and {Prialnik}, D. and {Shara}, M.~M. and {Kovetz}, A.},
 doi = {10.1086/428435},
 eprint = {arXiv:astro-ph/0503143},
 journal = {\apj},
 month = {April},
 pages = {398-410},
 title = {{An Extended Grid of Nova Models. II. The Parameter Space of Nova Outbursts}},
 volume = {623},
 year = {2005}
}

@ARTICLE{Weibel59,
       author = {{Weibel}, Erich S.},
        title = "{Spontaneously Growing Transverse Waves in a Plasma Due to an Anisotropic Velocity Distribution}",
      journal = {\prl},
         year = 1959,
        month = feb,
       volume = {2},
       number = {3},
        pages = {83-84},
          doi = {10.1103/PhysRevLett.2.83},
       adsurl = {https://ui.adsabs.harvard.edu/abs/1959PhRvL...2...83W}
}

@Article{Skilling75,
  author        = {{Skilling}, J.},
  journal       = {MNRAS},
  title         = {{Cosmic ray streaming. I - Effect of Alfven waves on particles}},
  year          = {1975},
  month         = sep,
  pages         = {557-566},
  volume        = {172},
  url           = {http://adsabs.harvard.edu/abs/1975MNRAS.172..557S},
}

\bsp   
\label{lastpage}

\end{document}